\documentclass[11pt]{article}

\usepackage[T1]{fontenc}
\usepackage{algorithm}
\usepackage{algorithmic}
\usepackage{enumitem}
\usepackage[authoryear,round]{natbib}
\usepackage{caption}
\usepackage{amssymb, blkarray, amsmath,xcolor,xspace,colortbl,rotating} %
\usepackage[raggedrightboxes]{ragged2e} 
\usepackage{textcomp}
\usepackage{appendix}  
\usepackage{bm}  
\usepackage{boxedminipage}  
\usepackage{color}  
\usepackage{endnotes}  
\usepackage{ragged2e}  
\usepackage[onehalfspacing]{setspace}  
\usepackage{tabulary}  
\usepackage{varioref}  
\usepackage{wrapfig}  
\usepackage{graphicx}
\usepackage{subcaption}
\usepackage{float}
\usepackage[hidelinks]{hyperref}
\usepackage[margin=1in]{geometry}
\DeclareGraphicsExtensions {.pdf,.svg,.eps,.ps,.png,.jpg,.jpeg}
\newtheorem {theorem}{Theorem}

\newtheorem {assumption}{Assumption}

\newtheorem {definition}{Definition}
\newtheorem {example}{Example}

\newtheorem {lemma}{Lemma}

\newtheorem {proposition}{Proposition}
\newtheorem {remark}{Remark}

\newenvironment {proof}[1][Proof]{\noindent \textbf {#1.} }{\ \rule {0.5em}{0.5em}}

\newcommand{\Hfunc}{\mathcal{H}} 
\newcommand{\Wtrans}{W} 
\newcommand{\util}{u} 
\newcommand{\cost}{c} 

\newcommand{\norm}[1]{\left\lVert #1 \right\rVert}

\begin{document}

\title{Computing and Learning Stationary Mean Field Equilibria with Low-Dimensional Interactions: Algorithms and Applications}

\author{Bar Light\protect\thanks{Business School and Institute of Operations Research and Analytics, National University of Singapore, Singapore. e-mail: \textsf{barlight@nus.edu.sg} }} 
\maketitle

\thispagestyle{empty}

\noindent \noindent \textsc{Abstract}:
\begin{quote}
Mean field equilibrium (MFE) has emerged as a computationally tractable solution concept for large dynamic games. However, computing MFE remains challenging due to nonlinearities and the absence of contraction properties, limiting its reliability for counterfactual analysis and comparative statics. This paper studies dynamic models in which agents interact through a small number of functions of the population distribution, with particular emphasis on the scalar case. Such low-dimensional interactions naturally arise in a wide range of applications in economics and operations. The main contribution of this paper is to introduce iterative algorithms that leverage this structure and have provide global convergence guarantees for computing and learning MFE under mild assumptions. Unlike existing approaches, our algorithms do not require monotonicity or contraction properties. We also provide model-free algorithms that learn an approximate MFE from simulation using reinforcement learning methods, without requiring prior knowledge of payoff or transition functions. Beyond computation, we establish existence of stationary MFE for non-compact state spaces. For scalar interactions, we also derive analytical comparative statics without requiring monotonicity of the equilibrium mapping. We apply our methods to classical models of dynamic competition, including capacity competition; to heterogeneous-agent macroeconomic models; and to models motivated by online marketplaces and learning, including inventory competition, ridesharing, and social learning.
 The applications illustrate how changes in market parameters affect equilibrium outcomes and how the algorithms can be used for reliable counterfactual analysis.
\end{quote}



\newpage

\section{Introduction}
The rise of digital marketplaces, e-commerce platforms, and blockchain-based systems has led to increasingly complex interactions among large numbers of strategic agents that compete and learn over time. These environments often require dynamic decision-making in response to competition and aggregate market conditions, making strategic behavior difficult to model and analyze. Mean field equilibrium (MFE) has gained prominence as a scalable and computationally tractable alternative to Markov perfect equilibrium (MPE) in such large dynamic games with strategic agents.  While MPE has traditionally served as the benchmark for such settings, its applicability is often limited in practice due to significant computational challenges. In MPE, agents are required to condition their strategies on the exact state of every competitor, leading to an exponential growth in the state space as the number of agents increases. This ``curse of dimensionality" makes the analysis and computation of MPE infeasible in many applications. Moreover, in large populations, the assumption that agents can precisely track and respond to every competitor’s state becomes increasingly unrealistic. 

MFE offers an approximation to MPE that addresses these limitations.\footnote{In an MFE, each agent optimizes her expected discounted payoff under the assumption that the distribution of competitors' states is fixed. The equilibrium distribution of states is then characterized as an invariant distribution of the stochastic process generated by agents' policies. This approximation significantly simplifies the computational burden. The literature has also established that MFE can closely approximate MPE in large games. See \cite{huang2006large}, \cite{lasry2007mean}, and \cite{weintraub2008markov}.}
Despite its advantages, computing MFE poses significant challenges. The nonlinearities inherent in state dynamics and agents' policies, combined with the lack of global contraction properties in many typical applications, often lead to the failure of traditional fixed-point methods to converge. This non-convergence reinforces the need for computational techniques that reliably compute MFEs.  Such techniques are crucial for applications where MFEs are used for counterfactual analysis of policy interventions and systemic changes.  Without reliable computation, MFE-based policy evaluations can produce misleading results.  Providing algorithms that ensure convergence to an MFE is therefore essential for the robustness and credibility of comparative statics and policy evaluation in these applications.

In this paper, we study dynamic models in which agents' payoffs and state transitions depend on the population through a fixed finite number of aggregate variables, such as average production, average wealth, or the fraction of available agents. When there is only one such aggregate variable, we refer to it as a scalar interaction. 
 This structure includes many important applications in economics and operations, including capacity competition models, heterogeneous-agent macroeconomic models, and various platform settings where market conditions such as aggregate demand or supply are captured by a scalar.\footnote{For example, in capacity competition models, firms’  payoffs depend on the aggregate production; in heterogeneous-agent settings, macro-level variables such as average wealth constitute natural scalar aggregates.}

The main contribution of this paper is the development of adaptive algorithms that converge iteratively to approximate MFE  under mild  assumptions. Unlike many existing methods to find algorithms that converge to an MFE, our method does not rely on monotonicity or contraction conditions, nor does it impose the uniqueness of MFE, which is often assumed as a consequence of contraction properties. This allows our approach to be applied in settings where standard fixed-point iteration fails to find an equilibrium, ensuring that MFE can still be computed and learned in a broader range of models.  To the best of our knowledge, these are the first algorithms with global convergence guarantees to an MFE in the class studied in this paper.

We first provide an algorithm for computing an MFE in a full-information setting in which the payoff and transition functions are known. We then provide a data-driven version of the algorithm that uses simulation for settings where the underlying payoff and transition functions are unknown. By integrating reinforcement learning techniques such as Q-learning, alongside Monte Carlo sampling for finding approximate invariant distributions, the algorithm learns the MFE from simulated trajectories. 
Our algorithms exploit the low dimensionality of the interaction. When the interaction is scalar, we update the interaction using a bisection procedure. When it has several dimensions, we develop a simplicial-grid algorithm that searches over the interaction space while learning from simulation how each interaction translates into the interaction generated by the population. We show that the scalar algorithm converges through successive bisection updates, while in the multidimensional case the approximation becomes increasingly accurate as the simplicial grid is refined.

We apply our algorithms to classical models of dynamic competition, including capacity competition and quality ladder models, and show how changes in demand affect equilibrium capacity distributions. We also show that natural fixed-point iteration can fail to converge to an equilibrium even though  agents interact through only a scalar variable in these applications. We then apply our low-dimensional simplicial algorithm to a heterogeneous-agent macroeconomic model with two groups. Households value their wealth relative to the mean wealth of their own group, while macroeconomic variables such as wages and interest rates depend on aggregate capital and therefore on the mean wealth of both groups. We also study applications motivated by online marketplaces and social learning. In the inventory model, we examine how holding costs and transaction fees affect inventories and platform revenue. In a ridesharing model, we study how the payoff from a long trip affects driver availability. In a social learning model, we study how the precision of private information affects the distribution of beliefs.

In addition to the computational results, we provide new theoretical results that exploit the finite-dimensional interaction structure. In particular, we establish existence of a stationary MFE when the individual state space may be non-compact and the interaction has any fixed finite dimension. For scalar interactions, we also derive analytical comparative statics without requiring the interaction generated by the stationary population to be monotone in the  interaction.

The computation and learning of MFE in discrete-time games have been studied extensively in recent years; see \cite{lauriere2022learning} for a survey. \cite{guo2019learning}, \cite{xie2021learning}, \cite{guo2023general}, and \cite{anahtarci2023learning} establish convergence using contraction properties. \cite{perrin2020fictitious} introduce fictitious play under Lasry--Lions uniqueness conditions \citep{lasry2007mean}; see also \cite{cui2024learning}. \cite{subramanian2019reinforcement} study a policy-gradient approach to a local version of MFE. \cite{weintraub2010computational} design algorithms for industry dynamics models, while \cite{adlakha2013mean} analyze supermodular models in which monotone fixed-point iteration can be used. \cite{saldi2023linear} develop algorithms for linear MFE games. Our approach instead uses the low dimension of the interaction to obtain global convergence without these restrictions.

Mean field games have also been applied in operations and economics to auction theory \citep{iyer2014mean,balseiro2015repeated}, dynamic oligopoly \citep{weintraub2008markov,adlakha2015equilibria,aid2025stationary}, heterogeneous-agent macroeconomic models \citep{aiyagari1994uninsured,acemoglu2012}, matching markets \citep{kanoria2021facilitating,arnosti2021managing,besbes2023signaling}, spatial competition \citep{yang2018mean}, experimentation in equilibrium \citep{wager2021experimenting}, contract theory \citep{carmona2021finite}, and blockchain \citep{li2024mean}. We contribute to this literature by providing algorithms and comparative statics for models in which the population distribution affects agents through a small number of aggregate variables.

\section{The Model} \label{Section: model}

This section introduces the mean field model with a fixed, low-dimensional interaction that we use throughout the paper. The model and the definition of an MFE are similar to \cite{adlakha2013mean} and \cite{light2022mean}; the distinguishing feature is that the population distribution affects payoffs and transitions through a small number of aggregate variables.

\textit{Time.} We study a discrete-time setting and index periods by $t=1,2,\ldots$.

\textit{Agents.} There is a continuum of ex-ante identical agents of measure $1$. We use $i$ to denote an agent. The continuum assumption simplifies the exposition because a single agent does not affect the population state. 

\textit{States.} The state of agent $i$ at time $t$ is $x_{i,t}\in X$. In our algorithms, $X$ is finite. We let $\mathcal P(X)$ denote the probability measures on $X$ and use $s_t\in\mathcal P(X)$ for the distribution of agents' states at time $t$. We call $s_t$ the population state.

\textit{Actions.} The action of agent $i$ at time $t$ is $a_{i,t}\in A$, where $A\subseteq\mathbb R^q$. The feasible actions in state $x$ form a nonempty compact set $\Gamma(x)\subseteq A$. We assume that $\Gamma$ is continuous.\footnote{A correspondence is continuous when it is both upper and lower hemicontinuous.}

\textit{Low-dimensional interaction.} We assume that each agent's payoff and transition probabilities depend on the population distribution only through a fixed, finite number of real-valued functions. If the population state is $s$, the interaction is $M(s)$, where $M:\mathcal P(X)\to\mathcal M\subseteq\mathbb R^n$  and $\mathcal M$ is compact. The components of the interaction may be means, moments, or other aggregate market outcomes. When $n=1$, we refer to $M(s)$ as a scalar interaction and write $\mathcal M=[a,b]$.

For example, in a quantity competition model, demand may depend on total production. If a seller in state $x$ produces $\bar q(x)$, then the scalar interaction is $M(s)=\sum_{x\in X}\bar q(x)s(x)$.

\textit{States' dynamics.} If the current state is $x$, the agent chooses $a$, and the interaction is fixed at $m$, let $P(\cdot\mid x,a,m)$ denote the distribution of the next state. A simulator may generate the next state from a shock representation $w(x,a,m,\zeta)$. 

\textit{Payoff.} If an agent is in state $x$, chooses action $a$, and faces population state $s$, her single-period payoff is $\pi(x,a,M(s))$. Agents discount future payoffs by $0<\beta<1$.

In our algorithmic analysis, $X$ is finite. There, we assume that $M$ is continuous,  $\pi(x,a,m)$ is bounded and continuous in $(a,m)$ on the graph of $\Gamma$, and that $P(y\mid x,a,m)$ is continuous in $(a,m)$ for every $x,y\in X$. In our existence result in Section \ref{Section:Existence} we consider possibly non-compact state spaces and provide the needed assumptions in that section.\footnote{In some of the learning algorithms, conditional on all preceding
observations and the current $(x,a,m)$, the
simulated next state has distribution $P(\cdot\mid x,a,m)$ and
the observed payoff is random and has conditional mean $\pi(x,a,m)$. In this case, 
we assume that the conditional second moments of the payoff
observations are uniformly bounded.}

\begin{remark}
For simplicity, we assume that agents are ex-ante homogeneous. However, the model can be readily extended to an ex-ante heterogeneous setting in which each agent has a permanent type, without changing the analysis. We use this extension in the heterogeneous-agent macroeconomic application in Section \ref{Sec:VectorMacroApplication}. In addition, the population state can be represented as a joint distribution over states and actions.   In this
case, the population state is represented by the joint distribution of
states and actions, and $M$ is defined on this distribution.\footnote{In particular, for a stationary policy $\sigma$ and state distribution $s$, we then define the
associated state-action distribution by
$
\mu_{s,\sigma}(x,a)=s(x)\sigma(a\mid x)
$
 and when the interaction depends only on
states, $M(\mu_{s,\sigma})$ means that $M$ is applied to the state marginal
$s$.} We use this
extension in our inventory model. 
We omit the details of these extensions here; see \cite{light2022mean} for a detailed discussion on such extenstions.
\end{remark}

\begin{remark}
The framework can be adapted to finite-agent settings as in \cite{weintraub2008markov}. In that case, MFE approximates Markov perfect equilibrium. 
\end{remark}

To illustrate the notation, we present a dynamic inventory competition model.

\begin{example} \label{Example: Inventory}
We consider a market with many retailers. A retailer's state is her current inventory level. In each period, she chooses an order-up-to level and then faces random demand. Following competitive inventory models such as \cite{lippman1997competitive,mahajan2001inventory,liu2007dynamic,olsen2014markov}, demand for one retailer depends on the inventory decisions of the others.\footnote{The simulations use a stockout-based substitution model based on \cite{olsen2014markov}. When a customer's first-choice retailer stocks out, part of the unmet demand is reallocated to other retailers. Static versions of this mechanism are studied in \cite{lippman1997competitive}.}

Let $X=\{0,\ldots,\overline x\}$ be the inventory state space. At state $x$, the retailer chooses $a\in\Gamma(x)=\{x,\ldots,\overline x\}$. Demand is $D(\zeta,M(s))$, where $\zeta$ is an idiosyncratic shock. The next inventory level is
\begin{equation*}
 x'=(a-D(\zeta,M(s)))_+,
\end{equation*}
where $y_+=\max\{y,0\}$.

The retailer earns revenue from sales and pays production, holding, and shortage costs. If $r$ is the unit price, $h$ is the holding cost, $b$ is the shortage cost, $\tau$ is the retailer's share of sales revenue, and $c(a-x)$ is the production cost, the single-period payoff is
\begin{align*}
\pi(x,a,M(s))
={}&
\tau r\mathbb E_\zeta\min\{a,D(\zeta,M(s))\}-c(a-x)\\
&-h\mathbb E_\zeta\bigl(a-D(\zeta,M(s))\bigr)_+
-b\mathbb E_\zeta\bigl(D(\zeta,M(s))-a\bigr)_+.
\end{align*}
Section \ref{Section:InventorySimulationDetails} gives the precise scalar interaction and the numerical specification used in the simulations.
\end{example}

\subsection{Mean Field Equilibrium}

Informally, an MFE consists of a policy and a population state satisfying two conditions. First, each agent's policy maximizes her expected discounted payoff when she treats the population state as fixed. Second, the fixed population state is an invariant distribution of the state dynamics generated by that policy. 

For a fixed interaction $m$, let $\sigma(da\mid x,m)$ be a probability measure supported on $\Gamma(x)$. Starting from state $x$, its expected discounted payoff is
\begin{equation*}
V_\sigma(x,m)
=
\mathbb E_\sigma\left[\sum_{t=1}^{\infty}\beta^{t-1}\pi(x(t),a(t),m)\right].
\end{equation*}
Let $V(x,m)=\sup_\sigma V_\sigma(x,m)$. Standard discounted dynamic programming arguments (see, e.g., \cite{bertsekas1978stochastic}) imply
\begin{equation}
\label{Equation:BellmanKernel}
V(x,m)
=
\max_{a\in\Gamma(x)}
\left\{
\pi(x,a,m)+\beta\sum_{y\in X}P(y\mid x,a,m)V(y,m)
\right\}.
\end{equation}
We denote the set of maximizers in Equation \eqref{Equation:BellmanKernel} by $G(x,m)$. Thus, an optimal stationary policy assigns probability only to actions in $G(x,m)$. The transition matrix induced by a stationary policy is
\begin{equation}
\label{Equation:RandomizedTransition}
L_{m,\sigma}(x,y)
=
\int_{\Gamma(x)}P(y\mid x,a,m)\,\sigma(da\mid x,m).
\end{equation}

The interpretation is the same as in the usual definition of MFE. An agent conjectures a fixed population state $s$ and solves the decision problem at $m=M(s)$. In equilibrium, the invariant distribution generated by the resulting policy reproduces that same population state.

\begin{definition}
\label{Def MFE}
A stationary policy $\sigma$ and a population state $s\in\mathcal P(X)$ constitute an MFE if:

1. \textit{Optimality:} $V_\sigma(x,M(s))=V(x,M(s))$ for every $x\in X$.

2. \textit{Consistency:} $s$ is an invariant distribution of $L_{M(s),\sigma}$, i.e.,
\begin{equation*}
s(y)=\sum_{x\in X}L_{M(s),\sigma}(x,y)s(x),
\qquad y\in X.
\end{equation*}
\end{definition}

For $\varepsilon\geq0$, an $\varepsilon$-MFE satisfies the same consistency condition and replaces optimality by $V_\sigma(x,M(s))\geq V(x,M(s))-\varepsilon$ for every state.

\section{Algorithms for Computing MFE} \label{Sec:AlgorithmsMFE}

Finding an MFE directly from Definition \ref{Def MFE} is difficult because optimality and consistency must hold simultaneously. The policy is derived while the agent treats the population state as fixed, but that population state must also be an invariant distribution of the dynamics generated by the policy. This circular dependence turns the MFE conditions into a nonlinear fixed-point problem.

A common numerical method to find MFE starts from a population state $s_0$. At iteration $t$, it computes an optimal policy $g_t$ for $s_t$ and updates the population state by one period:
\begin{equation} \label{Eq:fixed}
s_{t+1}(y)=\sum_{x\in X}L_{M(s_t),g_t}(x,y)s_t(x).
\end{equation}
If this sequence converges, its limit is an MFE. However, convergence generally requires contraction or monotonicity conditions that need not hold in applications.

\begin{example} \label{Example:non-convergence}
Assume that $X=\{1,2\}$ and there is one action. Let $M(s)=s(\{2\})$, let $\zeta$ be uniform on $[0,1]$, and set $w(x,a,M(s),\zeta)=1+1_{\{M(s)\leq\zeta\}}$ for both states. The probability of moving to state $1$ equals the fraction of agents in state $2$. If $s_0(\{1\})=\alpha$, Equation \eqref{Eq:fixed} gives $s_t(\{1\})=1-\alpha$ for odd $t$ and $s_t(\{1\})=\alpha$ for even $t$. Thus, the iteration does not converge unless $\alpha=1/2$, which is already the MFE.
\end{example}

Example \ref{Example:non-convergence}  illustrates how, unless the initial population state is an MFE, the fixed-point iteration method fails to converge to the MFE for any other initial population states even in a two-state case. We also show that such non-convergence and cycling can happen in standard dynamic oligopoly models such as the ones we consider in Section \ref{Subsec:Quality ladder}. This  is perhaps unsurprising, as convergence of fixed-point iteration typically requires strong monotonicity and contraction properties for nonlinear operators like those encountered in mean field games, assumptions often made in the learning MFE literature, as discussed in the introduction.\footnote{Note that by standard arguments, slowing the fixed-point iteration by considering \[s_{t+1}(y)
=
(1-\alpha)s_t(y)
+\alpha\sum_{x\in X}L_{M(s_t),g_t}(x,y)s_t(x)\] for some \(\alpha \in (0,1)\) does not resolve the issue, as we can construct simple two-state examples that still exhibit non-convergence under this modification. For example, we could extend Example \ref{Example:non-convergence} and consider \( P_{[0,1]} ((1+\lambda)0.5 - \lambda s_t(\{2\})) \) for some \(\lambda\), where \(P_{[0,1]}\) is the projection onto the set \([0,1]\), as the probability of transitioning to state 2 from either state 1 or 2. This would cause non-convergence due to non-contractiveness for various values of \(\lambda\), which depend on the choice of \(\alpha\).}
 This observation motivates us to explore alternative algorithms  that provide  convergence guarantees for MFE and can be applied in a wide range of applications of interest.

\subsection{Entropy Regularization}
\label{sec:single-value}
In practice, when computing the MFE, we usually consider a finite action space, either discretized via a grid or naturally occurring in the application. In this case, ties can cause an abrupt change in the selected optimal policy as $m$ changes making the policy non-continuous. We address this issue  by adding a small entropy term to the agent's single-period payoff.\footnote{Regularization for stationary mean field games
are studied by \cite{anahtarci2023regularized}. In our algorithms,
regularization ensures continuity of the optimal policy, while
global convergence of the interaction search does not require
the population response to be a contraction.}  
More formally, 
suppose that every $\Gamma(x)$ is finite and nonempty. For a stationary policy $\sigma$, define
\begin{equation*}
H(\sigma(\cdot\mid x,m))
=
-\sum_{a\in\Gamma(x)}\sigma(a\mid x,m)\log\sigma(a\mid x,m),
\end{equation*}
where $0\log0=0$. Fix $\lambda>0$ and add $\lambda H(\sigma(\cdot\mid x,m))$ to the single-period payoff.

For a fixed interaction $m$, the value function for this regularized problem satisfies
\begin{equation}
\label{Eq:RegularizedBellman}
V_\lambda(x,m)
=
\lambda\log\left(
\sum_{a\in\Gamma(x)}
\exp\left(
\frac{\pi(x,a,m)+\beta\int_XV_\lambda(y,m)P(dy\mid x,a,m)}{\lambda}
\right)
\right).
\end{equation}
Define
$
Q_{\lambda,m}^*(x,a)
=
\pi(x,a,m)+\beta\int_XV_\lambda(y,m)P(dy\mid x,a,m).$
The policy that solves the regularized problem is
\begin{equation}
\label{Eq:LogitPolicy}
\sigma_\lambda(a\mid x,m)
=
\frac{\exp(Q_{\lambda,m}^*(x,a)/\lambda)}
{\sum_{\widetilde a\in\Gamma(x)}\exp(Q_{\lambda,m}^*(x,\widetilde a)/\lambda)}.
\end{equation}
The operator on the right-hand side of Equation \eqref{Eq:RegularizedBellman} is a $\beta$-contraction in the sup norm. Hence $V_\lambda$, $Q_{\lambda,m}^*$, and the policy in Equation \eqref{Eq:LogitPolicy} are uniquely determined; see Proposition 2 and Theorem 1 in \cite{geist2019theory}. Continuity of the payoff and transition probabilities implies continuity of these functions in $m$.

We call a stationary policy $\sigma$ and population state $s$ a \emph{regularized MFE} if
$\sigma(\cdot\mid x)=\sigma_\lambda(\cdot\mid x,M(s))$ for every state $x$ and $s$ is invariant under the transition matrix induced by this policy at $M(s)$. Thus, a regularized MFE satisfies the consistency condition in Definition \ref{Def MFE}, while agents optimize the payoff that includes the entropy term. The following result is standard.  All the proofs are deferred to the Appendix.

\begin{proposition}
\label{Prop:logit-approximation}
For every interaction $m$, the policy in Equation \eqref{Eq:LogitPolicy} satisfies
\begin{equation*}
V_{\sigma_\lambda}(x,m)
\geq
V(x,m)-\varepsilon_\lambda
\end{equation*}
for every state, where
$
\varepsilon_\lambda
=\lambda\log A_{\max}/(1-\beta)
$ 
and $A_{\max}=\max_x|\Gamma(x)|$. 
Consequently, a regularized MFE is an $\varepsilon_\lambda$-MFE under Definition \ref{Def MFE}.
\end{proposition}

\subsection{Scalar Interactions: Known Payoffs and Transitions} \label{Section:Value}

We first consider a setting with scalar interactions in which the payoff and transition functions are known.
Let $\sigma_\lambda(\cdot\mid x,m)$ be the logit policy in Equation \eqref{Eq:LogitPolicy} and let $s_\lambda^m$ be its unique invariant distribution. Define
\begin{equation}
\label{Eq:ScalarRegularizedResponse}
F_\lambda(m)=M(s_\lambda^m),
\qquad
f_\lambda(m)=m-F_\lambda(m).
\end{equation}
When the interaction depends on actions, $M(s_\lambda^m)$ in Equation \eqref{Eq:ScalarRegularizedResponse} is evaluated under the stationary state-action distribution obtained from $s_\lambda^m$ and $\sigma_\lambda(\cdot\mid\cdot,m)$. A zero of $f_\lambda$ therefore gives a regularized MFE.

\begin{assumption} \label{Assumption:Unique}
For every $m\in[a,b]$, the Markov chain $L_{m,\sigma_\lambda}$
has a unique invariant distribution $s_\lambda^m$.
\end{assumption}

Assumption \ref{Assumption:Unique} guarantees that the policy associated with a fixed scalar generates one long-run population distribution. This condition is standard in reinforcement learning and Markov-chain analysis and can usually be checked directly from the transition structure; see \cite{bertsekas2019reinforcement} by establishing that $L_{m,\sigma_\lambda}$ is irreducible and aperiodic.

Each iteration of Algorithm \ref{alg:value_iteration_mfe} performs two standard calculations. First, it fixes the scalar interaction at $m_t$ and solves the discounted dynamic program with the entropy term, yielding the policy $\sigma_\lambda(\cdot\mid\cdot,m_t)$. Second, it computes the invariant distribution of the Markov chain induced by this policy. Because the transition probabilities are known and the state space is finite, the second calculation reduces to solving a linear system.

These two calculations determine $F_\lambda(m_t)$ in Equation \eqref{Eq:ScalarRegularizedResponse}. If $f_\lambda(m_t)>0$, the conjectured scalar is larger than the scalar implied by the invariant distribution, so the algorithm retains the lower half of the interval. If $f_\lambda(m_t)<0$, it retains the upper half. Thus, every iteration halves the current interval. The proof of Theorem \ref{Theorem:Value} shows that continuity of $f_\lambda$ preserves an equilibrium in the retained interval.

\begin{algorithm}[H]
\caption{Adaptive Value Function Iteration for Approximate MFE with Scalar Interaction}
\label{alg:value_iteration_mfe}

\textbf{Input:} Bounds $a,b$, regularization parameter $\lambda>0$, and tolerance $\delta\geq0$.

\vspace{1.5mm}
Repeat until $|f_\lambda(m_t)|\leq\delta$:
\begin{enumerate}[leftmargin=1em,itemsep=0.1ex]
    \item Set $m_t=(a+b)/2$.
    \item Solve Equation \eqref{Eq:RegularizedBellman} at $m_t$ and derive the policy in Equation \eqref{Eq:LogitPolicy}.
    \item Compute the unique invariant distribution $s_\lambda^{m_t}$ of $L_{m_t,\sigma_\lambda}$.
    \item Evaluate $F_\lambda(m_t)$ from the policy and invariant distribution obtained in Steps 2 and 3, as specified in Equation \eqref{Eq:ScalarRegularizedResponse}, and set $f_\lambda(m_t)=m_t-F_\lambda(m_t)$. Set $b=m_t$ if $f_\lambda(m_t)>\delta$ and set $a=m_t$ if $f_\lambda(m_t)<-\delta$.
\end{enumerate}
\end{algorithm}

\begin{theorem} \label{Theorem:Value}
Under the continuity assumptions in Section \ref{Section: model}, suppose that Assumption \ref{Assumption:Unique} holds and set $\delta=0$. Then the sequence $\{m_t\}$ generated by Algorithm \ref{alg:value_iteration_mfe} converges to some $m_\lambda^*$ satisfying $f_\lambda(m_\lambda^*)=0$. The policy $\sigma_\lambda(\cdot\mid\cdot,m_\lambda^*)$ and population state $s_\lambda^{m_\lambda^*}$ constitute a regularized MFE.
\end{theorem}

Proposition \ref{Prop:logit-approximation} then shows that the same policy and population state form an approximate MFE under Definition \ref{Def MFE}. 

\begin{remark}[Unique optimal policies]
The entropy term is needed only to avoid discontinuities caused by multiple optimal actions. If the dynamic program without this term has a unique optimal policy that is continuous in $m$, Algorithm \ref{alg:value_iteration_mfe} can use Equation \eqref{Equation:BellmanKernel} directly and the resulting limit is then an (exact) MFE. Section \ref{Subsec:Quality ladder} proves the required uniqueness and continuity for the dynamic oligopoly model.
\end{remark}

As an application of Algorithm~\ref{alg:value_iteration_mfe}, we derive numerical comparative statics results to analyze how changes in the per-unit holding cost influence the equilibrium distribution of retailers' inventories, and we provide an analysis of how the platform's revenue changes with the transaction fees and holding costs it charges retailers in the dynamic inventory competition described in Example~\ref{Example: Inventory}.

E-commerce marketplaces like Amazon, which often provide warehousing and fulfillment services for retailers, typically charge holding costs for inventory storage. Our numerical results below show that higher holding costs shift the equilibrium inventory distribution toward lower levels, as retailers reduce inventory to lower their storage expenses. This also increases the frequency of stockouts and therefore raises the demand received through substitution from other retailers.
Hence, a platform can use lower holding costs to encourage higher inventory levels. Higher inventories may reduce stockouts, shorten delivery times, and increase transaction volume, which generates revenue through transaction fees. We present the simulation results below and provide the full specification in Appendix \ref{Section:InventorySimulationDetails}.

\begin{figure}[H]
    \centering
    \begin{subfigure}[b]{0.38\textwidth}
        \includegraphics[width=\textwidth, height=3.8cm]{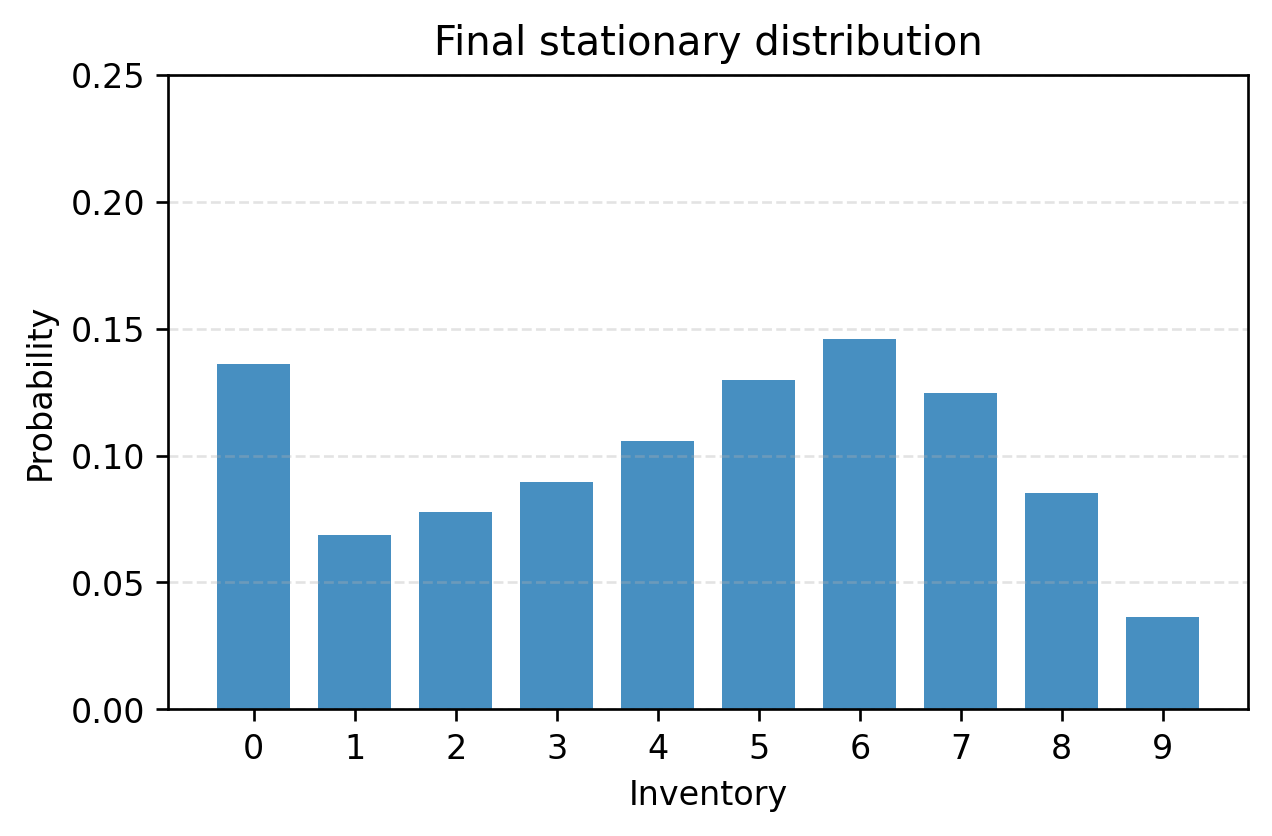}
        \caption{Holding cost = 2}
        \label{fig:holding_cost_2}
    \end{subfigure}
    \hfill
     \begin{subfigure}[b]{0.38\textwidth}
        \includegraphics[width=\textwidth, height=3.8cm]{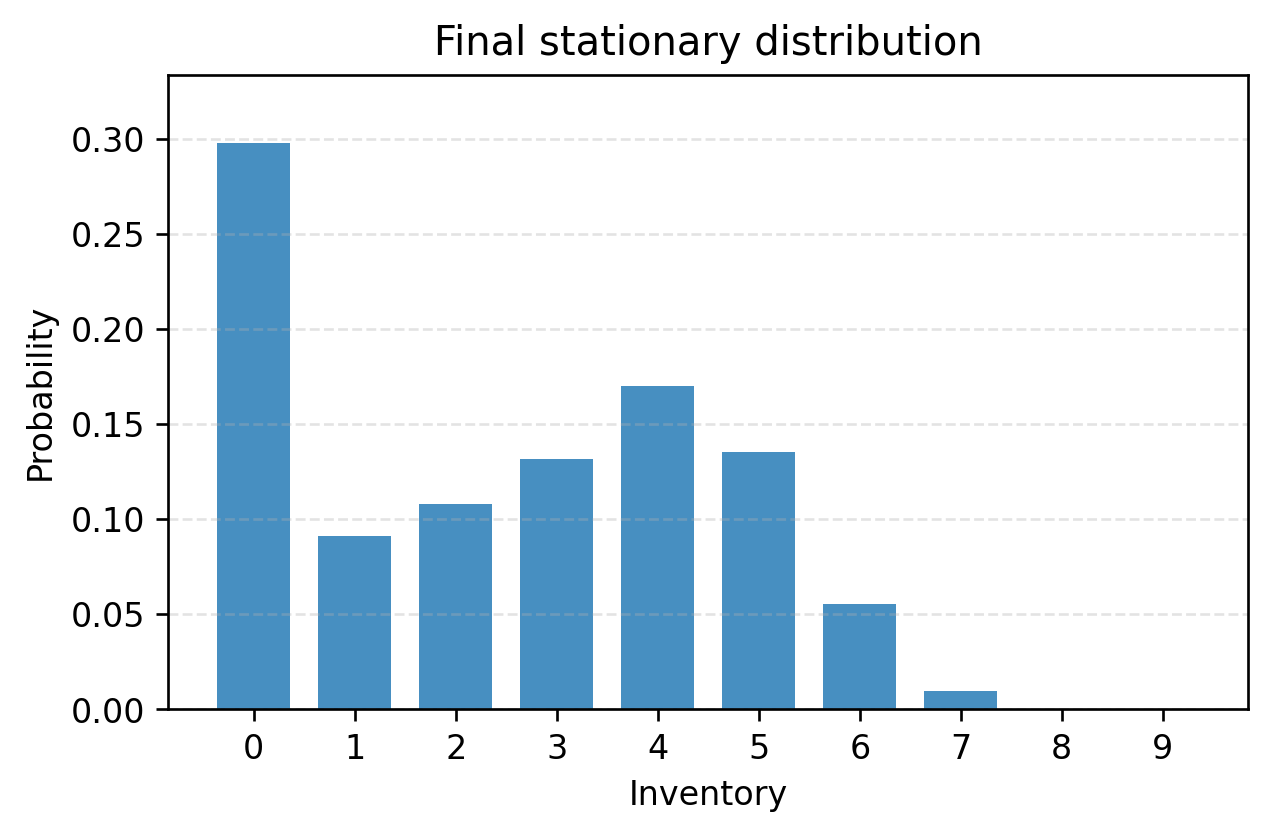}
        \caption{Holding cost = 8}
        \label{fig:holding_cost_8}
    \end{subfigure}
    \caption{Approximate equilibrium inventory distributions under different holding costs using Algorithm \ref{alg:value_iteration_mfe}. As holding costs increase, retailers hold less inventory.}
    \label{fig:holding_cost_comparative_statics}
\end{figure}

We then conduct a market-design exercise in which the platform selects the inventory holding cost $h$ and the transaction fee $1-\tau$ to maximize its revenue. For each pair $(h,\tau)$ on a grid, we compute an approximate MFE using Algorithm \ref{alg:value_iteration_mfe}. Platform revenue consists of commission revenue, equal to the platform's share of retailer sales, and holding-fee revenue. Both components depend on the inventory and sales levels generated in equilibrium.

\begin{figure}[]
    \centering
    \includegraphics[width=0.8\linewidth, height=6cm]{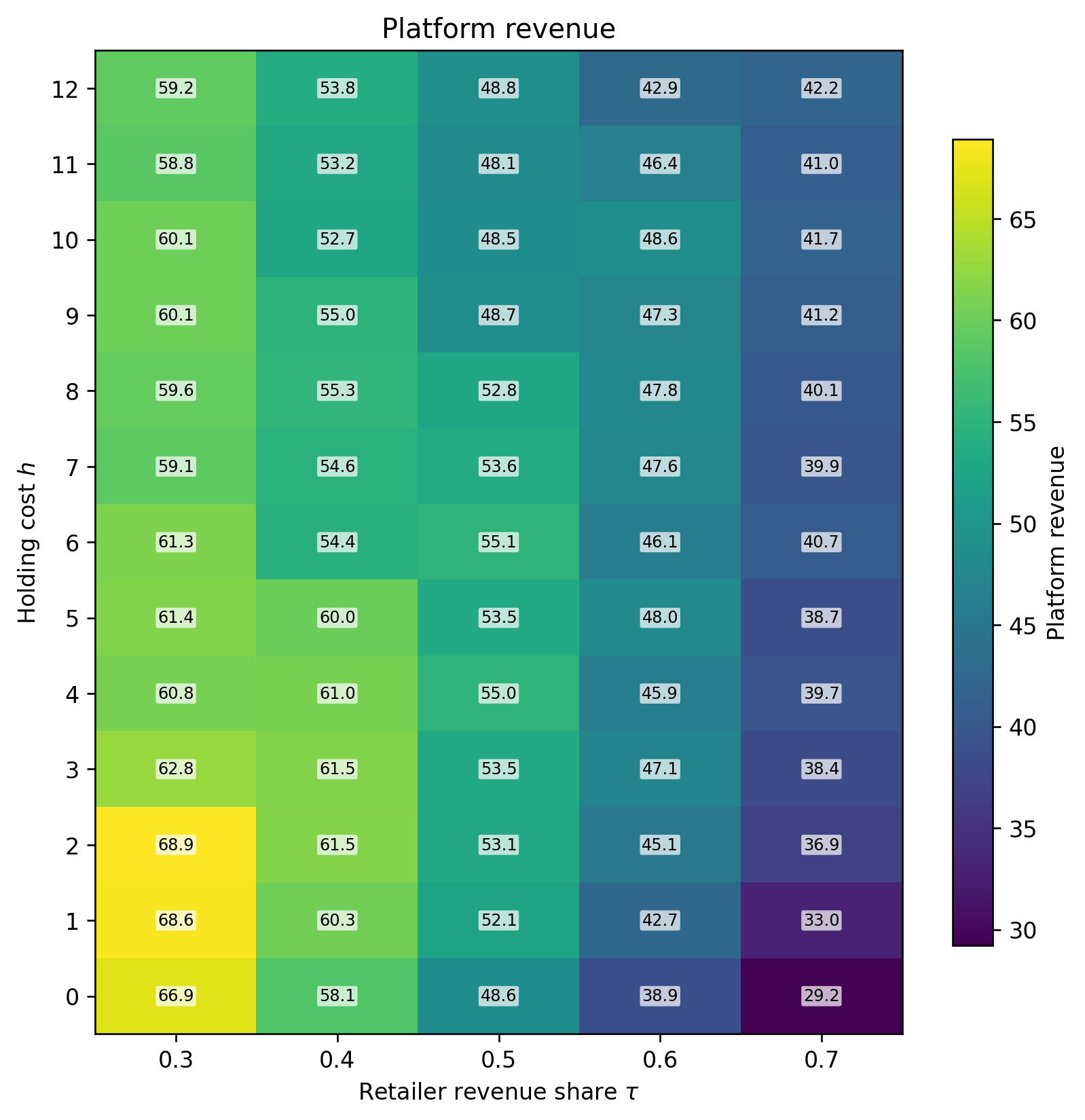}
    \caption{Platform revenue as a function of inventory holding cost $h$ (vertical axis) and retailer revenue share $\tau$ (horizontal axis). Each cell displays the platform's equilibrium revenue for a given $(h,\tau)$ pair. Lighter colors indicate higher revenue.}
    \label{fig:PlatformRevenueHeatmap}
\end{figure}

Figure \ref{fig:PlatformRevenueHeatmap} summarizes the results.
Higher platform revenue is generally concentrated toward the lower-left
region, which corresponds to higher transaction fees and relatively low
holding costs. Lower storage costs encourage retailers to hold more
inventory, reducing stockouts and supporting transaction volume.
Thus, a platform that controls both fees may find it profitable to
combine a relatively high transaction fee with inexpensive storage.

This exercise requires computing an approximate MFE for $65$ pairs of fees. Algorithm \ref{alg:value_iteration_mfe} typically reaches the required tolerance within $10$--$15$ bisection iterations. 

\subsection{Scalar Interactions: Unknown Payoffs and Transitions} \label{Sec:Q-learning}

We next introduce an algorithm that does not require knowledge of the payoff or transition functions. It differs from Algorithm \ref{alg:value_iteration_mfe} in two steps. First, at a fixed scalar $m_t$, it uses Q-learning to estimate the solution of the regularized dynamic program from simulated rewards and transitions. Second, it estimates the invariant distribution by simulating a separate state trajectory under the policy obtained from the Q-function. Thus, it replaces the dynamic-program and invariant-distribution calculations in Algorithm \ref{alg:value_iteration_mfe} with estimates based on observations from a simulator.
Equation \eqref{Eq:RegularizedQUpdate} in the Appendix gives the Q-learning update and Equation \eqref{Eq:LogitPolicy} converts the limiting Q-function into the randomized policy used to estimates the invariant distribution generated by that policy.  The following assumption is used to guarantee the convergence of the Q-learning algorithm.

\begin{assumption}
\label{Assumption:ScalarExploration}
For every fixed $m\in[a,b]$, every feasible state-action pair is
updated infinitely often, and its step sizes satisfy 
\begin{equation*}
\sum_{h=0}^{\infty}\gamma_h(x,a)=\infty,
\qquad
\sum_{h=0}^{\infty}\gamma_h(x,a)^2<\infty.
\end{equation*}
\end{assumption}

\begin{algorithm}[H]
\caption{Adaptive Q-Learning for Approximate MFE with Scalar Interaction}
\label{alg:bisection_q_learning}

\textbf{Input:} Bounds $a,b$ samples $K,H$, regularization parameter $\lambda>0$, and tolerance $\delta>0$.

\vspace{2mm}
Repeat until $|\widehat f_\lambda(m_t)|\leq\delta$:
\begin{enumerate}[leftmargin=1em,itemsep=0.1ex]
    \item \textbf{Set:} $m_t=(a+b)/2$.
    \item \textbf{Q-learning:} Apply Equation \eqref{Eq:RegularizedQUpdate} at $m_t$ for $H$ updates and use the resulting Q-function in Equation \eqref{Eq:LogitPolicy} to obtain $\widehat\sigma_t$.
    \item \textbf{Sampling:} Starting from an arbitrary state, simulate $K$ transitions under $L_{m_t,\widehat\sigma_t}$ and set
    \begin{equation*}
    \widehat s_t(y)=\frac{1}{K}\sum_{k=1}^{K}1_{\{x_k=y\}},
    \qquad y\in X.
    \end{equation*}
    \item \textbf{Evaluate and update:} Define
    \begin{equation*}
    \widehat\mu_t(x,a)=\widehat s_t(x)\widehat\sigma_t(a\mid x)
    \end{equation*}
    and set $\widehat F_t(m_t)=M(\widehat\mu_t)$, where $M$ is applied only to the state marginal when the interaction depends only on states. Set $\widehat f_\lambda(m_t)=m_t-\widehat F_t(m_t)$. Set $b=m_t$ if $\widehat f_\lambda(m_t)>\delta$ and set $a=m_t$ if $\widehat f_\lambda(m_t)<-\delta$.
\end{enumerate}
\end{algorithm}

Algorithm \ref{alg:bisection_q_learning} uses finite values of $H$ and $K$. The asymptotic result considers the following sequence. At iteration $t$, keep $m_t$ fixed while the Q-functions in Equation \eqref{Eq:RegularizedQUpdate} converge to $Q_{\lambda,m_t}^*$. Use Equation \eqref{Eq:LogitPolicy} to form the corresponding policy, keep that policy fixed, and continue the separate population simulation until its empirical distribution converges to $s_\lambda^{m_t}$. The bisection interval is then updated using $f_\lambda(m_t)=m_t-F_\lambda(m_t)$. Appendix \ref{Section:Finite-time} studies the errors when only $H$ Q-learning updates and $K$ population observations are used.

\begin{theorem}
\label{thm:Q_convergence}
Under the continuity assumptions in Section \ref{Section: model}, suppose that Assumptions \ref{Assumption:Unique} and \ref{Assumption:ScalarExploration} hold and set $\delta=0$. At iteration $t$, keep $m_t$ fixed and let the Q-functions in Equation \eqref{Eq:RegularizedQUpdate} converge. Form the policy using Equation \eqref{Eq:LogitPolicy}, keep that policy fixed, and let the empirical population distribution in Step 3 converge before updating the bisection interval. The resulting sequence $\{m_t\}$ converges almost surely to some $m_\lambda^*$ satisfying $f_\lambda(m_\lambda^*)=0$. The policy $\sigma_\lambda(\cdot\mid\cdot,m_\lambda^*)$ and population state $s_\lambda^{m_\lambda^*}$ constitute a regularized MFE.
\end{theorem}

\begin{remark} \label{Remark:PolicyGrad}
Because the scalar is held fixed while the agent's decision problem is solved, Q-learning in Step 2 can be replaced by another method. If that method converges to the policy in Equation \eqref{Eq:LogitPolicy} for every fixed scalar, the  proof of Theorem \ref{thm:Q_convergence} is unchanged. Suitable double Q-learning and policy-gradient methods may therefore be used under the assumptions required by their convergence results; see \cite{hasselt2010double,agarwal2021theory,bhandari2024global}.
\end{remark}

We apply Algorithm \ref{alg:bisection_q_learning} to the inventory competition model using the same parameter values as in Section \ref{Section:Value}. Appendix \ref{Section:InventorySimulationDetails} shows the numerical comparative statics.

\subsection{Low-Dimensional Interactions}
\label{sec:Multi}

We now consider models in which agents respond to several aggregate variables. Let the interaction be $M(s)\in\mathcal M\subseteq\mathbb R^n$, where $n$ is fixed and small. The set $\mathcal M=\prod_{\ell=1}^n[a_\ell,b_\ell]$ is a compact box. The state space and feasible action sets remain finite, and the payoff and transition probabilities satisfy the continuity assumptions in Section \ref{Section: model} with $m\in\mathcal M$.

When $n=1$, Algorithm \ref{alg:value_iteration_mfe} or Algorithm \ref{alg:bisection_q_learning} uses the sign of $f_\lambda(m)$ to identify how we should update the interaction. There is generally no corresponding update in the multi-dimensional case. Our approach in this section is to approximate the function that maps an interaction vector  into the interaction implied by the resulting invariant distribution, and then find a fixed point of that approximation using the simplicial approach in \cite{scarf1967fixedpoint}. The challenge here is that the function $F_\lambda$ is not known and must be estimated from simulator observations at selected vertices.

For a fixed $m\in\mathcal M$, let $\sigma_\lambda(\cdot\mid\cdot,m)$ be the policy in Equation \eqref{Eq:LogitPolicy} and let $s_\lambda^m$ be its unique invariant distribution. If the interaction depends on actions, define
$
\mu_\lambda^m(x,a)=s_\lambda^m(x)\sigma_\lambda(a\mid x,m)$ and 
if it depends only on states, $M(\mu_\lambda^m)$ means that $M$ is applied to the state marginal. Define
\begin{equation}
\label{Eq:VectorResponse}
F_\lambda(m)=M(\mu_\lambda^m),
\qquad
f_\lambda(m)=m-F_\lambda(m).
\end{equation}
The policy $\sigma_\lambda(\cdot\mid\cdot,m)$ and population state
$s_\lambda^m$ constitute a regularized MFE exactly when
$F_\lambda(m)=m$. The algorithm estimates $F_\lambda$ only at selected
grid vertices and retains the estimates obtained at each vertex across
visits. The following assumption corresponds to Assumptions
\ref{Assumption:Unique} and \ref{Assumption:ScalarExploration} in the
scalar case.

\begin{assumption}
\label{Assumption:Vector}
The following conditions hold:
\begin{enumerate}[label=(\roman*),leftmargin=2em,itemsep=0.2ex]
    \item the Markov chain $L_{m,\sigma_\lambda}$ has a unique invariant
    distribution $s_\lambda^m$ for every $m\in\mathcal M$;
    \item when Q-learning is used, at every grid vertex selected
    infinitely often, every feasible state-action pair is updated
    infinitely often and its step sizes satisfy the two conditions in
    Assumption \ref{Assumption:ScalarExploration}.
\end{enumerate}
\end{assumption}

We now describe how we approximate $F_\lambda$ on a simplicial grid. 
Fix a conforming partition $\mathcal T_h$ of $\mathcal M$ into simplices whose diameters are at most $h>0$, and denote its vertices by $m^1,\ldots,m^J$. At vertex $m^j$, let $\widehat F_t^j\in\mathcal M$ be the current estimate of $F_\lambda(m^j)$. Linear interpolation of these vertex values within every simplex defines a continuous function $\widehat F_t:\mathcal M\to\mathcal M$.
Consider a simplex with vertices $m^{j_0},\ldots,m^{j_n}$. A point in this simplex is a fixed point of the interpolated function if there are weights $\theta_0,\ldots,\theta_n$ satisfying
\begin{equation}
\label{Eq:SimplicialFixedPoint}
\theta_\ell\geq0,
\qquad
\sum_{\ell=0}^n\theta_\ell=1,
\qquad
\sum_{\ell=0}^n\theta_\ell\left(\widehat F_t^{j_\ell}-m^{j_\ell}\right)=0.
\end{equation}
For any fixed simplex, Equation \eqref{Eq:SimplicialFixedPoint}
is a linear feasibility problem in the $n+1$ weights. Equivalently,
it asks whether zero belongs to the convex hull of the vectors
$
\widehat F_t^{j_\ell}-m^{j_\ell}
$, $ \ell=0,\ldots,n.
$ 
If the equation is feasible, then
$\widetilde m=\sum_{\ell=0}^n\theta_\ell m^{j_\ell}$ is a fixed point
of the linear interpolation of $\widehat F_t$ on that simplex.
Because each vertex value $\widehat F_t^j$ lies in $\mathcal M$,
the interpolated function maps $\mathcal M$ into itself. Brouwer's
fixed-point theorem therefore implies that at least one simplex
satisfies Equation \eqref{Eq:SimplicialFixedPoint}.

\begin{algorithm}[H]
\caption{Online Simplicial Learning for Approximate MFE with Low-Dimensional Interactions}
\label{alg:online_simplicial_mfe}

\textbf{Input:} A simplicial partition $\mathcal T_h$, regularization parameter $\lambda>0$, and integers $H,K\geq1$.

\vspace{1mm}
Choose any $m^0\in\mathcal M$ and set $\widehat F_0^j=m^0$ at every grid vertex. At a vertex, initialize the policy estimate and population trajectory when that vertex is first selected.

\vspace{1mm}
For $t=1,2,\ldots$:
\begin{enumerate}[leftmargin=1em,itemsep=0.3ex]
    \item \textbf{Select a simplex.} Find vertices $m^{j_{t,0}},\ldots,m^{j_{t,n}}$ and weights satisfying Equation \eqref{Eq:SimplicialFixedPoint} for $\widehat F_{t-1}$. Let $\mathcal I_t=\{j_{t,0},\ldots,j_{t,n}\}$.
    \item \textbf{Update the estimates at its vertices.} For every $j\in\mathcal I_t$, use $H$ additional observations to continue the stored Q-learning sequence or the stored reward and transition estimates, and denote the resulting policy by $\widehat\sigma_t^j$. Continue a separate population trajectory for $K$ transitions under
$L_{m^j,\widehat\sigma_t^j}$. If $N_j(t)$ is the total number of population observations collected at $m^j$, set
    \begin{equation*}
    \widehat s_t^j(y)=\frac{1}{N_j(t)}\sum_{k=1}^{N_j(t)}1_{\{x_k^j=y\}},
    \qquad y\in X.
    \end{equation*}
    Define $\widehat\mu_t^j(x,a)=\widehat s_t^j(x)\widehat\sigma_t^j(a\mid x)$ and set $\widehat F_t^j=M(\widehat\mu_t^j)$. Keep the estimates at every unselected vertex unchanged.
    \item \textbf{Define the current output.} Choose
    \begin{equation*}
    \widehat j_t\in\operatorname*{argmin}_{j\in\mathcal I_t}\norm{m^j-\widehat F_t^j},
    \end{equation*}
    using a fixed tie-breaking rule, and set 
    $\widehat m_t=m^{\widehat j_t}$,
    $\widehat\sigma_t=\widehat\sigma_t^{\widehat j_t}$, and
    $\widehat s_t=\widehat s_t^{\widehat j_t}$.
\end{enumerate}
\end{algorithm}

Let
\begin{equation*}
\mathcal E_\lambda
=
\{(m,\sigma_\lambda(\cdot\mid\cdot,m),s_\lambda^m):F_\lambda(m)=m\}.
\end{equation*}
For two elements $(m,\sigma,s)$ and $(m',\sigma',s')$, measure their distance by
\begin{equation}
\label{Eq:VectorTripleDistance}
\norm{m-m'}+
\max_{x\in X}\norm{\sigma(\cdot\mid x)-\sigma'(\cdot\mid x)}_1+
\norm{s-s'}_1.
\end{equation}
Thus, closeness to $\mathcal E_\lambda$ means that the interaction, policy, and population state are all close to some regularized MFE.

\begin{theorem}[Low-dimensional interactions]
\label{thm:online_simplicial_mfe}
Suppose Assumption \ref{Assumption:Vector} holds and Step 2 of
Algorithm \ref{alg:online_simplicial_mfe} uses Q-learning.
Then, for every $\epsilon>0$, there exists $h_\epsilon>0$ such that,
if $h\leq h_\epsilon$, the interaction, policy, and population state
provided in Step 3 are within $\epsilon$ of $\mathcal E_\lambda$
at every sufficiently large iteration, with probability one.
\end{theorem}

Note that Theorem \ref{thm:online_simplicial_mfe} does not require a
unique equilibrium.  
Also note that the number of grid points depends on the interaction
dimension. If every coordinate is divided into $q$ intervals, the
standard triangulation of the resulting grid has $(q+1)^n$ vertices
and $n!q^n$ simplices. To find the simplex in Step 1, one can test
Equation \eqref{Eq:SimplicialFixedPoint} successively over the
simplices. Each test is a linear feasibility problem with only $n+1$
variables, and Brouwer's theorem guarantees that at least one test is
feasible. In practice, the simplex selected at the previous iteration can be
tested first; if it is infeasible, a complete search over the remaining
simplices can be used. Once a simplex is selected, only its $n+1$
vertices are updated, and these calculations can be performed in
parallel. Thus, the algorithm is practical when $n$ is small, which the setting targeted in this paper.

\section{Existence of MFE in General State Spaces} \label{Section:Existence}

In this section we establish existence when the individual state space is allowed to be non-compact.  In particular we assume the following. 
Let $X$ be a proper Polish space with metric $d$; that is, $X$ is complete and separable and every closed bounded subset is compact. Let the finite action set be $A$, let $\Gamma(x)$ be nonempty and continuous, and let $\mathcal M\subseteq\mathbb R^n$ be compact and convex. The scalar case corresponds to $\mathcal M=[a,b]$. The interaction function $M$ takes the probability measures considered below to $\mathcal M$. The payoff is bounded and continuous. The transition kernel is weakly continuous: for every bounded continuous function $v$, the map
\begin{equation*}
(x,a,m)\longmapsto\int_Xv(y)P(dy\mid x,a,m)
\end{equation*}
is continuous on the graph of $\Gamma$. Definition \ref{Def MFE} extends to this setting by replacing the finite sums with integrals. We write $s_k\Rightarrow s$ when $s_k$ converges weakly to $s$.

For $\lambda>0$ and $m\in\mathcal M$, let $\sigma_\lambda(\cdot\mid x,m)$ be the policy obtained from the entropy-regularized dynamic program, and let $s_\lambda^m$ be its invariant distribution. In assumption \ref{Assumption: exist}, Condition (i) identifies the long-run population distribution for each $m$ and $\lambda$,  Condition (ii) prevents these distributions from placing increasing mass arbitrarily far from a fixed state as $m$ varies and $\lambda$ converges to zero, and Condition (iii) ensures that the aggregate interaction is preserved when the invariant distributions converge.

\begin{assumption} \label{Assumption: exist}
There are $\bar\lambda>0$, $p\geq1$, $C<\infty$, and $x_0\in X$ such that:
\begin{enumerate}[label=(\roman*),leftmargin=2em,itemsep=0.2ex]
    \item for every $m\in\mathcal M$ and $0<\lambda\leq\bar\lambda$, the transition kernel induced by $\sigma_\lambda(\cdot\mid\cdot,m)$ has a unique invariant distribution $s_\lambda^m$;
    \item we have 
    \begin{equation*}
    \sup_{0<\lambda\leq\bar\lambda}\sup_{m\in\mathcal M}
    \int_Xd(x,x_0)^p s_\lambda^m(dx)\leq C;
    \end{equation*}
    \item if $s_k\Rightarrow s$ and $\sup_k\int_Xd(x,x_0)^p s_k(dx)\leq C$, then $M(s_k)\to M(s)$.
\end{enumerate}
\end{assumption}

Condition (iii) is immediate when $M$ is bounded and continuous under weak convergence. It also covers unbounded moment interactions. For example, if $M(s)=\int_Xk(x)s(dx)$, $k$ is continuous, and $|k(x)|\leq C_k(1+d(x,x_0)^r)$ for some $r<p$, then Condition (ii) implies the convergence required in Condition (iii).\footnote{\cite{wiecek2024multiple} prove an existence result in a more general
setting that does not require a finite-dimensional interaction, but their
stationary existence theorem requires assumptions that are not implied by
ours. In particular, their result requires setwise continuity of the
transition kernel. By exploiting the finite-dimensional interaction, we
require only weak continuity. This difference can be important in
applications: for example, a deterministic transition
$x'=g(x,a,m)$ with continuous $g$ is weakly continuous but generally
not setwise continuous.}

\begin{theorem} \label{thm:existence}
Suppose Assumption \ref{Assumption: exist} holds. Then there exist a stationary randomized policy $\sigma^*$ and a population state $s^*$ that constitute an MFE.
\end{theorem}

Note that Theorem \ref{thm:existence} proves the existence of an exact MFE so it concerns the model in Definition \ref{Def MFE}, without the entropy term.

\section{Analytical Comparative Statics Results} \label{Section:ComparativeStaticsMain}

The algorithms in Section \ref{Sec:AlgorithmsMFE}  can be used to compute numerical comparative statics. We now leverage the scalar interaction structure to prove analytical comparative statics result. 

Let $(Z,\geq_Z)$ be a partially ordered parameter space, and let $z\in Z$ be a model parameter. For example, $z$ may be the discount factor, a cost or demand parameter, or a parameter affecting the transition probabilities. For every $m\in[a,b]$ and $z\in Z$, suppose in this section that the optimal policy $g(\cdot,m,z)$ is unique and the induced Markov chain has a unique invariant distribution $s^{m,z}$. Define
\begin{equation*}
F(m,z)=M(s^{m,z}).
\end{equation*}

Suppose $X$ has a closed partial order. We write $s_2\succeq_{SD}s_1$ if $s_{2}$ stochastically dominates $s_{1}$, i.e., $s_2(B)\geq s_1(B)$ for every measurable upper set $B$, where $B$ is an upper set if $x_1\in B$ and $x_2\geq x_1$ imply $x_2\in B$.

Proposition \ref{Prop:comp} shows that it is enough to compare the invariant distributions at each fixed scalar $m$ to derive comparative statics results. In particular, if a higher value of $z$ shifts these distributions upward and $M$ is increasing under stochastic dominance, then both the lowest and highest equilibrium scalar interactions increase. The proof uses only continuity in $m$ and the total order on $[a,b]$ and applies Theorem 1 in \cite{milgrom1994comparing}. We include the argument for completeness in the Appendix. 

\begin{proposition} \label{Prop:comp}
Suppose $F(\cdot,z)$ is continuous and maps $[a,b]$ into itself for every $z\in Z$. If $s^{m,z_2}\succeq_{SD}s^{m,z_1}$ whenever $z_2\geq_Zz_1$ and $M$ is increasing with respect to stochastic dominance, then the lowest $\underline m(z)$ and highest $\overline m(z)$ equilibrium scalar interactions exist and are non-decreasing in $z$.
\end{proposition}

Importantly, to apply Proposition \ref{Prop:comp}, $F(m,z)$ need not be monotone in $m$. Hence, the proposition 
 differs from comparative statics based on supermodularity, such as \cite{adlakha2013mean}, because it does not assume that the MFE mapping is increasing in the population state. The result in \cite{light2022mean} also avoids that monotonicity but relies on uniqueness of MFE. Proposition \ref{Prop:comp} does not require equilibrium uniqueness and is similar in spirit to the aggregator approach in \cite{acemoglu2012}.\footnote{The proposition derives comparative statics result for the equilibrium scalar interaction. In the capacity competition model, for example, this scalar is average production, which is the economically meaningful outcome in the model.}

The main challenge in applying Proposition \ref{Prop:comp} is to prove the required order for the invariant distributions. This typically requires monotonicity of the policy and of the transition kernel. In Section \ref{Subsec:Quality ladder} we prove these steps for classical dynamic oligopoly models with a discrete, unbounded state space.

\section{Classical Dynamic Oligopoly Models}\label{Subsec:Quality ladder}

In this section, we study dynamic capacity competition and quality ladder models, both widely used in operations, economics, and  industrial organization \citep{pakes1994computing,weintraub2008markov}. A firm's state represents an economically important characteristic, such as production capacity or product quality. Each period, firms earn profits from a static product-market game and choose investments that affect their future states.

\textit{States.} The state of firm $i$ at time $t$ is $x_{i,t}\in X$, where $X=\{0,1,\ldots\}$ is the unbounded set of non-negative integers. This state space is standard in the analysis of these models \citep{weintraub2008markov}.

\textit{Actions.} Firm $i$ chooses investment $a_{i,t}\in A=[\underline a,\overline a]$, where $0<\underline a<\overline a$. We normalize $\overline a=1$.

\textit{States' dynamics.} Investment may raise the firm's state, while depreciation may lower it. A successful investment raises the state by one level. Its probability is $a/(1+a)$. Independently, depreciation occurs with probability $\delta\in(0,1)$. These dynamics are commonly used in quality ladder and capacity models; see, for example, \cite{weintraub2010computational}.
For $x\geq1$, the transition probabilities are
\begin{equation} \label{Eq:Oligopoly}
W(x,a,y)=
\begin{cases}
\dfrac{(1-\delta)a}{1+a}, & y=x+1,\\[1.5mm]
\dfrac{1-\delta+\delta a}{1+a}, & y=x,\\[1.5mm]
\dfrac{\delta}{1+a}, & y=x-1,\\[1.5mm]
0, & \text{otherwise}.
\end{cases}
\end{equation}
At the lower boundary,
\begin{equation*}
W(0,a,1)=\frac{(1-\delta)a}{1+a},
\qquad
W(0,a,0)=1-W(0,a,1).
\end{equation*}

\textit{Payoff.} The investment cost is $c(a)$. The single-period operating profit is $u(x,M(s))$ that is typically generated by a static game between the firms. The firm's single-period payoff is
\begin{equation*}
\pi(x,a,M(s))=u(x,M(s))-c(a).
\end{equation*}

We next describe two standard static competition models that produce this payoff structure.

The first is a capacity competition model based on \cite{besanko2004capacity} and \cite{besanko2010lumpy}. A firm in state $x$ has capacity $\bar q(x)$. Firms compete in quantities under inverse demand $P(Q)$ and, for simplicity, have zero marginal cost. Given competitors' output $Q_{-i}$, firm $i$ solves
\begin{equation*}
\max_{0\leq q_i\leq\bar q(x_i)}P(q_i+Q_{-i})q_i.
\end{equation*}
In the limiting model with many firms and negligible individual market power, the scalar interaction is average production. The static game determines $u(x,M(s))$.

The second is a quality ladder model in which $x$ represents product quality \citep{pakes1994computing,ericson1995markov}. Suppose consumers choose among differentiated products under a logit demand model. Consumer $j$ obtains utility
\begin{equation*}
u_{ijt}
=
\theta_1\ln(x_{it}+1)+\theta_2\ln(Y-p_{it})+v_{ijt},
\end{equation*}
where $\theta_1,\theta_2>0$, $p_{it}$ is price, $Y$ is income, and the shocks $v_{ijt}$ are i.i.d. Gumbel. The pricing game has a unique pure-strategy equilibrium under the conditions in \cite{caplin1991aggregation}. In a commonly used large-market limit, the resulting operating profit has the form
\begin{equation*}
u(x,M(s))=\frac{\widetilde c(x+1)^{\theta_1}}{M(s)},
\qquad
M(s)=\sum_{y\in X}(y+1)^{\theta_1}s(y),
\end{equation*}
where  $\widetilde c>0$ collects the remaining demand and cost parameters; see \cite{besanko1990logit}. We assume $\theta_1\geq1$. 
Note that in this model the population distribution enters profits through a scalar interaction. Using our results, we show that an MFE exists, that investment increases with the firm's state, and that changes in demand and patience shift the lowest and highest equilibrium scalar interactions. Since the state space is discrete, we use discrete convexity to prove our results. Recall that for a function $q$ on the non-negative integers, discrete convexity means
\begin{equation*}
q(x+2)-2q(x+1)+q(x)\geq0
\end{equation*}
for every $x\geq0$.

For computational and structural estimation purposes, these models are often studied with discrete state spaces and transition probabilities of the form in Equation \eqref{Eq:Oligopoly}. To the best of our knowledge, analytical comparative statics for these discrete models with an unbounded discrete state space are not available.\footnote{\cite{light2022mean} derive comparative statics for related models under uniqueness of MFE and continuous state transitions. Those assumptions differ from the discrete transition structure considered here.}

We will use the following assumption. 

\begin{assumption} \label{Assumption:CapacityCompt}
Assume that:
\begin{enumerate}[label=(\roman*),leftmargin=2em,itemsep=0.2ex]
    \item $\delta>1/2$;
    \item $c$ is continuous, increasing, and convex;
    \item for every parameter $z$ under consideration, $u(x,m,z)$ is discretely convex and strictly increasing in $x$ and is continuous in $(m,z)$;
    \item $M(s)=\sum_{x=0}^{\infty}k(x)s(x)$, where $k$ is non-negative and increasing and $k(x)\leq C_k(1+x^{p_k})$ for some $C_k>0$ and $p_k\geq0$. The interval $[a,b]$ is chosen so that $M(s^{m,g})\in[a,b]$ for every $m\in[a,b]$ and every stationary policy $g$ where $s^{m,g}$ denotes the invariant distribution induced by policy $g$ at interaction $m$;
    \item there are $C_u<\infty$ and $p_u\geq0$ such that $|u(x,m,z)|\leq C_u(1+x^{p_u})$ uniformly over the parameter region. Moreover, if $(m_n,z_n)\to(m,z)$, then
    \begin{equation*}
    \sup_{x\geq0}
    \frac{|u(x,m_n,z_n)-u(x,m,z)|}{1+x^{p_u}}
    \to0.
    \end{equation*}
\end{enumerate}
\end{assumption}

Condition (i) gives a  downward drift so the firms' states does not explode. Condition (ii) is a standard increasing marginal investment costs. Condition (iii) says that a higher state raises operating profit. Conditions (iv) and (v) control the growth of the interaction and payoff on the unbounded state space. They hold in the capacity and quality specifications described before Assumption \ref{Assumption:CapacityCompt}. 

Recall that $u$ has increasing differences in $(x,z)$ if, whenever $x_2\geq x_1$ and $z_2\geq_Zz_1$,
\begin{equation*}
u(x_2,m,z_2)-u(x_2,m,z_1)
\geq
u(x_1,m,z_2)-u(x_1,m,z_1).
\end{equation*}
This condition means that the parameter raises the return from being in a higher state.

\begin{theorem} \label{Thm:Comparative}
Suppose Assumption \ref{Assumption:CapacityCompt} holds. Then:
\begin{enumerate}[label=(\roman*),leftmargin=2em,itemsep=0.2ex]
    \item an MFE exists;
    \item the investment policy $g(x,m)$ is unique and non-decreasing in $x$;
    \item if $(Z,\geq_Z)$ is partially ordered and $u(x,m,z)$ has increasing differences in $(x,z)$, then the lowest $\underline m(z)$ and highest $\overline m(z)$ equilibrium scalar interactions are non-decreasing in $z$;
    \item the lowest $\underline m(\beta)$ and highest $\overline m(\beta)$ equilibrium scalar interactions are non-decreasing in the discount factor $\beta$.
\end{enumerate}
\end{theorem}

As an application, consider the capacity specification used in Appendix \ref{Sec:OligSimulations}, where $\bar q(x)=x$ and
\begin{equation*}
u(x,m)=x(\alpha-\gamma m).
\end{equation*}
We impose $\alpha>\gamma b$, so operating profit is strictly increasing in $x$ throughout $[a,b]$. The payoff has increasing differences in $(x,\alpha)$ and in $(x,-\gamma)$. Part (iii) of Theorem \ref{Thm:Comparative} therefore implies that the lowest and highest equilibrium average production levels increase with the demand intercept $\alpha$ and decrease with the demand slope $\gamma$.

\IfFileExists{vector_mfe_results_macros.tex}{
\newcommand{\QOneModerate}{1.142}
\newcommand{\QTwoModerate}{1.690}
\newcommand{\QOneHigher}{0.944}
\newcommand{\QTwoHigher}{1.931}
\newcommand{\GapModerate}{0.548}
\newcommand{\GapHigher}{0.987}
\newcommand{\CapitalModerate}{1.416}
\newcommand{\CapitalHigher}{1.438}
\newcommand{\RateModerate}{3.08\%}
\newcommand{\RateHigher}{2.99\%}
\newcommand{\BetaRModerate}{0.979}
\newcommand{\BetaRHigher}{0.978}
\newcommand{\GiniModerate}{0.237}
\newcommand{\GiniHigher}{0.363}
\newcommand{\PostResidualModerate}{0.0016}
\newcommand{\PostResidualHigher}{0.0036}
\newcommand{\ExactResidualModerate}{0.0008}
\newcommand{\ExactResidualHigher}{0.0059}
\newcommand{\ReferenceQOneModerate}{1.143}
\newcommand{\ReferenceQTwoModerate}{1.690}
\newcommand{\ReferenceQOneHigher}{0.942}
\newcommand{\ReferenceQTwoHigher}{1.933}
\newcommand{\CandidateDistanceModerate}{0.0004}
\newcommand{\CandidateDistanceHigher}{0.0030}
\newcommand{\UpperMassMaximum}{0.0008}
\newcommand{\LogitTemperature}{0.010}
\newcommand{\LogitValueBound}{0.589}

}{
    \PackageError{vector-MFE application}
    {Missing vector_mfe_results_macros.tex}
    {Run the supplied code and copy the generated file beside the manuscript source.}
}

\section{A Heterogeneous-Agent Macro Application}
\label{Sec:VectorMacroApplication}

We now apply Algorithm \ref{alg:online_simplicial_mfe} to a model with two interaction variables. 
The model is a finite-state version of the two-neighborhood
economy in \cite{acemoglu2012}. Households face an incomplete-markets
savings problem as in \cite{aiyagari1994uninsured} and value their wealth
relative to average wealth in their own group, following
\cite{cole1992social}. We now provide the details.

\textit{Types and states.}
There are two permanent types $j\in\{1,2\}$ with fixed population shares $\alpha_1=\alpha_2=1/2$. A household's state is $(x,z,j)$, where $x\in X$ is wealth and $z\in Z_d=\{1-d,1+d\}$ is idiosyncratic labor productivity. There are no aggregate shocks, and productivity shocks are independent across households. Conditional on type, productivity follows
\begin{equation*}
P_1=
\begin{pmatrix}
0.97&0.03\\
0.27&0.73
\end{pmatrix},
\qquad
P_2=
\begin{pmatrix}
0.73&0.27\\
0.03&0.97
\end{pmatrix}.
\end{equation*}

\textit{Interaction and prices.}
Let $s_j$ be the conditional state distribution of type $j$. The interaction variables are the two type-specific mean wealth levels
\begin{equation*}
Q_j(s)=
\sum_{x\in X}\sum_{z\in Z_d}x s_j(x,z),
\qquad j\in\{1,2\},
\end{equation*}
and write $Q(s)=(Q_1(s),Q_2(s))$. Aggregate capital is
\begin{equation*}
K(s)=\alpha_1Q_1(s)+\alpha_2Q_2(s).
\end{equation*}
A competitive firm produces according to $Y=BK^\vartheta L^{1-\vartheta}$. We normalize aggregate effective labor to one. A vector $Q=(Q_1,Q_2)$ therefore implies
\begin{equation*}
w(K)=(1-\vartheta)BK^\vartheta,
\qquad
R(K)=1-\delta_k+\vartheta BK^{\vartheta-1},
\qquad
K=\alpha_1Q_1+\alpha_2Q_2,
\end{equation*}
where $R(K)$ is the gross return on wealth.

\textit{Actions, dynamics, and payoff.}
The action $a$ is selected from a finite grid of saving rates in $[0,\overline a]$, where $\overline a<1$. Consumption and desired next-period wealth are
\begin{equation}
\label{Eq:MacroBudgetVector}
c=(1-a)\bigl(R(K)x+w(K)z\bigr),
\qquad
\widetilde x'
=a\bigl(R(K)x+w(K)z\bigr).
\end{equation}
Equation \eqref{Eq:MacroBudgetVector} implies $c+\widetilde x'=R(K)x+w(K)z$, and every feasible action leaves positive consumption. If $\widetilde x'$ lies between two adjacent wealth grid points, the next wealth level is chosen by a lottery whose conditional mean is $\widetilde x'$. Values outside the grid are projected to the nearest endpoint.

The part of a type-$j$ household's single-period payoff that excludes the entropy term is
\begin{equation}
\label{Eq:MacroPayoffVector}
\pi_j(x,z,a,Q)
=
\log c+
\eta\log\left(1+\frac{x}{Q_j}\right).
\end{equation}
The second term captures the household's position relative to the mean wealth of its own group. Current saving affects future wealth and therefore changes this relative position in later periods.

Households use the policy in Equation \eqref{Eq:LogitPolicy} on the finite action grid. In the simulations, $\lambda=\LogitTemperature$. For a fixed e $Q$, let $s_j^Q$ be the invariant distribution generated for type $j$ and define
\begin{equation*}
F_j(Q)=
\sum_{x\in X}\sum_{z\in Z_d}x s_j^Q(x,z),
\qquad
F(Q)=(F_1(Q),F_2(Q)).
\end{equation*}
The two group means are consistent when $F(Q)=Q$. Factor prices depend on $Q$ only through aggregate capital, but Equation \eqref{Eq:MacroPayoffVector} also makes a type-$j$ household respond directly to $Q_j$. Consequently, the same aggregate capital but different group means can generate different saving policies.

In the simulations we consider $d=0.25$ with $d=0.50$ while keeping the two transition matrices fixed. Because the groups have equal size, the stationary population places probability $1/2$ on each of $1-d$ and $1+d$. Average labor productivity is therefore one for both values of $d$. Hence, increasing $d$ raises income dispersion without changing its population mean: the type-specific mean productivities change from $(0.8,1.2)$ to $(0.6,1.4)$.

For the simulations, we set $\beta=0.95$, $\vartheta=0.36$, $\delta_k=0.07$, $B=0.35$, and $\eta=0.20$.\footnote{The wealth grid has $15$ points between $0.01$ and $5$, with more points near the borrowing constraint, and the action grid has $19$ saving rates between $0$ and $0.96$. We first apply Algorithm \ref{alg:online_simplicial_mfe} on a grid with $40$ intervals in each interaction coordinate. We then use a grid with $400$ intervals and initialize its vertex estimates at the value obtained from the coarser grid. At every  pair of group means, the
simulator is queried at each state-action pair. The code averages the
observed payoffs and transition outcomes and solves Equation
\eqref{Eq:RegularizedBellman} after replacing the true payoff and
transition probabilities by these averages. Appendix
\ref{sec:SavingsEstimates} verifies that this policy estimator has the
convergence property used in the proof of Theorem
\ref{thm:online_simplicial_mfe}. }

At every selected vertex and for each type, the code retains the sample averages of rewards and transition probabilities, the table of action values obtained from those averages, and a separate simulated population trajectory. When the same vertex is selected again, the code adds observations to the stored averages, resolves the Bellman equation from the updated averages, and extends the stored population trajectory. The stored quantities at unselected vertices are unchanged. 

When $d=0.25$, the resulting interaction is approximately $(\QOneModerate,\QTwoModerate)$; when $d=0.50$, it is approximately $(\QOneHigher,\QTwoHigher)$. The wealth gap $Q_2-Q_1$ increases from $\GapModerate$ to $\GapHigher$, and as expected the Gini coefficient of the pooled wealth distribution increases from $\GiniModerate$ to $\GiniHigher$ while aggregate capital changes only from $\CapitalModerate$ to $\CapitalHigher$ and the corresponding net interest rates are $\RateModerate$ and $\RateHigher$.

\begin{figure}[H]
    \centering
    \begin{subfigure}[b]{0.48\textwidth}
        \includegraphics[width=\textwidth,height=4.2cm]
        {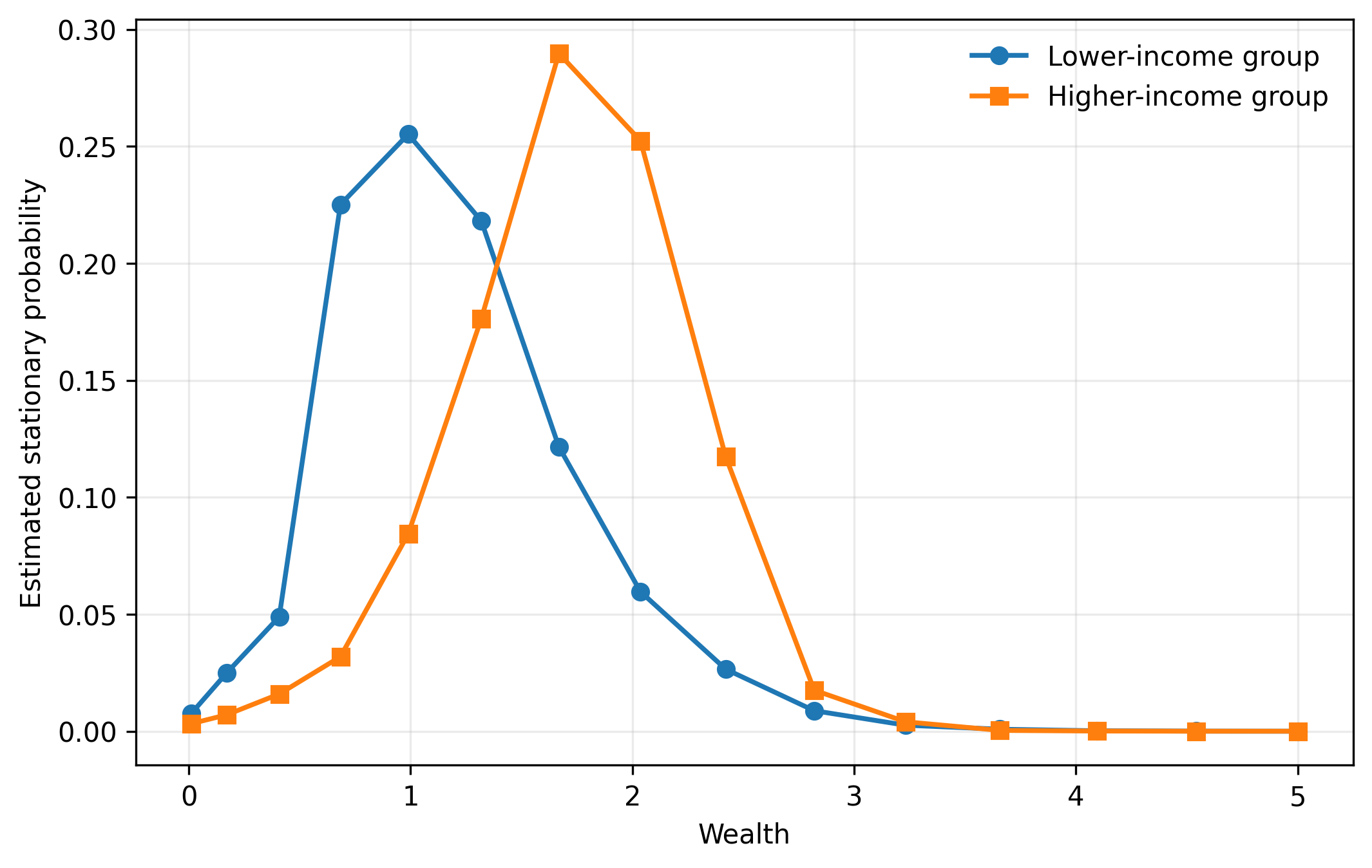}
        \caption{$d=0.25$ and
        $(Q_1,Q_2)=(\QOneModerate,\QTwoModerate)$.}
        \label{fig:VectorMacroModerate}
    \end{subfigure}
    \hfill
    \begin{subfigure}[b]{0.48\textwidth}
        \includegraphics[width=\textwidth,height=4.2cm]
        {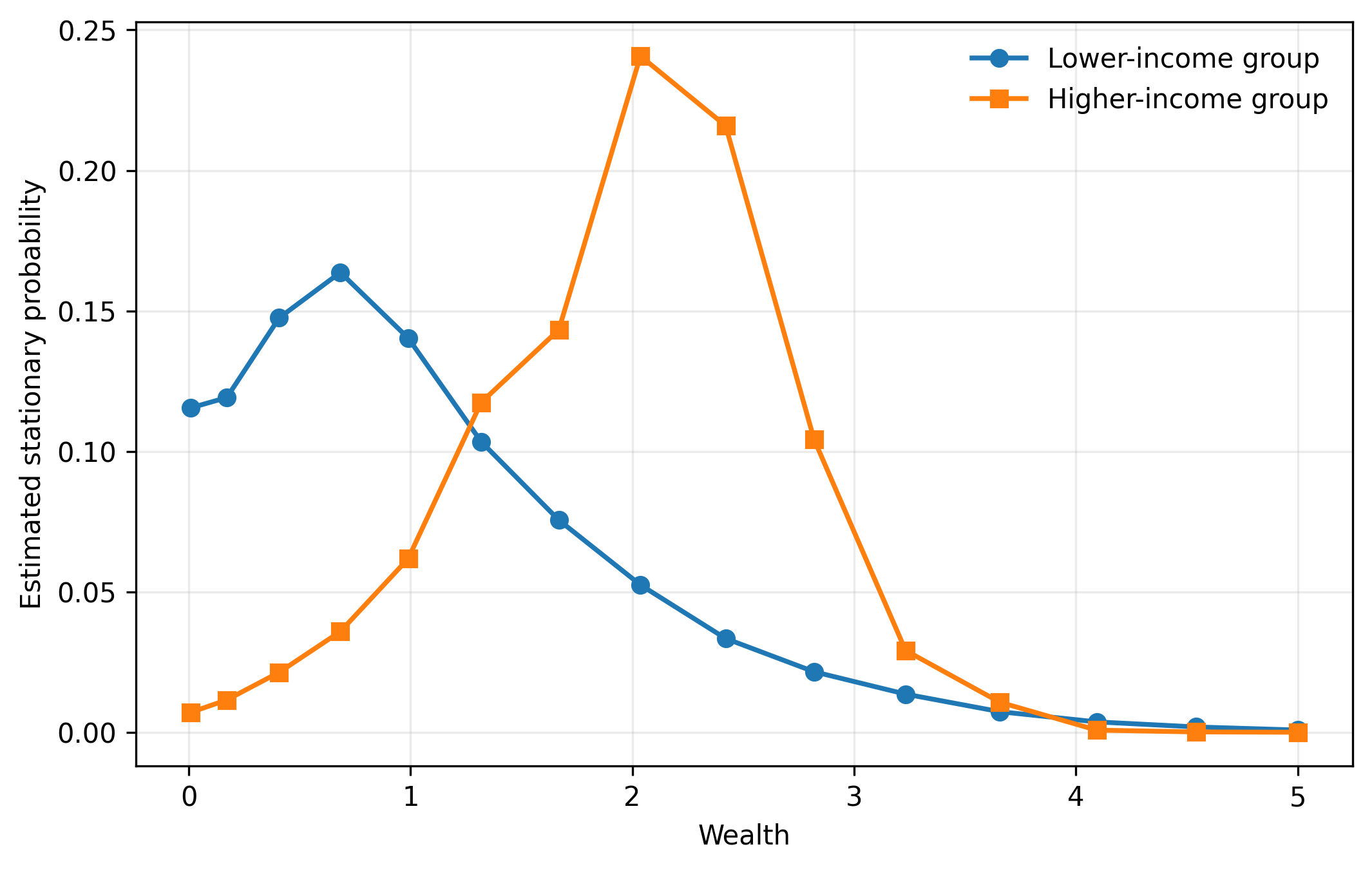}
        \caption{$d=0.50$ and
        $(Q_1,Q_2)=(\QOneHigher,\QTwoHigher)$.}
        \label{fig:VectorMacroHigher}
    \end{subfigure}
    \caption{Stationary wealth distributions under Algorithm \ref{alg:online_simplicial_mfe}. Increasing $d$ raises income dispersion without changing mean labor productivity. The estimated values of $\norm{Q-F(Q)}$ after the final simulation are $\PostResidualModerate$ and $\PostResidualHigher$.}
    \label{fig:VectorMacroDistributions}
\end{figure}

\section{Additional Applications} \label{Section:applications}

This section presents two additional models with scalar interactions. In both models, the finite action set is handled with the entropy term described in Section \ref{sec:single-value}.\footnote{In the social-learning numerical specification, the full support of the private signal make
every transition probability positive, so
Assumption~\ref{Assumption:Unique} holds.
In the ridesharing application it can be easily checked that Assumption~\ref{Assumption:Unique} holds.}

\subsection{Dynamic Ridesharing Model} \label{Section:Ridesharing}

We consider a ridesharing platform that assigns ride requests to drivers over time. A driver's state determines whether she is available, the request she has received, and, if she is currently serving a rider, the remaining trip duration. An available driver may accept or reject a request. Accepting makes her unavailable for the duration of the trip, while rejecting allows her to remain available and wait for another request. The probability of receiving a request depends on the fraction of available drivers, so drivers interact through a scalar population statistic.\footnote{There is a growing literature on strategic driver behavior in ridesharing platforms; see, for example, \cite{bimpikis2019spatial,guda2019your,besbes2021surge,garg2022driver,ma2022spatio}.}

\textit{States.} The state of driver $i$ at time $t$ is $x_{i,t}=(x_{i,t,1},x_{i,t,2})\in X_1\times X_2$. Let $X_1=\{0,1,\ldots,K'\}$ and $X_2=\{0,1,\ldots,K\}$. The first component describes the availability and remaining trip duration. A value $x_{i,t,1}=0$ means that the driver is available, while $x_{i,t,1}=d>0$ means that she remains unavailable for $d$ periods. The second component describes the current request type. A value $x_{i,t,2}=j\geq1$ denotes a request of type $j$, while $x_{i,t,2}=0$ means that no request has arrived.

\textit{Actions.} An available driver with request $j\geq1$ chooses $a_{i,t}\in\{0,1\}$, where $1$ means accept and $0$ means reject. If the driver is unavailable or has no request, her feasible action set is $\{0\}$.

\textit{States' dynamics.} If an available driver accepts a request of type $j$, she becomes unavailable for $d_j\in\{1,\ldots,K'\}$ periods. If she rejects the request or receives no request, she remains available. For an unavailable driver, the remaining duration falls by one each period until it reaches zero.\footnote{The model can incorporate entry and exit, spatial locations, driver characteristics, or request-specific attributes by expanding the state space.}

Let
\begin{equation*}
M(s)=\sum_{y_2\in X_2}s(0,y_2)
\end{equation*}
be the fraction of available drivers. A driver receives no request with probability $f(M(s))$. Conditional on receiving a request, its type has probabilities $\bar p_1,\ldots,\bar p_K$, where $\bar p_j\geq0$ and $\sum_{j=1}^K\bar p_j=1$. Thus,
\begin{equation*}
\Pr(x_{i,t+1,2}=j)=
\begin{cases}
f(M(s_t)), & j=0,\\
(1-f(M(s_t)))\bar p_j, & j=1,\ldots,K.
\end{cases}
\end{equation*}
We assume that $f$ is increasing: when more drivers are available, each driver is more likely to receive no request.

\textit{Payoff.} If an available driver accepts a request of type $j$, she receives payoff $u_j$. Otherwise, her payoff is zero. A driver may therefore reject a current request in the hope of receiving a more valuable request later.

Loosely motivated by the shift in ridesharing platforms from multiplicative pricing schemes which amplify earnings for long trips during surge pricing to additive pricing schemes which add a surge bonus that is independent of the trip length \citep{garg2022driver}, we use our algorithms to compute and learn MFE and analyze two distinct payoff structures. One payoff structure assigns higher rewards for longer trips, which incentivizes drivers to strategically ``cherry-pick'' rides by rejecting lower-paying requests in anticipation of more rewarding ones. We numerically show in Section \ref{Sec:RideSimulations} that higher payoffs for long trips result in greater equilibrium availability of drivers as they strategically reject lower-paying requests in equilibrium even though their equilibrium probability of being matched decreases as more drivers become available in equilibrium.

\subsection{Social Learning Model} \label{Section:SocialLearning}

In this section, we study a mean field model of social learning, where agents interact and update their beliefs about an unknown state \( \theta \) over time.\footnote{Similar social learning settings, in which many agents face uncertainty about an unknown state, observe private signals, and interact, have been extensively studied. A static solution concept that focuses on asymptotic steady states to simplify the challenging task of computing equilibrium has been proposed recently (e.g., \cite{mossel2020social} and \cite{dasaratha2023learning}). } Each agent has access to private signals and aggregate social information, and they strategically allocate effort to improve the precision of their private signals. This model captures the interaction between private learning and social learning dynamic. In particular, each agent's belief evolves according to a DeGroot-style  linear updating rule that incorporates their prior belief, the average belief of the population, and a noisy private signal. The precision of the private signal is endogenous, as agents can exert effort to reduce its variance, though at a cost. An important feature of the model is the weight function $k(a)$, which determines how much weight is placed on the private signal relative to social learning and increases with effort. Hence, effort not only enhances signal precision but also   allows an agent to rely more on their own signal rather than the population. The agent's payoff depends on their belief about 
$\theta$, the actual state $\theta$, and the cost of effort. 

\textit{States.} The state $x_{i,t}\in X\subset[0,1]$ is agent $i$'s belief about $\theta$. The set $X$ is a finite grid.

\textit{Actions.} The agent chooses effort $a_{i,t}\in\{0,1,\ldots,\overline a\}$. Greater effort improves the precision of the private signal but is costly.

\textit{States' dynamics.} The next belief is
\begin{equation*}
x_{i,t+1}
=
f\left(
 c(1-k(a_{i,t}))x_{i,t}
 +(1-c)(1-k(a_{i,t}))M(s_t)
 +k(a_{i,t})\zeta_{i,t}
\right),
\end{equation*}
where $c\in(0,1)$, $k(a)\in[0,1]$ is increasing, and
\begin{equation*}
M(s_t)=\sum_{x\in X}xs_t(x)
\end{equation*}
is the average belief. The private signal satisfies $\zeta_{i,t}\sim\mathcal N(\theta,\sigma^2(a_{i,t}))$, where $\sigma^2(a)$ decreases with effort. The function $f$ maps the expression in parentheses to the nearest point in $X$, using a fixed rule when two grid points are equally close.

\textit{Payoff.} The single-period payoff is
\begin{equation*}
\pi(x,a,M(s))=u(x,\theta)-da,
\end{equation*}
where $d>0$ is the unit cost of effort. Although the payoff does not depend directly on the population state, the transition probabilities depend on the average belief through $M(s)$.

\begin{figure}[H]
    \centering
    \begin{subfigure}[b]{0.4\textwidth}
        \includegraphics[width=\textwidth, height=4cm]{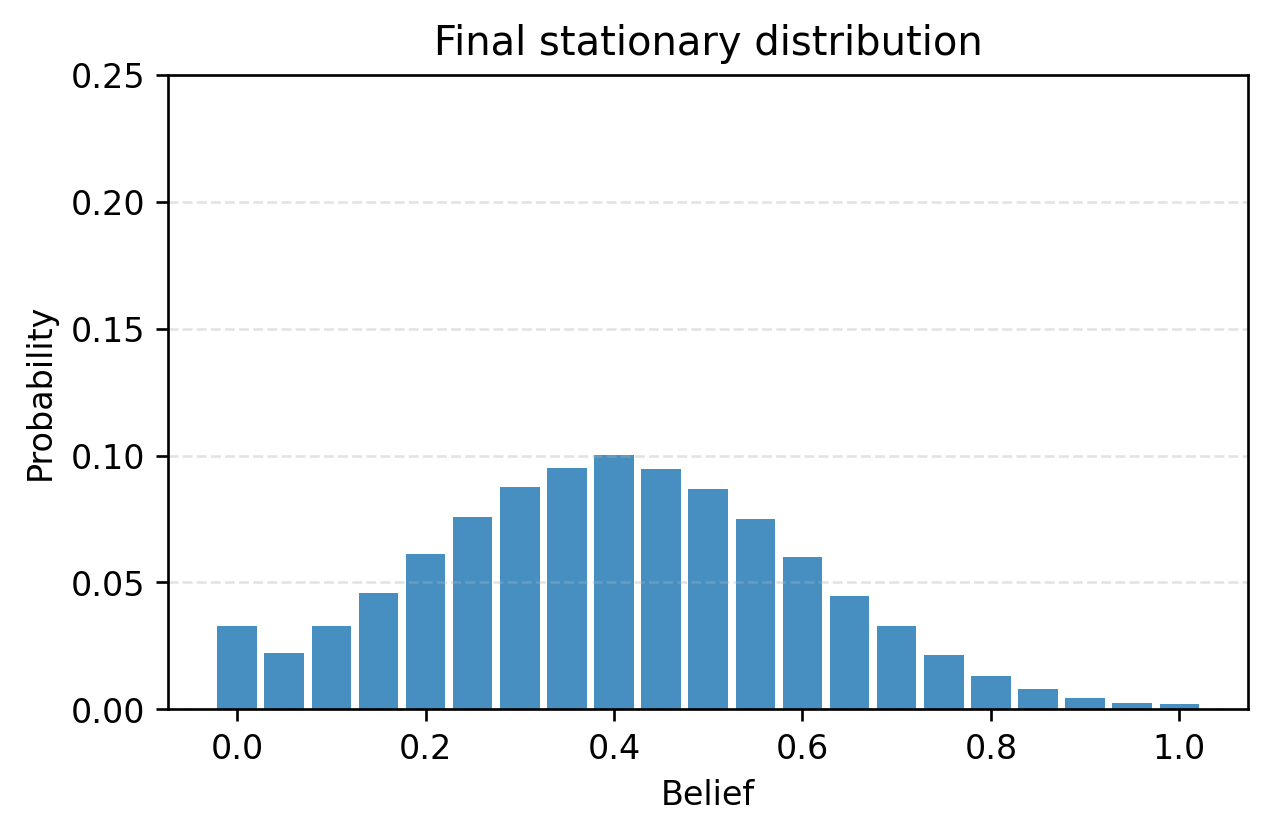}
        \caption{Precision $\gamma=5$}
        \label{fig:precision5}
    \end{subfigure}
    \hfill
    \begin{subfigure}[b]{0.4\textwidth}
        \includegraphics[width=\textwidth, height=4cm]{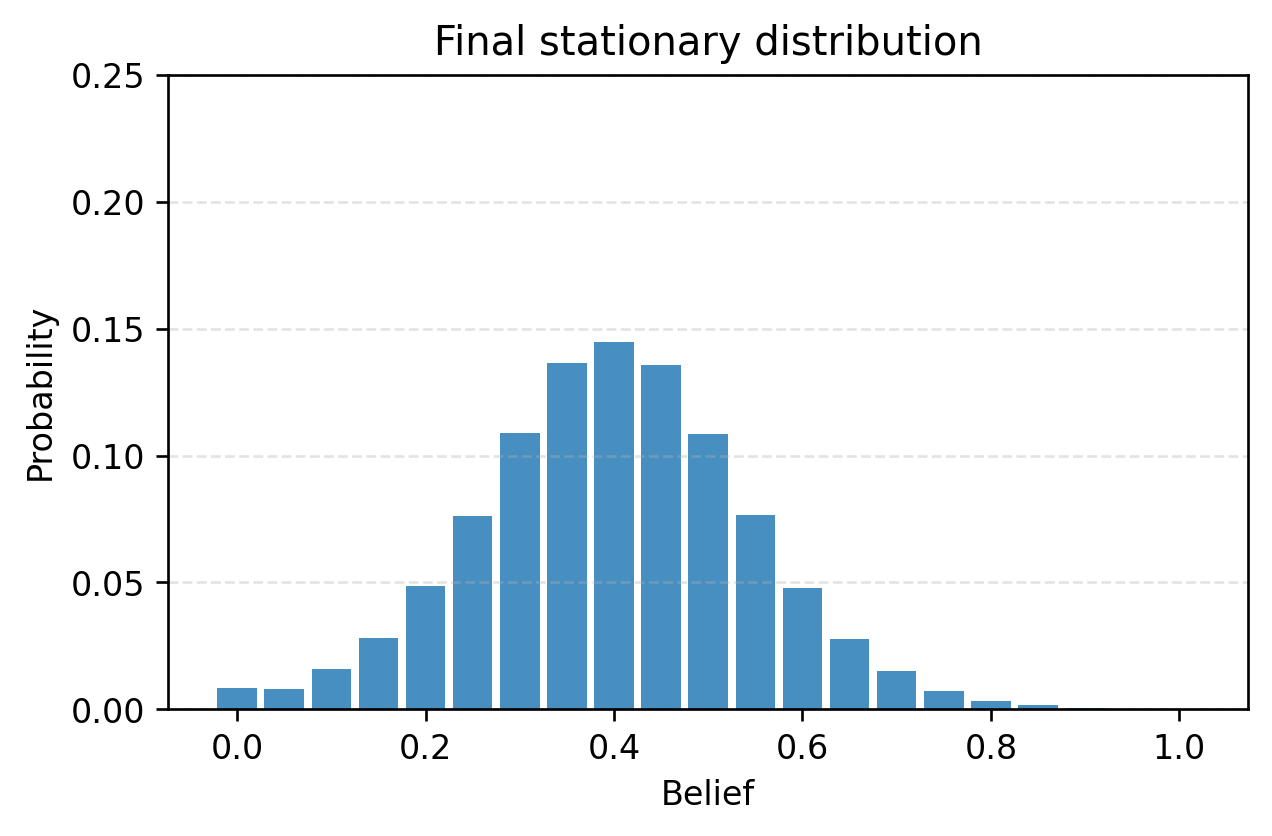}
        \caption{Precision $\gamma=15$}
        \label{fig:precision15}
    \end{subfigure}
    \caption{Approximate equilibrium belief distributions under two levels of signal precision. The variances are approximately $0.0390$ for $\gamma=5$ and $0.0208$ for $\gamma=15$. The corresponding mean beliefs are approximately $0.3992$ and $0.3986$.}
    \label{fig:sociallearning}
\end{figure}

The unknown state makes the model-free algorithm natural. Appendix \ref{Section:InventorySimulationDetails} describes the Q-learning and population-simulation parameters, and Figure \ref{fig:sociallearning} compares two levels of signal precision. The average belief remains close to $\theta$, but lower precision produces a more dispersed belief distribution. This precision parameter may represent, for example, the quality of information supplied by a platform or the quality of disclosure in a financial market.\footnote{Platform interventions that affect information quality are studied in \cite{candogan2022social,acemoglu2024model}.}

In the simulations, $\theta=0.4$ and beliefs range from $0$ to $1$ in increments of $0.05$, so $|X|=21$. Effort takes values from $0$ to $5$, $c=0.4$, and
\begin{equation*}
k(a)=\frac{0.5+a}{1.5+a},
\qquad
\sigma^2(a)=\frac{1}{3+\gamma a}.
\end{equation*}
The payoff from belief accuracy is $u(x,\theta)=-20(\theta-x)^2$ and $d=0.1$. We compare $\gamma=5$ and $\gamma=15$. The population distribution is estimated from $K=200{,}000$ observations and the stopping tolerance is $10^{-3}$.

\section*{Conclusions}

This paper develops algorithms for stationary MFE when agents interact through a scalar or a low-dimensional vector of aggregate variables. We develop algorithm for scalar interactions and for several interaction variables that converge to an approximate MFE without requiring  contraction or monotonicity conditions on the MFE operator. 
We also use prove existence of an exact MFE on non-compact state spaces by using the interaction structure and analytical comparative statics results.  

The applications illustrate the different roles of the results. The dynamic oligopoly model has a unique continuous-action policy, so it yields an exact MFE and exact comparative statics without regularization. The inventory, ridesharing, and social learning models illustrate the scalar algorithms with finite actions. The heterogeneous-agent savings model shows how the simplicial algorithm handles two interactions.

There are several promising directions for future research. First, understanding the performance of the algorithms in computing an MFE in  large state spaces using function approximation  remains open. Second, while our algorithms guarantee convergence to an MFE, the specific equilibrium to which they converge in models with multiple MFE remains unclear. This has important implications for market design and counterfactual analysis. 

\bibliographystyle{elsarticle-harv}
\bibliography{learning}

@article{acemoglu2012,
  author  = {Acemoglu, Daron and Jensen, Martin Kaae},
  title   = {Robust Comparative Statics in Large Dynamic Economies},
  journal = {Journal of Political Economy},
  year    = {2015},
  volume  = {123},
  number  = {3},
  pages   = {587--640},
  doi     = {10.1086/680685}
}

@article{acemoglu2024model,
  author  = {Acemoglu, Daron and Ozdaglar, Asuman and Siderius, James},
  title   = {A Model of Online Misinformation},
  journal = {The Review of Economic Studies},
  year    = {2024},
  volume  = {91},
  number  = {6},
  pages   = {3117--3150},
  doi     = {10.1093/restud/rdad111}
}

@article{adlakha2013mean,
  author  = {Adlakha, Sachin and Johari, Ramesh},
  title   = {Mean Field Equilibrium in Dynamic Games with Strategic Complementarities},
  journal = {Operations Research},
  year    = {2013},
  volume  = {61},
  number  = {4},
  pages   = {971--989},
  doi     = {10.1287/opre.2013.1192}
}

@article{adlakha2015equilibria,
  author  = {Adlakha, Sachin and Johari, Ramesh and Weintraub, Gabriel Y.},
  title   = {Equilibria of Dynamic Games with Many Players: Existence, Approximation, and Market Structure},
  journal = {Journal of Economic Theory},
  year    = {2015},
  volume  = {156},
  pages   = {269--316},
  doi     = {10.1016/j.jet.2013.07.002}
}

@article{agarwal2021theory,
  title={On the theory of policy gradient methods: Optimality, approximation, and distribution shift},
  author={Agarwal, Alekh and Kakade, Sham M and Lee, Jason D and Mahajan, Gaurav},
  journal={Journal of Machine Learning Research},
  volume={22},
  number={98},
  pages={1--76},
  year={2021}
}

@article{aid2025stationary,
  author  = {A{\"i}d, Ren{\'e} and Basei, Matteo and Ferrari, Giorgio},
  title   = {A Stationary Mean-Field Equilibrium Model of Irreversible Investment in a Two-Regime Economy},
  journal = {Operations Research},
  year    = {2025},
  volume  = {73},
  number  = {5},
  pages   = {2351--2374},
  doi     = {10.1287/opre.2023.0250}
}

@article{aiyagari1994uninsured,
  author  = {Aiyagari, S. Rao},
  title   = {Uninsured Idiosyncratic Risk and Aggregate Saving},
  journal = {The Quarterly Journal of Economics},
  year    = {1994},
  volume  = {109},
  number  = {3},
  pages   = {659--684},
  doi     = {10.2307/2118417}
}

@article{anahtarci2023learning,
  title={Learning mean-field games with discounted and average costs},
  author={Anahtarci, Berkay and Kariksiz, Can Deha and Saldi, Naci},
  journal={Journal of Machine Learning Research},
  volume={24},
  number={17},
  pages={1--59},
  year={2023}
}

@article{arnosti2021managing,
  title={Managing congestion in matching markets},
  author={Arnosti, Nick and Johari, Ramesh and Kanoria, Yash},
  journal={Manufacturing \& Service Operations Management},
  volume={23},
  number={3},
  pages={620--636},
  year={2021}
}

@article{balseiro2015repeated,
  title={Repeated auctions with budgets in ad exchanges: Approximations and design},
  author={Balseiro, Santiago R and Besbes, Omar and Weintraub, Gabriel Y},
  journal={Management Science},
  volume={61},
  number={4},
  pages={864--884},
  year={2015}
}

@book{bertsekas1978stochastic,
  author    = {Bertsekas, Dimitri P. and Shreve, Steven E.},
  title     = {Stochastic Optimal Control: The Discrete-Time Case},
  publisher = {Academic Press},
  address   = {New York},
  year      = {1978}
}

@book{bertsekas1996neuro,
  author    = {Bertsekas, Dimitri P. and Tsitsiklis, John N.},
  title     = {Neuro-Dynamic Programming},
  publisher = {Athena Scientific},
  address   = {Belmont, MA},
  year      = {1996}
}

@book{bertsekas2019reinforcement,
  author    = {Bertsekas, Dimitri P.},
  title     = {Reinforcement Learning and Optimal Control},
  publisher = {Athena Scientific},
  address   = {Belmont, MA},
  year      = {2019}
}

@article{besanko1990logit,
  author  = {Besanko, David and Perry, Martin K. and Spady, Richard H.},
  title   = {The Logit Model of Monopolistic Competition: Brand Diversity},
  journal = {The Journal of Industrial Economics},
  year    = {1990},
  volume  = {38},
  number  = {4},
  pages   = {397--415},
  doi     = {10.2307/2098347}
}

@article{besanko2004capacity,
  author  = {Besanko, David and Doraszelski, Ulrich},
  title   = {Capacity Dynamics and Endogenous Asymmetries in Firm Size},
  journal = {The RAND Journal of Economics},
  year    = {2004},
  volume  = {35},
  number  = {1},
  pages   = {23--49}
}

@article{besanko2010lumpy,
  title={Lumpy Capacity Investment and Disinvestment Dynamics},
  author={Besanko, David and Doraszelski, Ulrich and Lu, Lauren Xiaoyuan and Satterthwaite, Mark},
  journal={Operations Research},
  volume={58},
  number={4-part-2},
  pages={1178--1193},
  year={2010}
}

@article{besbes2021surge,
  title={Surge pricing and its spatial supply response},
  author={Besbes, Omar and Castro, Francisco and Lobel, Ilan},
  journal={Management Science},
  volume={67},
  number={3},
  pages={1350--1367},
  year={2021}
}

@inproceedings{besbes2023signaling,
  author    = {Besbes, Omar and Fonseca, Yuri and Lobel, Ilan and Zheng, Fanyin},
  title     = {Signaling Competition in Two-Sided Markets},
  booktitle = {Proceedings of the 24th ACM Conference on Economics and Computation},
  year      = {2023},
  pages     = {293},
  doi       = {10.1145/3580507.3597737}
}

@article{bhandari2024global,
  author  = {Bhandari, Jalaj and Russo, Daniel},
  title   = {Global Optimality Guarantees for Policy Gradient Methods},
  journal = {Operations Research},
  year    = {2024},
  volume  = {72},
  number  = {5},
  pages   = {1906--1927},
  doi     = {10.1287/opre.2021.0014}
}

@article{bimpikis2019spatial,
  title={Spatial pricing in ride-sharing networks},
  author={Bimpikis, Kostas and Candogan, Ozan and Saban, Daniela},
  journal={Operations Research},
  volume={67},
  number={3},
  pages={744--769},
  year={2019}
}

@article{candogan2022social,
  author  = {Candogan, Ozan and Immorlica, Nicole and Light, Bar and Anunrojwong, Jerry},
  title   = {Social Learning under Platform Influence: Consensus and Persistent Disagreement},
  journal = {arXiv preprint arXiv:2202.12453},
  year    = {2022},
  doi     = {10.48550/arXiv.2202.12453}
}

@article{caplin1991aggregation,
  author  = {Caplin, Andrew and Nalebuff, Barry},
  title   = {Aggregation and Imperfect Competition: On the Existence of Equilibrium},
  journal = {Econometrica},
  year    = {1991},
  volume  = {59},
  number  = {1},
  pages   = {25--59},
  doi     = {10.2307/2938239}
}

@article{carmona2021finite,
  author  = {Carmona, Ren{\'e} and Wang, Peiqi},
  title   = {Finite-State Contract Theory with a Principal and a Field of Agents},
  journal = {Management Science},
  year    = {2021},
  volume  = {67},
  number  = {8},
  pages   = {4725--4741},
  doi     = {10.1287/mnsc.2020.3760}
}

@article{cole1992social,
  author  = {Cole, Harold L. and Mailath, George J. and Postlewaite, Andrew},
  title   = {Social Norms, Savings Behavior, and Growth},
  journal = {Journal of Political Economy},
  year    = {1992},
  volume  = {100},
  number  = {6},
  pages   = {1092--1125}
}

@inproceedings{cui2024learning,
  author    = {Cui, Kai and Dayan{\i}kl{\i}, G{\"o}k{\c{c}}e and Lauri{\`e}re, Mathieu and Geist, Matthieu and Pietquin, Olivier and Koeppl, Heinz},
  title     = {Learning Discrete-Time Major-Minor Mean Field Games},
  booktitle = {Proceedings of the AAAI Conference on Artificial Intelligence},
  year      = {2024},
  volume    = {38},
  number    = {9},
  pages     = {9616--9625},
  doi       = {10.1609/aaai.v38i9.28818}
}

@article{dasaratha2023learning,
  title={Learning from neighbours about a changing state},
  author={Dasaratha, Krishna and Golub, Benjamin and Hak, Nir},
  journal={Review of Economic Studies},
  volume={90},
  number={5},
  pages={2326--2369},
  year={2023}
}

@article{ericson1995markov,
  author  = {Ericson, Richard and Pakes, Ariel},
  title   = {Markov-Perfect Industry Dynamics: A Framework for Empirical Work},
  journal = {The Review of Economic Studies},
  year    = {1995},
  volume  = {62},
  number  = {1},
  pages   = {53--82},
  doi     = {10.2307/2297841}
}

@article{even2003learning,
  author  = {Even-Dar, Eyal and Mansour, Yishay},
  title   = {Learning Rates for {Q}-Learning},
  journal = {Journal of Machine Learning Research},
  year    = {2003},
  volume  = {5},
  pages   = {1--25}
}

@article{garg2022driver,
  title={Driver surge pricing},
  author={Garg, Nikhil and Nazerzadeh, Hamid},
  journal={Management Science},
  volume={68},
  number={5},
  pages={3219--3235},
  year={2022}
}

@inproceedings{geist2019theory,
  title     = {A Theory of Regularized Markov Decision Processes},
  author    = {Geist, Matthieu and Scherrer, Bruno and Pietquin, Olivier},
  booktitle = {Proceedings of the 36th International Conference on Machine Learning},
  pages     = {2160--2169},
  year      = {2019},
  volume    = {97},
  series    = {Proceedings of Machine Learning Research},
  publisher = {PMLR}
}

@article{guda2019your,
  title={Your uber is arriving: Managing on-demand workers through surge pricing, forecast communication, and worker incentives},
  author={Guda, Harish and Subramanian, Upender},
  journal={Management Science},
  volume={65},
  number={5},
  pages={1995--2014},
  year={2019}
}

@inproceedings{guo2019learning,
  author    = {Guo, Xin and Hu, Anran and Xu, Renyuan and Zhang, Junzi},
  title     = {Learning Mean-Field Games},
  booktitle = {Advances in Neural Information Processing Systems},
  year      = {2019},
  volume    = {32},
  pages     = {4966--4976}
}

@article{guo2023general,
  title={A general framework for learning mean-field games},
  author={Guo, Xin and Hu, Anran and Xu, Renyuan and Zhang, Junzi},
  journal={Mathematics of Operations Research},
  volume={48},
  number={2},
  pages={656--686},
  year={2023}
}

@inproceedings{hasselt2010double,
  author    = {van Hasselt, Hado},
  title     = {Double {Q}-Learning},
  booktitle = {Advances in Neural Information Processing Systems},
  year      = {2010},
  volume    = {23},
  pages     = {2613--2621}
}

@article{huang2006large,
  author  = {Huang, Minyi and Malham{\'e}, Roland P. and Caines, Peter E.},
  title   = {Large Population Stochastic Dynamic Games: Closed-Loop {McKean--Vlasov} Systems and the {Nash} Certainty Equivalence Principle},
  journal = {Communications in Information and Systems},
  year    = {2006},
  volume  = {6},
  number  = {3},
  pages   = {221--252},
  doi     = {10.4310/CIS.2006.v6.n3.a5}
}

@article{iyer2014mean,
  author  = {Iyer, Krishnamurthy and Johari, Ramesh and Sundararajan, Mukund},
  title   = {Mean Field Equilibria of Dynamic Auctions with Learning},
  journal = {Management Science},
  year    = {2014},
  volume  = {60},
  number  = {12},
  pages   = {2949--2970},
  doi     = {10.1287/mnsc.2014.2018}
}

@article{kanoria2021facilitating,
  title={Facilitating the search for partners on matching platforms},
  author={Kanoria, Yash and Saban, Daniela},
  journal={Management Science},
  volume={67},
  number={10},
  pages={5990--6029},
  year={2021}
}

@article{lasry2007mean,
  author  = {Lasry, Jean-Michel and Lions, Pierre-Louis},
  title   = {Mean Field Games},
  journal = {Japanese Journal of Mathematics},
  year    = {2007},
  volume  = {2},
  number  = {1},
  pages   = {229--260},
  doi     = {10.1007/s11537-007-0657-8}
}

@article{lauriere2022learning,
  author  = {Lauri{\`e}re, Mathieu and Perrin, Sarah and Geist, Matthieu and Pietquin, Olivier},
  title   = {Learning Mean Field Games: A Survey},
  journal = {arXiv preprint arXiv:2205.12944},
  year    = {2022},
  doi     = {10.48550/arXiv.2205.12944}
}

@inproceedings{li2020sample,
  author    = {Li, Gen and Wei, Yuting and Chi, Yuejie and Gu, Yuantao and Chen, Yuxin},
  title     = {Sample Complexity of Asynchronous {Q}-Learning: Sharper Analysis and Variance Reduction},
  booktitle = {Advances in Neural Information Processing Systems},
  year      = {2020},
  volume    = {33},
  pages     = {7031--7043}
}

@article{li2024mean,
  author  = {Li, Zongxi and Reppen, A. Max and Sircar, Ronnie},
  title   = {A Mean Field Games Model for Cryptocurrency Mining},
  journal = {Management Science},
  year    = {2024},
  volume  = {70},
  number  = {4},
  pages   = {2188--2208},
  doi     = {10.1287/mnsc.2023.4798}
}

@article{li2024q,
  author  = {Li, Gen and Cai, Changxiao and Chen, Yuxin and Wei, Yuting and Chi, Yuejie},
  title   = {Is {Q}-Learning Minimax Optimal? A Tight Sample Complexity Analysis},
  journal = {Operations Research},
  year    = {2024},
  volume  = {72},
  number  = {1},
  pages   = {222--236},
  doi     = {10.1287/opre.2023.2450}
}

@article{light2021stochastic,
  author  = {Light, Bar},
  title   = {Stochastic Comparative Statics in Markov Decision Processes},
  journal = {Mathematics of Operations Research},
  year    = {2021},
  volume  = {46},
  number  = {2},
  pages   = {797--810},
  doi     = {10.1287/moor.2020.1086}
}

@article{light2022mean,
  author  = {Light, Bar and Weintraub, Gabriel Y.},
  title   = {Mean Field Equilibrium: Uniqueness, Existence, and Comparative Statics},
  journal = {Operations Research},
  year    = {2022},
  volume  = {70},
  number  = {1},
  pages   = {585--605},
  doi     = {10.1287/opre.2020.2090}
}

@article{lippman1997competitive,
  title={The competitive newsboy},
  author={Lippman, Steven A and McCardle, Kevin F},
  journal={Operations Research},
  volume={45},
  number={1},
  pages={54--65},
  year={1997}
}

@article{liu2007dynamic,
  title={Dynamic competitive newsvendors with service-sensitive demands},
  author={Liu, Liming and Shang, Weixin and Wu, Shaohua},
  journal={Manufacturing \& Service Operations Management},
  volume={9},
  number={1},
  pages={84--93},
  year={2007}
}

@article{ma2022spatio,
  title={Spatio-temporal pricing for ridesharing platforms},
  author={Ma, Hongyao and Fang, Fei and Parkes, David C},
  journal={Operations Research},
  volume={70},
  number={2},
  pages={1025--1041},
  year={2022}
}

@article{mahajan2001inventory,
  title={Inventory competition under dynamic consumer choice},
  author={Mahajan, Siddharth and Van Ryzin, Garrett},
  journal={Operations Research},
  volume={49},
  number={5},
  pages={646--657},
  year={2001}
}

@article{milgrom1994comparing,
  author  = {Milgrom, Paul and Roberts, John},
  title   = {Comparing Equilibria},
  journal = {American Economic Review},
  year    = {1994},
  volume  = {84},
  number  = {3},
  pages   = {441--459}
}

@article{mossel2020social,
  title={Social learning equilibria},
  author={Mossel, Elchanan and Mueller-Frank, Manuel and Sly, Allan and Tamuz, Omer},
  journal={Econometrica},
  volume={88},
  number={3},
  pages={1235--1267},
  year={2020}
}

@article{olsen2014markov,
  title={On Markov equilibria in dynamic inventory competition},
  author={Olsen, Tava Lennon and Parker, Rodney P},
  journal={Operations Research},
  volume={62},
  number={2},
  pages={332--344},
  year={2014}
}

@article{pakes1994computing,
  author  = {Pakes, Ariel and McGuire, Paul},
  title   = {Computing Markov-Perfect Nash Equilibria: Numerical Implications of a Dynamic Differentiated Product Model},
  journal = {The RAND Journal of Economics},
  year    = {1994},
  volume  = {25},
  number  = {4},
  pages   = {555--589}
}

@article{paulin2015concentration,
  author  = {Paulin, Daniel},
  title   = {Concentration Inequalities for Markov Chains by Marton Couplings and Spectral Methods},
  journal = {Electronic Journal of Probability},
  year    = {2015},
  volume  = {20},
  number  = {79},
  pages   = {1--32},
  doi     = {10.1214/EJP.v20-4039}
}

@inproceedings{perrin2020fictitious,
  author    = {Perrin, Sarah and P{\'e}rolat, Julien and Lauri{\`e}re, Mathieu and Geist, Matthieu and Elie, Romuald and Pietquin, Olivier},
  title     = {Fictitious Play for Mean Field Games: Continuous Time Analysis and Applications},
  booktitle = {Advances in Neural Information Processing Systems},
  year      = {2020},
  volume    = {33},
  pages     = {13199--13213}
}

@inproceedings{qu2020finite,
  author    = {Qu, Guannan and Wierman, Adam},
  title     = {Finite-Time Analysis of Asynchronous Stochastic Approximation and {Q}-Learning},
  booktitle = {Proceedings of the 33rd Conference on Learning Theory},
  series    = {Proceedings of Machine Learning Research},
  year      = {2020},
  volume    = {125},
  pages     = {3185--3205}
}

@article{saldi2023linear,
  author  = {Saldi, Naci},
  title   = {Linear Mean-Field Games with Discounted Cost},
  journal = {Mathematics of Operations Research},
  year    = {2025},
  volume  = {51},
  number  = {1},
  pages   = {358--397},
  doi     = {10.1287/moor.2023.0148}
}

@article{scarf1967fixedpoint,
  author  = {Scarf, Herbert E.},
  title   = {The Approximation of Fixed Points of a Continuous Mapping},
  journal = {SIAM Journal on Applied Mathematics},
  year    = {1967},
  volume  = {15},
  number  = {5},
  pages   = {1328--1343},
  doi     = {10.1137/0115116}
}

@book{seneta2006non,
  author    = {Seneta, Eugene},
  title     = {Non-Negative Matrices and Markov Chains},
  publisher = {Springer},
  address   = {New York},
  year      = {2006}
}

@article{shardlow2000perturbation,
  title={A perturbation theory for ergodic Markov chains and application to numerical approximations},
  author={Shardlow, Tony and Stuart, Andrew M},
  journal={SIAM Journal on Numerical Analysis},
  volume={37},
  number={4},
  pages={1120--1137},
  year={2000}
}

@inproceedings{subramanian2019reinforcement,
  title={Reinforcement learning in stationary mean-field games},
  author={Subramanian, Jayakumar and Mahajan, Aditya},
  booktitle={Proceedings of the 18th International Conference on Autonomous Agents and MultiAgent Systems},
  pages={251--259},
  year={2019}
}

@book{topkis2011supermodularity,
  author    = {Topkis, Donald M.},
  title     = {Supermodularity and Complementarity},
  publisher = {Princeton University Press},
  address   = {Princeton, NJ},
  year      = {1998}
}

@article{wager2021experimenting,
  title={Experimenting in equilibrium},
  author={Wager, Stefan and Xu, Kuang},
  journal={Management Science},
  volume={67},
  number={11},
  pages={6694--6715},
  year={2021}
}

@article{weintraub2008markov,
  title={Markov perfect industry dynamics with many firms},
  author={Weintraub, Gabriel Y and Benkard, C Lanier and Van Roy, Benjamin},
  journal={Econometrica},
  volume={76},
  number={6},
  pages={1375--1411},
  year={2008}
}

@article{weintraub2010computational,
  title={Computational methods for oblivious equilibrium},
  author={Weintraub, Gabriel Y and Benkard, C Lanier and Van Roy, Benjamin},
  journal={Operations Research},
  volume={58},
  number={4-part-2},
  pages={1247--1265},
  year={2010}
}

@inproceedings{xie2021learning,
  author    = {Xie, Qiaomin and Yang, Zhuoran and Wang, Zhaoran and Minca, Andreea},
  title     = {Learning while Playing in Mean-Field Games: Convergence and Optimality},
  booktitle = {Proceedings of the 38th International Conference on Machine Learning},
  series    = {Proceedings of Machine Learning Research},
  year      = {2021},
  volume    = {139},
  pages     = {11436--11447}
}

@article{anahtarci2023regularized,
  author  = {Anahtarci, Berkay and Kariksiz, Can Deha and Saldi, Naci},
  title   = {{Q}-Learning in Regularized Mean-field Games},
  journal = {Dynamic Games and Applications},
  year    = {2023},
  volume  = {13},
  number  = {1},
  pages   = {89--117},
  doi     = {10.1007/s13235-022-00450-2}
}

@article{wiecek2024multiple,
  author  = {Wi{\k{e}}cek, Piotr},
  title   = {Multiple-Population Discrete-Time Mean Field Games
             with Discounted and Total Payoffs:
             The Existence of Equilibria},
  journal = {Dynamic Games and Applications},
  year    = {2024},
  volume  = {14},
  number  = {4},
  pages   = {997--1026},
  doi     = {10.1007/s13235-024-00567-6}
}

@article{yang2018mean,
  title={Mean Field Equilibria for Resource Competition in Spatial Settings},
  author={Yang, Pu and Iyer, Krishnamurthy and Frazier, Peter},
  journal={Stochastic Systems},
  volume={8},
  number={4},
  pages={307--334},
  year={2018}
}

\appendix

\section{Entropy Regularization }
\begin{proof}[Proof of Proposition \ref{Prop:logit-approximation}]
An optimal deterministic policy for the unregularized problem has zero entropy. Hence the optimal value of the regularized problem satisfies
$V_\lambda(x,m)\geq V(x,m)$ for every state $x$.

The policy $\sigma_\lambda$ attains $V_\lambda$. Since
$H(\sigma_\lambda(\cdot\mid x,m))\leq \log A_{\max}$ at every state, the discounted contribution of the entropy term is at most
$\lambda\log A_{\max}/(1-\beta)$. Therefore,
\[
V_{\sigma_\lambda}(x,m)
\geq
V_\lambda(x,m)-\frac{\lambda\log A_{\max}}{1-\beta}
\geq
V(x,m)-\frac{\lambda\log A_{\max}}{1-\beta}.
\]
At a regularized MFE, the population state is invariant under $\sigma_\lambda$, so consistency holds exactly and only the optimality condition in Definition \ref{Def MFE} is approximate.
\end{proof}

\subsection{Q-Learning at a Fixed Interaction}
\label{Section:AppendixProofs}

Fix $m$. At update $h$, suppose the observed state, action, payoff, and next state are $(x_h,a_h,\pi_h,x_{h+1})$. The Q-learning update for Equation \eqref{Eq:RegularizedBellman} is
\begin{equation}
\begin{aligned}
\label{Eq:RegularizedQUpdate}
Q_{h+1}(x_h,a_h,m)
={}&
(1-\gamma_h(x_h,a_h))Q_h(x_h,a_h,m)\\
&+\gamma_h(x_h,a_h)
\left[
\pi_h+
\beta\lambda\log\left(
\sum_{\widetilde a\in\Gamma(x_{h+1})}
\exp\left(\frac{Q_h(x_{h+1},\widetilde a,m)}{\lambda}\right)
\right)
\right].
\end{aligned}
\end{equation}
All other state-action values are unchanged at update $h$.

\begin{proposition}
\label{prop:Q}
Suppose every feasible state-action pair is updated infinitely often, the sampled payoff and next state have the conditional distribution specified by the model and bounded second moments, and
\begin{equation*}
\sum_{h=0}^{\infty}\gamma_h(x,a)=\infty,
\qquad
\sum_{h=0}^{\infty}\gamma_h(x,a)^2<\infty
\end{equation*}
for every feasible $(x,a)$, where the sums use the updates of that pair. Then $Q_h(\cdot,\cdot,m)$ converges almost surely to $Q_{\lambda,m}^*$.
\end{proposition}

\begin{proof}
The conditional expectation of the update in Equation \eqref{Eq:RegularizedQUpdate} applies the mapping
\begin{equation*}
(T_{\lambda,m}Q)(x,a)
=
\pi(x,a,m)+
\beta\mathbb E\left[
\lambda\log\sum_{\widetilde a\in\Gamma(x')}
\exp\left(\frac{Q(x',\widetilde a,m)}{\lambda}\right)
\,\bigg|\,x,a,m
\right].
\end{equation*}
For any two Q-functions,
\begin{equation*}
\norm{T_{\lambda,m}Q-T_{\lambda,m}Q'}_\infty
\leq
\beta\norm{Q-Q'}_\infty,
\end{equation*}
because the log-sum-exp expression changes by no more than the largest change in its arguments. Proposition \ref{prop:Q} therefore follows from the standard convergence result for asynchronous Q-learning with a contraction mapping and the stated visitation and step-size conditions; see \cite{bertsekas1996neuro}.
\end{proof}

\section{Proofs for the Scalar Algorithms}
\label{Section:ProofScalar}

\begin{proof}[Proof of Theorem \ref{Theorem:Value}]
We first show that $f_\lambda$ in Equation \eqref{Eq:ScalarRegularizedResponse} is continuous. Continuity of the payoff and transition probabilities, together with the contraction in Equation \eqref{Eq:RegularizedBellman}, implies that $Q_{\lambda,m}^*$ is continuous in $m$. Equation \eqref{Eq:LogitPolicy} therefore implies
\begin{equation*}
\sigma_\lambda(\cdot\mid x,m_k)
\longrightarrow
\sigma_\lambda(\cdot\mid x,m)
\end{equation*}
whenever $m_k\to m$.

Let $s_k=s_\lambda^{m_k}$. Because $X$ is finite, a subsequence of $\{s_k\}$ converges to some probability vector $\bar s$. The invariance equations give
\begin{equation*}
s_k(y)
=
\sum_{x\in X}s_k(x)L_{m_k,\sigma_\lambda}(x,y).
\end{equation*}
The transition matrix in Equation \eqref{Equation:RandomizedTransition} is continuous in $m$ and in the policy. Passing to the limit along the convergent subsequence shows that $\bar s$ is invariant under $L_{m,\sigma_\lambda}$. Assumption \ref{Assumption:Unique} implies $\bar s=s_\lambda^m$. Since every convergent subsequence has this same limit, $s_\lambda^{m_k}\to s_\lambda^m$. Continuity of $M$ now gives $F_\lambda(m_k)\to F_\lambda(m)$ and hence continuity of $f_\lambda$.

Because $F_\lambda(m)\in[a,b]$ for every $m\in[a,b]$,
\begin{equation*}
f_\lambda(a)=a-F_\lambda(a)\leq0,
\qquad
f_\lambda(b)=b-F_\lambda(b)\geq0.
\end{equation*}
Algorithm \ref{alg:value_iteration_mfe} retains the left half of the current interval when $f_\lambda(m_t)>0$ and the right half when $f_\lambda(m_t)<0$. The two displayed endpoint inequalities and continuity imply that every retained interval contains a zero of $f_\lambda$. These intervals are nested and their lengths converge to zero. Their intersection therefore consists of one point $m_\lambda^*$, and continuity gives $f_\lambda(m_\lambda^*)=0$. By Equation \eqref{Eq:ScalarRegularizedResponse},
\begin{equation*}
m_\lambda^*=M(s_\lambda^{m_\lambda^*}).
\end{equation*}
The policy in Equation \eqref{Eq:LogitPolicy} is optimal for the model with the entropy term, and $s_\lambda^{m_\lambda^*}$ is invariant under that policy by its definition. Hence the policy and population state constitute a regularized MFE, as stated in Theorem \ref{Theorem:Value}. Proposition \ref{Prop:logit-approximation} gives the approximation guarantee under Definition \ref{Def MFE}.
\end{proof}

\begin{proof}[Proof of Theorem \ref{thm:Q_convergence}]
Fix an iteration $t$ and condition on all observations collected before $m_t$ is chosen. While $m_t$ is held fixed, Proposition \ref{prop:Q} gives
\begin{equation*}
Q_h(\cdot,\cdot,m_t)
\longrightarrow
Q_{\lambda,m_t}^*
\end{equation*}
almost surely. By Equation \eqref{Eq:LogitPolicy}, the policy derived from $Q_h$ converges to $\sigma_\lambda(\cdot\mid\cdot,m_t)$. After this convergence, keep that policy fixed in the separate population simulation. Assumption \ref{Assumption:Unique} and the ergodic theorem imply that the empirical state distribution in Step 3 converges almost surely to $s_\lambda^{m_t}$. Thus, the value used to update the interval is
\begin{equation*}
m_t-F_\lambda(m_t)=f_\lambda(m_t)
\end{equation*}
with conditional probability one.

There are only countably many bisection iterations. The intersection of these countably many events still has probability one. On that event, every update in Algorithm \ref{alg:bisection_q_learning} uses the same value $f_\lambda(m_t)$ as Algorithm \ref{alg:value_iteration_mfe}. Theorem \ref{Theorem:Value} therefore gives $m_t\to m_\lambda^*$ with $f_\lambda(m_\lambda^*)=0$, and the policy and invariant distribution at $m_\lambda^*$ satisfy the conclusion of Theorem \ref{thm:Q_convergence}.
\end{proof}

\section{Proof of the Low-Dimensional Simplicial Result}
\label{Section:ProofVectorMFE}

The population trajectory at a grid vertex is generated under policies that change as the policy estimate improves. The next lemma shows that this does not affect the limiting empirical distribution when the transition matrices themselves converge.

\begin{lemma}
\label{Lemma:PopulationUnderConvergingPolicies}
Let $X$ be finite. At transition $k$, suppose $X_k$ and the
stochastic matrix $P_k$ are known and that, conditional on all
information available before this transition, $X_{k+1}$ has
distribution $P_k(X_k,\cdot)$. If $P_k\to P$ entrywise almost surely and $P$ has a unique invariant distribution $s$, then
\begin{equation*}
\frac{1}{N}\sum_{k=1}^N1_{\{X_k=y\}}
\longrightarrow
s(y)
\end{equation*}
almost surely for every $y\in X$.
\end{lemma}

\begin{proof}
Fix a function $\varphi:X\to\mathbb R$, and let $\mathcal F_k$
contain all information available before transition $k$, including
$X_k$ and $P_k$. Then
\begin{equation*}
\mathbb E[\varphi(X_{k+1})\mid\mathcal F_k]
=(P_k\varphi)(X_k).
\end{equation*}
Hence the variables
\begin{equation*}
\varphi(X_{k+1})-(P_k\varphi)(X_k)
\end{equation*}
are bounded martingale differences, and their averages converge almost surely to zero. Also,
\begin{equation*}
\frac{1}{N}\sum_{k=1}^N\bigl(\varphi(X_{k+1})-\varphi(X_k)\bigr)
=\frac{\varphi(X_{N+1})-\varphi(X_1)}{N}
\longrightarrow0.
\end{equation*}
Since $X$ is finite and $P_k\to P$ entrywise,
\begin{equation*}
\max_{x\in X}|(P_k\varphi)(x)-(P\varphi)(x)|\longrightarrow0.
\end{equation*}
Adding the martingale-difference average, the telescoping average, and the difference between $P_k\varphi$ and $P\varphi$ shows that every subsequential limit $\mu$ of the empirical distributions satisfies
\begin{equation*}
\sum_{x\in X}\varphi(x)\mu(x)
=
\sum_{x\in X}(P\varphi)(x)\mu(x).
\end{equation*}
Thus $\mu P=\mu$. Uniqueness of the invariant distribution gives $\mu=s$. Because the probability simplex is compact, the full sequence of empirical distributions converges to $s$.
\end{proof}

\begin{proof}[Proof of Theorem \ref{thm:online_simplicial_mfe}]
The proof has four parts. We first establish continuity of the policy
and population distribution as functions of $m$. We then establish
convergence at every vertex selected infinitely often. The finiteness
of the grid makes this convergence uniform over the vertices selected
at a sufficiently large iteration. Finally, Equation
\eqref{Eq:SimplicialFixedPoint} converts that uniform convergence into
a bound on $\norm{f_\lambda(m^j)}$ at every vertex of the selected
simplex. 

\textit{Step 1: Continuity in the interaction.}
The contraction in Equation \eqref{Eq:RegularizedBellman} and continuity of the payoff and transition probabilities imply that $Q_{\lambda,m}^*$ is continuous in $m$. Equation \eqref{Eq:LogitPolicy} then implies continuity of $\sigma_\lambda(\cdot\mid\cdot,m)$.

Let $m_k\to m$ and write $s_k=s_\lambda^{m_k}$. A subsequence of $\{s_k\}$ converges to some probability vector $\bar s$ because $X$ is finite. By Equation \eqref{Equation:RandomizedTransition},
\begin{equation*}
L_{m_k,\sigma_\lambda}(x,y)
\longrightarrow
L_{m,\sigma_\lambda}(x,y)
\end{equation*}
for every $x,y\in X$. Passing to the limit in
\begin{equation*}
s_k(y)=\sum_{x\in X}s_k(x)L_{m_k,\sigma_\lambda}(x,y)
\end{equation*}
shows that $\bar s$ is invariant under $L_{m,\sigma_\lambda}$. Assumption \ref{Assumption:Vector} gives $\bar s=s_\lambda^m$. Therefore $s_\lambda^{m_k}\to s_\lambda^m$, and Equation \eqref{Eq:VectorResponse} implies that $F_\lambda$ and $f_\lambda$ are continuous.

The set $\{m\in\mathcal M:F_\lambda(m)=m\}$ is nonempty by Brouwer's fixed-point theorem and is compact by continuity. It follows that $\mathcal E_\lambda$ is nonempty and compact. The interpolated function $\widehat F_t$ is also a continuous map from $\mathcal M$ into itself. A fixed point of $\widehat F_t$ belongs to at least one simplex, and its barycentric coordinates in that simplex satisfy Equation \eqref{Eq:SimplicialFixedPoint}. Hence Step 1 of Algorithm \ref{alg:online_simplicial_mfe} is well defined.

\textit{Step 2: Convergence at a vertex selected infinitely often.}
Fix a vertex $m^j$ that is selected infinitely often. Each visit to
$m^j$ adds $H\geq1$ Q-learning updates, so the total number of updates
at this vertex diverges. Assumption \ref{Assumption:Vector} and
Proposition \ref{prop:Q} therefore imply
\begin{equation*}
\widehat Q_t^j\to Q_{\lambda,m^j}^*
\end{equation*}
almost surely along the iterations in which $m^j$ is selected.
Equation \eqref{Eq:LogitPolicy} then gives
$\widehat\sigma_t^j\to
\sigma_\lambda(\cdot\mid\cdot,m^j)$.

The transition matrices used in the stored population trajectory at
$m^j$ consequently converge to $L_{m^j,\sigma_\lambda}$. Each visit
also adds $K\geq1$ transitions to this trajectory, so Lemma
\ref{Lemma:PopulationUnderConvergingPolicies} and Assumption
\ref{Assumption:Vector} imply
$\widehat s_t^j\to s_\lambda^{m^j}$ almost surely. Finally,
$\widehat\mu_t^j(x,a)=\widehat s_t^j(x)\widehat\sigma_t^j(a\mid x)$
converges to $\mu_\lambda^{m^j}(x,a)$, and continuity of $M$ gives
$\widehat F_t^j\to F_\lambda(m^j)$.

Thus, along the iterations in which $m^j$ is selected,
\begin{equation}
\label{Eq:VertexConvergence}
\widehat\sigma_t^j\to\sigma_\lambda(\cdot\mid\cdot,m^j),
\qquad
\widehat s_t^j\to s_\lambda^{m^j},
\qquad
\widehat F_t^j\to F_\lambda(m^j)
\end{equation}
almost surely.

\textit{Step 3: Uniform convergence over the vertices selected at a sufficiently large iteration.}
Let
\begin{equation*}
\mathcal J_\infty
=
\{j:m^j\text{ is selected at infinitely many iterations}\}.
\end{equation*}
Every vertex outside $\mathcal J_\infty$ has a last iteration at which it is selected. Since the grid has finitely many vertices, after the largest of these finitely many iterations every selected simplex has all its vertices in $\mathcal J_\infty$.

The set $\mathcal J_\infty$ is finite. For any $\varepsilon>0$, Equation \eqref{Eq:VertexConvergence} therefore holds within $\varepsilon$ at every vertex of $\mathcal J_\infty$ once each of these finitely many vertices has received sufficiently many observations. The simplex at iteration $t$ is selected using the estimates available at iteration $t-1$. For every vertex selected infinitely often, the number of visits before iteration $t$ also converges to infinity. Since every sufficiently late selected simplex uses only vertices in $\mathcal J_\infty$, we obtain
\begin{equation}
\label{Eq:UniformVectorResponse}
\max_{j\in\mathcal I_t}
\norm{\widehat F_{t-1}^j-F_\lambda(m^j)}
\longrightarrow0
\end{equation}
almost surely. Applying the same finite-set argument to the first two convergences in Equation \eqref{Eq:VertexConvergence} gives
\begin{equation}
\label{Eq:UniformVectorPolicyPopulation}
\max_{j\in\mathcal I_t}
\left[
\max_{x\in X}
\norm{\widehat\sigma_t^j(\cdot\mid x)-\sigma_\lambda(\cdot\mid x,m^j)}_1
+
\norm{\widehat s_t^j-s_\lambda^{m^j}}_1
\right]
\longrightarrow0
\end{equation}
almost surely.

\textit{Step 4: Distance from the equilibrium set.}
For $r\geq0$, define
\begin{equation*}
\omega_f(r)
=
\sup\{\norm{f_\lambda(m)-f_\lambda(m')}:m,m'\in\mathcal M,
\ \norm{m-m'}\leq r\}.
\end{equation*}
Continuity on the compact set $\mathcal M$ implies $\omega_f(r)\to0$ as $r\downarrow0$.

At iteration $t$, let $m^{j_{t,0}},\ldots,m^{j_{t,n}}$ and $\theta_{t,0},\ldots,\theta_{t,n}$ satisfy Equation \eqref{Eq:SimplicialFixedPoint} for $\widehat F_{t-1}$. Fix any $j\in\mathcal I_t$. Using $f_\lambda(m)=m-F_\lambda(m)$ and Equation \eqref{Eq:SimplicialFixedPoint},
\begin{equation}
\begin{aligned}
\label{Eq:VectorResidualIdentity}
f_\lambda(m^j)
={}&
\sum_{\ell=0}^n\theta_{t,\ell}
\left[f_\lambda(m^j)-f_\lambda(m^{j_{t,\ell}})\right]\\
&+
\sum_{\ell=0}^n\theta_{t,\ell}
\left[\widehat F_{t-1}^{j_{t,\ell}}-F_\lambda(m^{j_{t,\ell}})\right].
\end{aligned}
\end{equation}
Every pair of vertices in the simplex is at distance at most $h$. Taking norms in Equation \eqref{Eq:VectorResidualIdentity} and using Equation \eqref{Eq:UniformVectorResponse} gives
\begin{equation}
\label{Eq:VectorFixedGridBound}
\limsup_{t\to\infty}
\max_{j\in\mathcal I_t}\norm{f_\lambda(m^j)}
\leq
\omega_f(h)
\end{equation}
almost surely.

For $m\in\mathcal M$, define
\begin{equation*}
z_\lambda(m)
=
(m,\sigma_\lambda(\cdot\mid\cdot,m),s_\lambda^m).
\end{equation*}
Fix $\epsilon>0$. If $\operatorname{dist}(z_\lambda(m),\mathcal E_\lambda)<\epsilon/2$ for every $m\in\mathcal M$, Equation \eqref{Eq:UniformVectorPolicyPopulation} gives the desired conclusion. Otherwise, the set
\begin{equation*}
\{m\in\mathcal M:
\operatorname{dist}(z_\lambda(m),\mathcal E_\lambda)\geq\epsilon/2\}
\end{equation*}
is nonempty and compact. It contains no zero of $f_\lambda$. Hence
\begin{equation}
\label{Eq:VectorPositiveResidualAwayFromEquilibrium}
\rho(\epsilon)
=
\min_{\operatorname{dist}(z_\lambda(m),\mathcal E_\lambda)\geq\epsilon/2}
\norm{f_\lambda(m)}
>0.
\end{equation}
Choose $h_\epsilon>0$ such that
\begin{equation*}
\omega_f(h_\epsilon)<\rho(\epsilon)/2.
\end{equation*}
For $h\leq h_\epsilon$, Equation \eqref{Eq:VectorFixedGridBound} implies that, at every sufficiently large iteration and every $j\in\mathcal I_t$,
\begin{equation*}
\norm{f_\lambda(m^j)}<\rho(\epsilon).
\end{equation*}
By Equation \eqref{Eq:VectorPositiveResidualAwayFromEquilibrium}, this inequality gives
\begin{equation*}
\operatorname{dist}(z_\lambda(m^j),\mathcal E_\lambda)<\epsilon/2.
\end{equation*}
By Equation \eqref{Eq:UniformVectorPolicyPopulation}, at every sufficiently large iteration,
\begin{equation*}
\max_{j\in\mathcal I_t}
\left[
\max_{x\in X}
\norm{\widehat\sigma_t^j(\cdot\mid x)-\sigma_\lambda(\cdot\mid x,m^j)}_1
+
\norm{\widehat s_t^j-s_\lambda^{m^j}}_1
\right]
<\epsilon/2.
\end{equation*}
For each $j\in\mathcal I_t$, the last two displayed inequalities and Equation \eqref{Eq:VectorTripleDistance} imply
\begin{equation*}
\operatorname{dist}
\left(
(m^j,\widehat\sigma_t^j,\widehat s_t^j),
\mathcal E_\lambda
\right)
<\epsilon.
\end{equation*}
Step 3 of Algorithm \ref{alg:online_simplicial_mfe} reports one of these vertices, so the same inequality holds for $(\widehat m_t,\widehat\sigma_t,\widehat s_t)$.
\end{proof}

Permanent types require no additional fixed-point argument. Following \cite{weintraub2008markov} and \cite{light2022mean}, stationarity is imposed separately for each type and the conditional distributions are combined using the fixed type shares. The continuity argument in Step 1 and the convergence arguments in Steps 2--4 apply componentwise to the finitely many conditional distributions.

\subsection{Policy Estimation in the Heterogeneous-Agent Macro Application}
\label{sec:SavingsEstimates}

At every selected interaction in Section \ref{Sec:VectorMacroApplication}, the simulator is queried at each state-action pair. The code forms sample averages of the payoff and transition probabilities and solves the regularized Bellman equation obtained by replacing the true quantities with these averages. Let $T_m=T_{\lambda,m}$ be the operator on Q-functions defined
in the proof of Proposition~\ref{prop:Q}, and let $\widehat T_N$
be the same operator with the payoffs and transition
probabilities replaced by their sample averages after $N$
observations at each state-action pair.

\begin{lemma}
\label{Lemma:SavingsPolicyConvergence}
Suppose the estimated payoffs converge uniformly to the true payoffs,
the estimated transition probabilities converge uniformly in total
variation, and
\begin{equation*}
\norm{\widehat Q_N-\widehat T_N\widehat Q_N}_\infty
\longrightarrow0.
\end{equation*}
Then $\widehat Q_N\to Q_{\lambda,m}^*$ in the sup norm.
\end{lemma}

\begin{proof}
Both $T_m$ and $\widehat T_N$ are $\beta$-contractions. Since $T_mQ_{\lambda,m}^*=Q_{\lambda,m}^*$,
\begin{equation}
\begin{aligned}
\label{Eq:SavingsContractionBound}
(1-\beta)\norm{\widehat Q_N-Q_{\lambda,m}^*}_\infty
\leq{}&
\norm{\widehat Q_N-\widehat T_N\widehat Q_N}_\infty\\
&+
\norm{\widehat T_NQ_{\lambda,m}^*-T_mQ_{\lambda,m}^*}_\infty.
\end{aligned}
\end{equation}
The first term on the right-hand side of Equation \eqref{Eq:SavingsContractionBound} converges to zero by assumption. The second converges to zero because the state and action spaces are finite and the estimated payoffs and transition probabilities converge uniformly. Therefore the left-hand side converges to zero.
\end{proof}

Equation \eqref{Eq:LogitPolicy} therefore gives convergence of the
policy estimates used in the heterogeneous-agent macro application.
Lemma \ref{Lemma:PopulationUnderConvergingPolicies} then gives
convergence of the corresponding population estimates. Hence the
vertexwise convergence established in Step 2 of the proof of Theorem
\ref{thm:online_simplicial_mfe} also holds for this implementation,
and the remainder of that proof is unchanged.

\section{Proofs of Existence and Comparative Statics}
\label{sec:existenceappendix}

\subsection{Existence of MFE}

\begin{proof}[Proof of Theorem \ref{thm:existence}]
The proof first constructs an equilibrium for every $\lambda>0$. It then lets $\lambda$ converge to zero and verifies separately the consistency and optimality conditions in Definition \ref{Def MFE}.

\textit{Step 1: For every $\lambda>0$, a regularized MFE exists.}
Fix $\lambda\in(0,\bar\lambda]$. Because $A$ is finite and $\Gamma$ is continuous, the feasible action set is locally constant. Indeed, for every $x\in X$, continuity of $\Gamma$ and the positive distance between distinct elements of $A$ give a neighborhood of $x$ on which $\Gamma$ equals $\Gamma(x)$. Consequently, for every $a\in A$, the set
\begin{equation*}
D_a=\{x\in X:a\in\Gamma(x)\}
\end{equation*}
is both open and closed.

For a bounded continuous function $v:X\times\mathcal M\to\mathbb R$ and a feasible action $a$, weak continuity of $P$ implies that
\begin{equation*}
(x,m)\longmapsto
\pi(x,a,m)+\beta\int_Xv(y,m)P(dy\mid x,a,m)
\end{equation*}
is continuous on $D_a\times\mathcal M$. To see this, if $(x_k,m_k)\to(x,m)$, then $P(\cdot\mid x_k,a,m_k)\Rightarrow P(\cdot\mid x,a,m)$ and $\delta_{m_k}\Rightarrow\delta_m$. Here $\delta_{m_k}$ denotes unit mass at $m_k$. The product measures $P(\cdot\mid x_k,a,m_k)\otimes\delta_{m_k}$ converge weakly on $X\times\mathcal M$, so the integrals of $v(y,m_k)$ converge. Since the feasible set is locally constant, the right-hand side of Equation \eqref{Eq:RegularizedBellman} is continuous whenever $v$ is continuous. The operator in Equation \eqref{Eq:RegularizedBellman} is a $\beta$-contraction in the sup norm. Starting from a bounded continuous function and applying this operator repeatedly therefore shows that its fixed point $V_\lambda(x,m)$ is bounded and continuous in $(x,m)$. Equation \eqref{Eq:LogitPolicy} then implies that $\sigma_\lambda(a\mid x,m)$ is continuous on $D_a\times\mathcal M$ and is zero outside $D_a$.

We next show that $s_\lambda^m$ varies continuously with $m$. Let $m_k\to m$ and write $s_k=s_\lambda^{m_k}$. For every $R>0$, Assumption \ref{Assumption: exist}(ii) gives
\begin{equation}
\label{Eq:ExistenceTightness}
s_k\bigl(\{x:d(x,x_0)>R\}\bigr)
\leq
\frac{C}{R^p}.
\end{equation}
Because closed balls are compact, Equation \eqref{Eq:ExistenceTightness} implies that $\{s_k\}$ is tight. Pass to a subsequence, still indexed by $k$, such that $s_k\Rightarrow\bar s$.

For a bounded continuous function $v:X\to\mathbb R$, define
\begin{equation*}
h_v(x,m)
=
\sum_{a\in\Gamma(x)}
\sigma_\lambda(a\mid x,m)
\int_Xv(y)P(dy\mid x,a,m).
\end{equation*}
Because each $\sigma_\lambda(a\mid x,m)$ is continuous and $P$ is weakly continuous, $h_v$ is bounded and continuous on $X\times\mathcal M$. Since $s_k\Rightarrow\bar s$ and $m_k\to m$,
\begin{equation}
\label{Eq:ExistenceKernelLimit}
\int_Xh_v(x,m_k)s_k(dx)
\longrightarrow
\int_Xh_v(x,m)\bar s(dx).
\end{equation}
The invariance of $s_k$ gives
\begin{equation*}
\int_Xv(y)s_k(dy)
=
\int_Xh_v(x,m_k)s_k(dx).
\end{equation*}
Taking limits and using Equation \eqref{Eq:ExistenceKernelLimit} shows that $\bar s$ is invariant under the policy $\sigma_\lambda(\cdot\mid\cdot,m)$. Assumption \ref{Assumption: exist}(i) therefore gives $\bar s=s_\lambda^m$. Every weakly convergent subsequence has the same limit, so $s_\lambda^{m_k}\Rightarrow s_\lambda^m$. Assumption \ref{Assumption: exist}(iii) now implies
\begin{equation*}
M(s_\lambda^{m_k})\longrightarrow M(s_\lambda^m).
\end{equation*}
Hence $m\mapsto M(s_\lambda^m)$ is a continuous function from $\mathcal M$ into itself. Brouwer's fixed-point theorem gives $m_\lambda\in\mathcal M$ such that
\begin{equation}
\label{Eq:RegularizedExistenceFixedPoint}
m_\lambda=M(s_\lambda^{m_\lambda}).
\end{equation}
When $\mathcal M=[a,b]$, the endpoint inequalities and the intermediate value theorem give the same conclusion.

\textit{Step 2: Select a convergent sequence as $\lambda\downarrow0$.}
Choose $\lambda_k\downarrow0$ and let $m_k=m_{\lambda_k}$, $s_k=s_{\lambda_k}^{m_k}$, and $\sigma_k=\sigma_{\lambda_k}(\cdot\mid\cdot,m_k)$. Compactness of $\mathcal M$ and Equation \eqref{Eq:ExistenceTightness}, now applied uniformly in $\lambda$, allow us to pass to a subsequence such that
\begin{equation}
\label{Eq:ExistenceSubsequence}
m_k\to m^*,
\qquad
s_k\Rightarrow s^*.
\end{equation}
The lower semicontinuity of $x\mapsto d(x,x_0)^p$ gives
\begin{equation*}
\int_Xd(x,x_0)^p s^*(dx)\leq C.
\end{equation*}
Assumption \ref{Assumption: exist}(iii), Equation \eqref{Eq:RegularizedExistenceFixedPoint}, and Equation \eqref{Eq:ExistenceSubsequence} therefore imply
\begin{equation}
\label{Eq:ExistenceConsistencyInteraction}
m^*=M(s^*).
\end{equation}

Define the stationary state-action measures
\begin{equation*}
\mu_k(dx,da)=s_k(dx)\sigma_k(da\mid x).
\end{equation*}
The sequence $\{\mu_k\}$ is tight because $A$ is finite and $\{s_k\}$ is tight. Pass to a further subsequence such that $\mu_k\Rightarrow\mu$. The state marginal of $\mu$ is $s^*$. The graph of $\Gamma$ is closed, and every $\mu_k$ assigns probability one to this graph, so $\mu$ does as well. Because $X$ and $A$ are standard Borel spaces, there is a measurable stationary policy $\sigma^*$ such that
\begin{equation}
\label{Eq:ExistenceLimitPolicy}
\mu(dx,da)=s^*(dx)\sigma^*(da\mid x).
\end{equation}
This policy is feasible for $s^*$-almost every state.

\textit{Step 3: The limiting population state is invariant under $\sigma^*$.}
Fix a bounded continuous function $v:X\to\mathbb R$. For each $a\in A$, define
\begin{equation*}
\psi_v(x,a,m)
=
\begin{cases}
\displaystyle \int_Xv(y)P(dy\mid x,a,m), & x\in D_a,\\
0, & x\notin D_a.
\end{cases}
\end{equation*}
The set $D_a$ is open and closed, so weak continuity of $P$ implies that $\psi_v$ is bounded and continuous on $X\times A\times\mathcal M$. Invariance of $s_k$ gives
\begin{equation*}
\int_Xv(y)s_k(dy)
=
\int_{X\times A}\psi_v(x,a,m_k)\mu_k(dx,da).
\end{equation*}
The weak convergence in Equation \eqref{Eq:ExistenceSubsequence}, the weak convergence $\mu_k\Rightarrow\mu$, and $m_k\to m^*$ allow us to pass to the limit:
\begin{equation}
\label{Eq:ExistenceInvariantLimit}
\int_Xv(y)s^*(dy)
=
\int_{X\times A}\psi_v(x,a,m^*)\mu(dx,da).
\end{equation}
Equations \ref{Eq:ExistenceLimitPolicy} and \ref{Eq:ExistenceInvariantLimit} show that $s^*$ is invariant under $\sigma^*$ at interaction $m^*$.

\textit{Step 4: The limiting policy is optimal without the entropy term.}
Let $V$ be the value function in Equation \eqref{Equation:BellmanKernel}. Equation \eqref{Equation:BellmanKernel} defines a $\beta$-contraction on bounded continuous functions by the same product-measure argument used for Equation \eqref{Eq:RegularizedBellman}. Hence $V(x,m)$ is bounded and continuous in $(x,m)$. For $a\in A$, define
\begin{equation*}
D(x,a,m)
=
\begin{cases}
V(x,m)-\pi(x,a,m)-\beta\displaystyle\int_XV(y,m)P(dy\mid x,a,m), & x\in D_a,\\
0, & x\notin D_a.
\end{cases}
\end{equation*}
The function $D$ is bounded and continuous. On the graph of $\Gamma$, Equation \eqref{Equation:BellmanKernel} gives $D(x,a,m)\geq0$, with equality exactly when $a\in G(x,m)$.

Let $V_{\sigma_k}(x,m_k)$ denote the payoff generated by $\sigma_k$ when the entropy term is omitted. Proposition \ref{Prop:logit-approximation} gives
\begin{equation}
\label{Eq:ExistenceValueLoss}
0\leq V(x,m_k)-V_{\sigma_k}(x,m_k)
\leq
\frac{\lambda_k\log A_{\max}}{1-\beta}
\end{equation}
for every state. Set
\begin{equation*}
d_k(x)=\int_A D(x,a,m_k)\sigma_k(da\mid x).
\end{equation*}
The Bellman equation for $V$ and the policy-evaluation equation for $V_{\sigma_k}$ imply
\begin{equation*}
V(\cdot,m_k)-V_{\sigma_k}(\cdot,m_k)
=
d_k+
\beta L_{m_k,\sigma_k}
\bigl(V(\cdot,m_k)-V_{\sigma_k}(\cdot,m_k)\bigr).
\end{equation*}
Integrating this equality with respect to the invariant distribution $s_k$ gives
\begin{align}
\label{Eq:IntegratedBellmanGap}
\int_{X\times A}D(x,a,m_k)\mu_k(dx,da)
&=(1-\beta)\int_X
\left[V(x,m_k)-V_{\sigma_k}(x,m_k)\right]s_k(dx)\notag\\
&\leq \lambda_k\log A_{\max},
\end{align}
where the inequality follows from Equation \eqref{Eq:ExistenceValueLoss}.
Since $D$ is bounded and continuous, Equation \eqref{Eq:IntegratedBellmanGap}, $\mu_k\Rightarrow\mu$, and $m_k\to m^*$ imply
\begin{equation*}
\int_{X\times A}D(x,a,m^*)\mu(dx,da)=0.
\end{equation*}
The integrand is nonnegative on the feasible graph. Hence Equation \eqref{Eq:ExistenceLimitPolicy} implies
\begin{equation}
\label{Eq:ExistenceOptimalAlmostEverywhere}
\sigma^*(G(x,m^*)\mid x)=1
\end{equation}
for $s^*$-almost every $x$.

It remains only to obtain optimality at states that have zero probability under $s^*$. Order the finite action set as $a_1,\ldots,a_{|A|}$ and, at each state, choose the first feasible action that belongs to $G(x,m^*)$. The sets on which a given $a_\ell$ is the first such action are measurable because $D(\cdot,a_\ell,m^*)$ is continuous. Replace $\sigma^*$ on the $s^*$-null set where Equation \eqref{Eq:ExistenceOptimalAlmostEverywhere} may fail by the point mass on this selected action. This modification does not change Equation \eqref{Eq:ExistenceInvariantLimit}.

After the modification, $\sigma^*$ assigns probability one to $G(x,m^*)$ at every state. Averaging Equation \eqref{Equation:BellmanKernel} over $\sigma^*(\cdot\mid x)$ therefore gives the policy-evaluation equation for $V(\cdot,m^*)$. That equation has a unique bounded solution because its right-hand side is a $\beta$-contraction. Hence
\begin{equation*}
V_{\sigma^*}(x,m^*)=V(x,m^*)
\end{equation*}
for every $x\in X$. This is the optimality condition in Definition \ref{Def MFE}. Equation \eqref{Eq:ExistenceInvariantLimit} gives invariance, and Equation \eqref{Eq:ExistenceConsistencyInteraction} gives $m^*=M(s^*)$. Hence $(\sigma^*,s^*)$ is an MFE.
\end{proof}

\subsection{Comparative Statics}
\label{Section:ComparativeAnalysis}

\begin{proof}[Proof of Proposition \ref{Prop:comp}]
For $m\in[a,b]$ and $z\in Z$, define
\begin{equation*}
\phi(m,z)=F(m,z)-m.
\end{equation*}
Because $F(m,z)\in[a,b]$,
\begin{equation*}
\phi(a,z)\geq0,
\qquad
\phi(b,z)\leq0.
\end{equation*}
Continuity of $F(\cdot,z)$ implies that the zero set of $\phi(\cdot,z)$ is nonempty and compact. Its lowest and highest elements, $\underline m(z)$ and $\overline m(z)$, therefore exist.

Let $z_2\geq_Zz_1$. The stochastic dominance assumption and monotonicity of $M$ give
\begin{equation}
\label{Eq:ComparativePhiOrder}
\phi(m,z_2)\geq\phi(m,z_1)
\end{equation}
for every $m\in[a,b]$. At $m=\overline m(z_1)$, Equation \eqref{Eq:ComparativePhiOrder} gives $\phi(m,z_2)\geq0$, while $\phi(b,z_2)\leq0$. The intermediate value theorem therefore gives a zero of $\phi(\cdot,z_2)$ in $[\overline m(z_1),b]$. Hence
\begin{equation*}
\overline m(z_2)\geq\overline m(z_1).
\end{equation*}

Similarly, at $m=\underline m(z_2)$, Equation \eqref{Eq:ComparativePhiOrder} gives $\phi(m,z_1)\leq0$, while $\phi(a,z_1)\geq0$. Thus $\phi(\cdot,z_1)$ has a zero in $[a,\underline m(z_2)]$, which implies
\begin{equation*}
\underline m(z_1)\leq\underline m(z_2).
\end{equation*}
\end{proof}

\section{Proofs for the Dynamic Oligopoly Results}
\label{Section:OligopolyProofs}

For a function $f$ on $X\times[a,b]$, define
\begin{equation}
\label{eq:H_def}
\Hfunc(x,a;m,f)
=
\util(x,m)-\cost(a)+\beta\sum_{y\in X}\Wtrans(x,a,y)f(y,m),
\end{equation}
and write
\begin{equation*}
E_f(x,a)=\sum_{y\in X}\Wtrans(x,a,y)f(y,m).
\end{equation*}
For a function on $\mathbb N_0$, let
\begin{equation*}
\Delta f(x)=f(x+1)-f(x),
\qquad
\Delta^2f(x)=\Delta f(x+1)-\Delta f(x).
\end{equation*}
Thus, $f$ is discretely convex when $\Delta^2f(x)\geq0$ for every $x$.

\begin{lemma} \label{Lemma:ErgodicCapacity}
Every stationary policy induces an irreducible, aperiodic, positive recurrent Markov chain. The invariant distributions have uniformly bounded moments of every finite order.
\end{lemma}

\begin{proof}
The lower action bound $\underline a>0$ makes every upward transition possible, depreciation makes every downward transition possible, and Equation \eqref{Eq:Oligopoly} assigns positive probability to remaining in the current state. Hence the chain is irreducible and aperiodic under every stationary policy.

First consider a deterministic policy $g$. Define $v(0)=1$ and, for $x\geq0$,
\begin{equation}
\label{Eq:BirthDeathWeights}
\frac{v(x+1)}{v(x)}
=
\frac{W(x,g(x),x+1)}{W(x+1,g(x+1),x)}.
\end{equation}
Because every action is at most one, Equation \eqref{Eq:Oligopoly} gives
\begin{equation}
\label{Eq:BirthDeathGeometricBound}
\frac{v(x+1)}{v(x)}
\leq
\frac{1-\delta}{\delta}
<1.
\end{equation}
The weights are summable and therefore normalize to an invariant distribution. Equation \eqref{Eq:BirthDeathGeometricBound} also gives a geometric upper bound on its tail. Irreducibility then implies positive recurrence and uniqueness.

For a randomized policy $\sigma$, set
\begin{equation*}
r_x=\int_A\frac{1}{1+a}\,\sigma(da\mid x).
\end{equation*}
The upward and downward transition probabilities are $(1-\delta)(1-r_x)$ and $\delta r_x$. Since $a\leq1$, $r_x\geq1/2$, and the ratio corresponding to Equation \eqref{Eq:BirthDeathWeights} satisfies
\begin{equation*}
\frac{(1-\delta)(1-r_x)}{\delta r_{x+1}}
\leq
\frac{1-\delta}{\delta}.
\end{equation*}
Thus, Equation \eqref{Eq:BirthDeathGeometricBound} holds uniformly over deterministic and randomized policies. A geometric tail has every finite moment, and the bound is uniform over policies and interactions.
\end{proof}

\begin{lemma}
\label{Lemma:WeightedDPCapacity}
Under Assumption \ref{Assumption:CapacityCompt}, the discounted control problem has a finite value function. The value function is continuous in the interaction and in the comparative-statics parameter.
\end{lemma}

\begin{proof}
The investment cost is bounded because $A$ is compact. Let $p_u$ be the exponent in Assumption \ref{Assumption:CapacityCompt}(v). For a constant $C>0$, define
\begin{equation*}
\omega_C(x)=C+(x+1)^{p_u}.
\end{equation*}
A transition changes the state by at most one, so
\begin{equation}
\label{Eq:WeightedDriftBound}
\sup_{x,a}
\frac{\sum_yW(x,a,y)\omega_C(y)}{\omega_C(x)}
\leq
\sup_{x\geq0}
\frac{C+(x+2)^{p_u}}{C+(x+1)^{p_u}}
=:\kappa_C.
\end{equation}
The right-hand side converges downward to one as $C\to\infty$. Choose $C$ so that $\beta\kappa_C<1$.

For $\norm{f}_{\omega_C}=\sup_x|f(x)|/\omega_C(x)$, Equation \eqref{Eq:WeightedDriftBound} gives
\begin{equation*}
\norm{Tf-Tg}_{\omega_C}
\leq
\beta\kappa_C\norm{f-g}_{\omega_C},
\end{equation*}
where $T$ is the Bellman operator associated with Equation \eqref{eq:H_def}. Assumption \ref{Assumption:CapacityCompt}(v) implies that the single-period payoff has finite $\omega_C$-norm. Hence $T$ has a unique fixed point with finite $\omega_C$-norm, and value iteration converges to it.

Now let $(m_n,z_n)\to(m,z)$ and denote the corresponding Bellman operators by $T_n$ and $T$. The transition probabilities in Equation \eqref{Eq:Oligopoly} do not depend on $(m,z)$. Assumption \ref{Assumption:CapacityCompt}(v) therefore gives
\begin{equation*}
\norm{T_nV(\cdot,m,z)-TV(\cdot,m,z)}_{\omega_C}
\longrightarrow0.
\end{equation*}
Using the contraction bound for $T_n$,
\begin{align*}
\norm{V(\cdot,m_n,z_n)-V(\cdot,m,z)}_{\omega_C}
&\leq
\beta\kappa_C
\norm{V(\cdot,m_n,z_n)-V(\cdot,m,z)}_{\omega_C}\\
&\quad+
\norm{T_nV(\cdot,m,z)-TV(\cdot,m,z)}_{\omega_C}.
\end{align*}
Since $\beta\kappa_C<1$, the value functions converge in the weighted norm. In particular, $V(x,m,z)$ is continuous in $(m,z)$ for every state. Lemma \ref{Lemma:G-singlevalued} proves that the maximizer in Equation \eqref{eq:H_def} is unique. Continuity of the objective and compactness of $A$ then imply continuity of the policy.
\end{proof}

\begin{lemma} \label{Lemma:IncreasingConvex}
The value function is discretely convex and strictly increasing in the state.
\end{lemma}

\begin{proof}
We show that the Bellman operator preserves increasing and discretely convex functions. Suppose $f$ has these two properties. For $x\geq1$, taking first differences gives
\begin{equation}
\label{Eq:ContinuationFirstDifference}
E_f(x+1,a)-E_f(x,a)
=
\frac{(1-\delta)a}{1+a}\Delta f(x+1)
+
\frac{1-\delta+\delta a}{1+a}\Delta f(x)
+
\frac{\delta}{1+a}\Delta f(x-1).
\end{equation}
The right-hand side is nonnegative. At $x=0$,
\begin{equation}
\label{Eq:ContinuationBoundaryFirstDifference}
E_f(1,a)-E_f(0,a)
=
\frac{(1-\delta)a}{1+a}\Delta f(1)
+
\frac{1-\delta+\delta a}{1+a}\Delta f(0)
\geq0.
\end{equation}
Equations \ref{Eq:ContinuationFirstDifference} and \ref{Eq:ContinuationBoundaryFirstDifference} show that $E_f(\cdot,a)$ is increasing.

For $x\geq1$, the second difference is
\begin{equation}
\label{Eq:ContinuationSecondDifference}
\Delta_x^2E_f(x,a)
=
\frac{(1-\delta)a}{1+a}\Delta^2f(x+1)
+
\frac{1-\delta+\delta a}{1+a}\Delta^2f(x)
+
\frac{\delta}{1+a}\Delta^2f(x-1).
\end{equation}
At the boundary,
\begin{equation}
\label{Eq:ContinuationBoundarySecondDifference}
\Delta_x^2E_f(0,a)
=
\frac{(1-\delta)a}{1+a}\Delta^2f(1)
+
\frac{1-\delta+\delta a}{1+a}\Delta^2f(0)
+
\frac{\delta}{1+a}\Delta f(0).
\end{equation}
All terms on the right-hand sides of Equations \ref{Eq:ContinuationSecondDifference} and \ref{Eq:ContinuationBoundarySecondDifference} are nonnegative. Thus $E_f(\cdot,a)$ is discretely convex.

Assumption \ref{Assumption:CapacityCompt}(iii) implies that $\Hfunc(\cdot,a;m,f)$ is increasing and discretely convex for every action. To see that maximization over $a$ preserves discrete convexity, linearly interpolate each sequence $\Hfunc(\cdot,a;m,f)$ between consecutive integers. Each interpolation is convex, and their pointwise maximum is convex. At every integer, this maximum equals $Tf$. Hence $Tf$ is discretely convex; it is also increasing because each function being maximized is increasing.

Start value iteration from the zero function. Equations \ref{Eq:ContinuationFirstDifference}--\ref{Eq:ContinuationBoundarySecondDifference} show by induction that every value-iteration function is increasing and discretely convex. Lemma \ref{Lemma:WeightedDPCapacity} gives pointwise convergence to $V$, so $V$ is increasing and discretely convex. It is strictly increasing because $u$ is strictly increasing in $x$: if $x_2>x_1$ and $a_1$ is optimal at $x_1$, then
\begin{equation*}
\Hfunc(x_2,a_1;m,V)
>
\Hfunc(x_1,a_1;m,V)=V(x_1,m),
\end{equation*}
and therefore $V(x_2,m)>V(x_1,m)$.
\end{proof}

\begin{lemma} \label{Lemma:G-singlevalued}
The optimal investment is unique.
\end{lemma}

\begin{proof}
For $x\geq1$, differentiating the continuation value twice with respect to $a$ gives
\begin{equation*}
\frac{\partial^2E_V(x,a)}{\partial a^2}
=
\frac{2}{(1+a)^3}
\left[
-(1-\delta)(V(x+1)-V(x))
-\delta(V(x)-V(x-1))
\right]
<0,
\end{equation*}
where the inequality follows from Lemma \ref{Lemma:IncreasingConvex}. At $x=0$,
\begin{equation*}
\frac{\partial^2E_V(0,a)}{\partial a^2}
=
-\frac{2(1-\delta)}{(1+a)^3}[V(1)-V(0)]
<0.
\end{equation*}
Thus the continuation value is strictly concave in investment. Since $-c(a)$ is concave, the objective in Equation \eqref{eq:H_def} is strictly concave and has a unique maximizer.
\end{proof}

\begin{lemma} \label{Lemma:IncreasingDiff}
The policy $g(x,m)$ is non-decreasing in $x$.
\end{lemma}

\begin{proof}
For $x\geq1$,
\begin{equation}
\label{Eq:ContinuationActionDerivative}
\frac{\partial E_V(x,a)}{\partial a}
=
\frac{1}{(1+a)^2}
\left[(1-\delta)V(x+1)+(2\delta-1)V(x)-\delta V(x-1)\right].
\end{equation}
The difference between the bracketed expressions in Equation \eqref{Eq:ContinuationActionDerivative} at $x+1$ and $x$ is
\begin{equation*}
(1-\delta)\Delta^2V(x)+\delta\Delta^2V(x-1)\geq0.
\end{equation*}
At the boundary, the difference between the derivatives at $x=1$ and $x=0$ is proportional to
\begin{equation*}
(1-\delta)\Delta^2V(0)+\delta\Delta V(0)\geq0.
\end{equation*}
Hence the objective in Equation \eqref{eq:H_def} has increasing differences in $(x,a)$.

To obtain monotonicity directly, take $x_2>x_1$ and let $a_i=g(x_i,m)$. If $a_2<a_1$, increasing differences imply
\begin{equation*}
\Hfunc(x_2,a_1;m,V)-\Hfunc(x_2,a_2;m,V)
\geq
\Hfunc(x_1,a_1;m,V)-\Hfunc(x_1,a_2;m,V)
>0,
\end{equation*}
where the strict inequality uses uniqueness at $x_1$. This contradicts optimality of $a_2$ at $x_2$. Therefore $a_2\geq a_1$.
\end{proof}

\begin{lemma}
\label{Lemma:ParameterMonotonicity}
If $u$ has increasing differences in $(x,z)$, the optimal investment is non-decreasing in $z$. It is also non-decreasing in $\beta$.
\end{lemma}

\begin{proof}
Fix $m$ and start finite-horizon value iteration from $V_0=0$. The preservation argument follows the argument used in Theorem 3 in \cite{light2021stochastic}; we give the calculations needed for the transition probabilities in Equation \eqref{Eq:Oligopoly}. We first consider a parameter $z$. Suppose $V_h(x,z)$ is increasing and discretely convex in $x$ and has increasing differences in $(x,z)$. The first difference in the continuation value is the nonnegative combination of first differences of $V_h$ displayed in the proof of Lemma \ref{Lemma:IncreasingConvex}. Since each $\Delta V_h(x,z)$ is non-decreasing in $z$, the continuation value has increasing differences in $(x,z)$.

Let $r(a)=a/(1+a)$. For $a_2\geq a_1$ and $x\geq1$, direct substitution in Equation \eqref{Eq:Oligopoly} gives
\begin{equation}
\label{Eq:ActionGainParameter}
E_{V_h}(x,a_2,z)-E_{V_h}(x,a_1,z)
=
[r(a_2)-r(a_1)]
\left[(1-\delta)\Delta V_h(x,z)+\delta\Delta V_h(x-1,z)\right].
\end{equation}
At $x=0$, the bracket in Equation \eqref{Eq:ActionGainParameter} is $(1-\delta)\Delta V_h(0,z)$. These expressions are non-decreasing in $z$. Hence the Bellman objective has increasing differences in $(a,z)$. The proof of Lemma \ref{Lemma:IncreasingDiff} gives increasing differences in $(x,a)$.

The standard maximization result for functions with these increasing-differences properties implies that $V_{h+1}$ has increasing differences in $(x,z)$; see \cite{topkis2011supermodularity}. The zero function satisfies the induction hypothesis, so the property holds for every finite horizon. Weighted-norm convergence from Lemma \ref{Lemma:WeightedDPCapacity} preserves the inequalities. Thus, the infinite-horizon objective has increasing differences in $(a,z)$. The direct contradiction argument in the proof of Lemma \ref{Lemma:IncreasingDiff}, with $z$ in place of $x$, and uniqueness from Lemma \ref{Lemma:G-singlevalued} imply that $g(x,m,z)$ is non-decreasing in $z$.

For the discount factor, apply the same induction with parameter $\beta$. The state difference in the continuation term and the gain in Equation \eqref{Eq:ActionGainParameter} are multiplied by $\beta$. By the induction hypothesis, the expressions being multiplied are nonnegative and non-decreasing in $\beta$. Their products with $\beta$ are therefore non-decreasing. The induction and weighted-norm limit used for $z$ therefore apply with $\beta$ in place of $z$, and imply that $g(x,m,\beta)$ is non-decreasing in $\beta$.
\end{proof}

\begin{proof}[Proof of Theorem \ref{Thm:Comparative}]
Lemma \ref{Lemma:ErgodicCapacity} gives a unique invariant distribution for every stationary policy and a geometric tail bound that is uniform over policies and interactions. Lemma \ref{Lemma:WeightedDPCapacity} gives a finite value function that is continuous in $m$, and Lemma \ref{Lemma:G-singlevalued} gives a unique optimal policy that is continuous in $m$.

Let $m_k\to m$ and let $s_k$ be the invariant distribution under $g(\cdot,m_k)$. The geometric tail bound makes $\{s_k\}$ tight. Any weak limit is invariant under the policy $g(\cdot,m)$ because the policy and transition probabilities are continuous. Uniqueness from Lemma \ref{Lemma:ErgodicCapacity} identifies the limit as the invariant distribution under $g(\cdot,m)$. Thus the invariant distribution is continuous in $m$. Polynomial growth of $k$ and the uniform geometric tail imply that
\begin{equation*}
F(m)=M(s^{m,g})
\end{equation*}
is continuous. Assumption \ref{Assumption:CapacityCompt}(iv) chooses $[a,b]$ to contain the range of $F$, so $F(a)\geq a$ and $F(b)\leq b$. The intermediate value theorem gives $m^*=F(m^*)$, proving part (i). Part (ii) follows from Lemmas \ref{Lemma:G-singlevalued} and \ref{Lemma:IncreasingDiff}.

For part (iii), Lemma \ref{Lemma:ParameterMonotonicity} implies that $g(x,m,z)$ is non-decreasing in $z$, and Lemma \ref{Lemma:IncreasingDiff} implies that it is non-decreasing in $x$. Equation \eqref{Eq:Oligopoly} is stochastically increasing in both the current state and the action. Hence, for $z_2\geq_Zz_1$, the transition kernel induced by $g(\cdot,m,z_2)$ stochastically dominates the kernel induced by $g(\cdot,m,z_1)$, and the kernel for $z_2$ is itself stochastically increasing in the current state.

Start both chains from the same distribution. After one transition, the distribution under $z_2$ stochastically dominates the distribution under $z_1$. Because the kernel under $z_2$ dominates the kernel under $z_1$ and is stochastically increasing in the current state, this order is preserved after every additional transition. Lemma \ref{Lemma:ErgodicCapacity} implies convergence to the two invariant distributions, and stochastic dominance is closed under weak convergence. Therefore
\begin{equation*}
s^{m,z_2}\succeq_{SD}s^{m,z_1}.
\end{equation*}
Since $k$ is increasing, $M$ is increasing under stochastic dominance. Proposition \ref{Prop:comp} proves part (iii).

Lemma \ref{Lemma:ParameterMonotonicity} implies that the policy is non-decreasing in $\beta$. Applying the result used in part (iii), now with two discount factors, orders the invariant distributions. Proposition \ref{Prop:comp} then proves part (iv).
\end{proof}

\section{Finite-Time Analysis of the Adaptive Learning Algorithms}
\label{Section:Finite-time}

Theorems \ref{thm:Q_convergence} and \ref{thm:online_simplicial_mfe} describe what happens when the policy and the invariant distribution are learned with arbitrarily many observations. In a computation, the number of Q-learning updates $H$ and the number of population observations $K$ are finite. Two errors then enter the value of $\widehat f_\lambda(m)$. The policy obtained after $H$ updates may differ from the policy in Equation \eqref{Eq:LogitPolicy}, and the empirical state distribution obtained from $K$ observations may differ from the invariant distribution generated by the learned policy.

We first consider any method that returns an interaction $m$ with a small value of $\norm{\widehat f_\lambda(m)}$. This result applies whether $m$ is chosen by bisection, by the simplicial algorithm, or by another numerical method. We then use Equation \eqref{Eq:SimplicialFixedPoint} to obtain a separate bound for every vertex of a selected simplex. Finally, we show how the total order on a scalar interval allows bisection to retain an interval containing an equilibrium and to reduce its length geometrically.

Throughout this section, $X$ and $A$ are finite and $\lambda>0$ is fixed. Set the probability of every infeasible action equal to zero. For two stationary policies, define
\begin{equation*}
d_\sigma(\sigma,\sigma')
=
\max_{x\in X}
\norm{\sigma(\cdot\mid x)-\sigma'(\cdot\mid x)}_1.
\end{equation*}
For a policy $\sigma$ and interaction $m$, let $s^{m,\sigma}$ denote the invariant distribution of $L_{m,\sigma}$ when it is unique.

Assumption \ref{Assumption:Unique} in the scalar case and Assumption \ref{Assumption:Vector} in the low-dimensional case give a unique invariant distribution under $\sigma_\lambda(\cdot\mid\cdot,m)$. To analyze a policy that is close to $\sigma_\lambda$, we also need nearby policies to induce nearby invariant distributions.

\begin{assumption}
\label{Assumption:FiniteTime}
There exists $\bar r>0$ such that, whenever $m\in\mathcal M$ and
\begin{equation*}
d_\sigma(\sigma,\sigma_\lambda(\cdot\mid\cdot,m))\leq\bar r,
\end{equation*}
the Markov chain $L_{m,\sigma}$ has a unique invariant distribution. Moreover, if
\begin{equation}
\label{Eq:InvariantDistributionPolicyContinuity}
\omega_{I,\lambda}(r)
=
\sup_{\substack{m\in\mathcal M,\ \sigma\\
d_\sigma(\sigma,\sigma_\lambda(\cdot\mid\cdot,m))\leq r}}
\norm{s^{m,\sigma}-s_\lambda^m}_1,
\end{equation}
then $\omega_{I,\lambda}(r)\to0$ as $r\downarrow0$.
\end{assumption}

In a finite-state model, a sufficient condition for
Assumption \ref{Assumption:FiniteTime} is continuity of the
transition matrix in $(m,\sigma)$ and irreducibility of
$L_{m,\sigma_\lambda}$ for every $m\in\mathcal M$. Irreducibility is preserved by a sufficiently small change in a finite transition matrix. Compactness of $\mathcal M$ makes the required change uniform in $m$, and the invariant distribution is continuous on this neighborhood of transition matrices.

The interaction may depend on actions as well as states. Define the stationary state-action distribution
\begin{equation*}
\mu_\lambda^m(x,a)
=
s_\lambda^m(x)\sigma_\lambda(a\mid x,m),
\end{equation*}
and let
\begin{equation*}
\mathcal D
=
\{\mu\in\mathcal P(X\times A):
\mu(x,a)=0\text{ if }a\notin\Gamma(x)\}.
\end{equation*}
We view $M$ as a continuous function on $\mathcal D$. When $M$ depends only on states, $M(\mu)$ means that $M$ is applied to the state marginal of $\mu$. Thus,
\begin{equation*}
F_\lambda(m)=M(\mu_\lambda^m),
\qquad
f_\lambda(m)=m-F_\lambda(m).
\end{equation*}
Let
\begin{equation*}
\mathcal Z_\lambda
=
\{m\in\mathcal M:f_\lambda(m)=0\}.
\end{equation*}
This set is nonempty and compact.

We next define three functions. Each one states directly how a small error in one object can affect another. For $r\geq0$, let
\begin{align}
\label{Eq:FiniteInteractionDistance}
b_\lambda(r)
&=
\sup_{\substack{m\in\mathcal M\\
\norm{f_\lambda(m)}\leq r}}
\operatorname{dist}(m,\mathcal Z_\lambda),\\
\label{Eq:FiniteInteractionContinuity}
\omega_M(r)
&=
\sup_{\substack{\mu,\mu'\in\mathcal D\\
\norm{\mu-\mu'}_1\leq r}}
\norm{M(\mu)-M(\mu')},\\
\label{Eq:FiniteResponseContinuity}
\omega_\lambda(r)
&=
\sup_{\substack{m,m'\in\mathcal M\\
\norm{m-m'}\leq r}}
\left[
 d_\sigma\left(
 \sigma_\lambda(\cdot\mid\cdot,m),
 \sigma_\lambda(\cdot\mid\cdot,m')
 \right)
 +
 \norm{s_\lambda^m-s_\lambda^{m'}}_1
\right].
\end{align}
Equation \eqref{Eq:FiniteInteractionDistance} measures how far an interaction with a small value of $\norm{f_\lambda(m)}$ can be from the equilibrium set. Equation \eqref{Eq:FiniteInteractionContinuity} describes the continuity of the interaction as the stationary state-action distribution changes. Equation \eqref{Eq:FiniteResponseContinuity} describes the continuity of the policy and invariant distribution as $m$ changes.

Continuity on the compact sets in Equations \ref{Eq:FiniteInteractionContinuity} and \ref{Eq:FiniteResponseContinuity} gives
\begin{equation}
\label{Eq:FiniteContinuityLimits}
\omega_M(r)\to0,
\qquad
\omega_\lambda(r)\to0
\end{equation}
as $r\downarrow0$. Also,
\begin{equation}
\label{Eq:FiniteDistanceLimit}
b_\lambda(r)\to0.
\end{equation}
To verify Equation \eqref{Eq:FiniteDistanceLimit}, suppose otherwise. Then there are $r_k\downarrow0$, $\epsilon>0$, and $m_k\in\mathcal M$ such that $\norm{f_\lambda(m_k)}\leq r_k$ but $\operatorname{dist}(m_k,\mathcal Z_\lambda)\geq\epsilon$. A convergent subsequence has a limit $m$. Continuity of $f_\lambda$ gives $f_\lambda(m)=0$, contradicting the distance inequality. This argument does not require a unique or isolated equilibrium.

Suppose that, at an interaction $m$, the policy obtained after the available observations is $\widehat\sigma$ and the empirical state distribution is $\widehat s$. Define
\begin{equation*}
\widehat\mu(x,a)
=
\widehat s(x)\widehat\sigma(a\mid x),
\qquad
\widehat F(m)=M(\widehat\mu),
\qquad
\widehat f_\lambda(m)=m-\widehat F(m).
\end{equation*}
Assume that
\begin{equation}
\label{Eq:FinitePolicyPopulationErrors}
d_\sigma\left(
\widehat\sigma,
\sigma_\lambda(\cdot\mid\cdot,m)
\right)
\leq\delta_H\leq\bar r,
\qquad
\norm{\widehat s-s^{m,\widehat\sigma}}_1
\leq\delta_K.
\end{equation}
The first inequality in Equation \eqref{Eq:FinitePolicyPopulationErrors} bounds the difference between the learned policy and the policy in Equation \eqref{Eq:LogitPolicy}. The second bounds the sampling error relative to the invariant distribution of the policy that is actually used to generate the population observations. By Assumption \ref{Assumption:FiniteTime}, the difference between this invariant distribution and $s_\lambda^m$ is at most $\omega_{I,\lambda}(\delta_H)$.

Define
\begin{equation}
\label{Eq:FiniteResponseErrorDefinition}
e_{H,K}
=
\omega_M\left(
\delta_H+
\delta_K+
\omega_{I,\lambda}(\delta_H)
\right).
\end{equation}

\begin{proposition}
\label{Thm:finite_bounds1}
Suppose Equation \eqref{Eq:FinitePolicyPopulationErrors} holds and
$
\norm{\widehat f_\lambda(m)}\leq\delta.
$
Then there exists $m^*\in\mathcal Z_\lambda$ such that, with
\begin{equation*}
d=b_\lambda(\delta+e_{H,K}),
\end{equation*}
we have
\begin{equation}
\label{Eq:FiniteGeneralBounds}
\norm{m-m^*}\leq d
\end{equation}
and
\begin{equation}
\label{Eq:FiniteGeneralPolicyPopulationBound}
d_\sigma\left(
\widehat\sigma,
\sigma_\lambda(\cdot\mid\cdot,m^*)
\right)
+
\norm{\widehat s-s_\lambda^{m^*}}_1
\leq
\delta_H+
\delta_K+
\omega_{I,\lambda}(\delta_H)+
\omega_\lambda(d).
\end{equation}
\end{proposition}

\begin{proof}
We first compare the estimated stationary state-action distribution with the distribution generated by the policy in Equation \eqref{Eq:LogitPolicy} at the same interaction. Adding and subtracting the distributions that combine $\widehat\sigma$ with $s^{m,\widehat\sigma}$ and with $s_\lambda^m$ gives
\begin{equation}
\begin{aligned}
\label{Eq:FiniteStateActionError}
\norm{\widehat\mu-\mu_\lambda^m}_1
&\leq
\norm{\widehat s-s^{m,\widehat\sigma}}_1
+
\norm{s^{m,\widehat\sigma}-s_\lambda^m}_1\\
&\quad+
\sum_{x\in X}s_\lambda^m(x)
\norm{\widehat\sigma(\cdot\mid x)-\sigma_\lambda(\cdot\mid x,m)}_1\\
&\leq
\delta_K+
\omega_{I,\lambda}(\delta_H)+
\delta_H.
\end{aligned}
\end{equation}
Equations \ref{Eq:FiniteInteractionContinuity}, \ref{Eq:FiniteResponseErrorDefinition}, and \ref{Eq:FiniteStateActionError} imply
\begin{equation}
\label{Eq:FiniteResponseError}
\norm{\widehat F(m)-F_\lambda(m)}
\leq
e_{H,K}.
\end{equation}
Therefore,
\begin{equation*}
\norm{f_\lambda(m)}
\leq
\norm{\widehat f_\lambda(m)}+
\norm{\widehat F(m)-F_\lambda(m)}
\leq
\delta+e_{H,K}.
\end{equation*}
Equation \eqref{Eq:FiniteInteractionDistance} gives an $m^*\in\mathcal Z_\lambda$ satisfying Equation \eqref{Eq:FiniteGeneralBounds}. Finally, apply the triangle inequality, Equation \eqref{Eq:InvariantDistributionPolicyContinuity}, and Equation \eqref{Eq:FiniteResponseContinuity} to obtain Equation \eqref{Eq:FiniteGeneralPolicyPopulationBound}.
\end{proof}

Proposition \ref{Thm:finite_bounds1} does not depend on how $m$ is chosen. It applies whenever any numerical method finds a small value of $\norm{\widehat f_\lambda(m)}$. It also allows multiple equilibria: the conclusion compares $(m,\widehat\sigma,\widehat s)$ with some $m^*\in\mathcal Z_\lambda$. 

The simplicial algorithm provides additional information even when no individual vertex has a small value of $\norm{m^j-\widehat F^j}$. Equation \eqref{Eq:SimplicialFixedPoint} says that the weighted average of these vectors is zero. The next proposition uses this equation directly.

\begin{proposition}
\label{Prop:FiniteSimplicial}
Suppose a simplex of diameter at most $h$ with vertices  $m^{j_0},\ldots,m^{j_n}$ and weights $\theta_0,\ldots,\theta_n$ satisfies Equation \eqref{Eq:SimplicialFixedPoint} for values $\widetilde F^{j_0},\ldots,\widetilde F^{j_n}$. If
\begin{equation*}
\max_{\ell=0,\ldots,n}
\norm{\widetilde F^{j_\ell}-F_\lambda(m^{j_\ell})}
\leq e,
\end{equation*}
then every vertex $m^j$ of this simplex satisfies
\begin{equation}
\label{Eq:FiniteSimplicialInteractionBound}
\operatorname{dist}(m^j,\mathcal Z_\lambda)
\leq
b_\lambda\left(
h+
\omega_M(\omega_\lambda(h))+
e
\right).
\end{equation}
\end{proposition}

\begin{proof}
If $\norm{m-m'}\leq h$, Equation \eqref{Eq:FiniteResponseContinuity} implies
\begin{equation*}
\norm{\mu_\lambda^m-\mu_\lambda^{m'}}_1
\leq
\omega_\lambda(h).
\end{equation*}
Therefore Equation \eqref{Eq:FiniteInteractionContinuity} gives
\begin{equation}
\label{Eq:FiniteFContinuity}
\norm{f_\lambda(m)-f_\lambda(m')}
\leq
h+
\omega_M(\omega_\lambda(h)).
\end{equation}
Fix a vertex $m^j$ of the simplex. Using Equation \eqref{Eq:SimplicialFixedPoint},
\begin{align*}
f_\lambda(m^j)
={}&
\sum_{\ell=0}^n\theta_\ell
\left[f_\lambda(m^j)-f_\lambda(m^{j_\ell})\right]\\
&+
\sum_{\ell=0}^n\theta_\ell
\left[\widetilde F^{j_\ell}-F_\lambda(m^{j_\ell})\right].
\end{align*}
Every pair of vertices in the simplex is at distance at most $h$. Equation \eqref{Eq:FiniteFContinuity} and the assumed bound on $\widetilde F^{j_\ell}$ therefore give
\begin{equation*}
\norm{f_\lambda(m^j)}
\leq
h+
\omega_M(\omega_\lambda(h))+
e.
\end{equation*}
Equation \eqref{Eq:FiniteInteractionDistance} now yields Equation \eqref{Eq:FiniteSimplicialInteractionBound}.
\end{proof}

If Equation \eqref{Eq:FinitePolicyPopulationErrors} holds at each selected vertex, Equation \eqref{Eq:FiniteResponseError} permits $e=e_{H,K}$ in Proposition \ref{Prop:FiniteSimplicial}. If the policy and state distribution reported at a vertex also satisfy Equation \eqref{Eq:FinitePolicyPopulationErrors}, Proposition \ref{Thm:finite_bounds1} gives the corresponding policy and population bounds, with $d$ equal to the right-hand side of Equation \eqref{Eq:FiniteSimplicialInteractionBound}. Thus, the grid diameter, the errors in the values used in Equation \eqref{Eq:SimplicialFixedPoint}, and the errors in the reported policy and state distribution enter separately.

We next use the total order on $[a,b]$ to obtain the additional interval bound from bisection.

\begin{proposition}
\label{Thm:finite_bounds}
Suppose $\mathcal M=[a,b]$, Equation \eqref{Eq:FinitePolicyPopulationErrors} holds at every evaluated interaction, and Algorithm \ref{alg:bisection_q_learning} updates the interval only when
\begin{equation*}
|\widehat f_\lambda(m_t)|>\delta,
\qquad
\delta\geq e_{H,K}.
\end{equation*}
Then every update retains an interval containing a zero of $f_\lambda$. If the algorithm stops at the midpoint $m_n$ of $[a_n,b_n]$, then
\begin{equation}
\label{Eq:FiniteStoppingDistance}
\operatorname{dist}(m_n,\mathcal Z_\lambda)
\leq
\min\left\{
b_\lambda(\delta+e_{H,K}),
\frac{b_n-a_n}{2}
\right\}.
\end{equation}
After $T$ updates without stopping, the midpoint of the retained interval is within $(b-a)2^{-(T+1)}$ of $\mathcal Z_\lambda$.
\end{proposition}

\begin{proof}
Equation \eqref{Eq:FiniteResponseError} gives
\begin{equation*}
|\widehat f_\lambda(m_t)-f_\lambda(m_t)|
\leq e_{H,K}.
\end{equation*}
Thus, $\widehat f_\lambda(m_t)>\delta$ implies $f_\lambda(m_t)>0$, and $\widehat f_\lambda(m_t)<-\delta$ implies $f_\lambda(m_t)<0$. Every update therefore uses the sign of the true function. Since $f_\lambda(a)\leq0\leq f_\lambda(b)$, continuity implies that the retained interval contains a zero.

If the algorithm stops at $m_n$, then
\begin{equation*}
|f_\lambda(m_n)|
\leq
|\widehat f_\lambda(m_n)|+
e_{H,K}
\leq
\delta+e_{H,K}.
\end{equation*}
Equation \eqref{Eq:FiniteInteractionDistance} gives the first term on the right-hand side of Equation \eqref{Eq:FiniteStoppingDistance}. The interval $[a_n,b_n]$ also contains a zero, and $m_n$ is its midpoint, which gives the second term. Every nonstopping update halves the interval. After $T$ updates, the retained interval has length $(b-a)2^{-T}$ and contains a zero, so its midpoint is within half this length of $\mathcal Z_\lambda$.
\end{proof}

At a stopping point, Proposition \ref{Thm:finite_bounds1} applies with $d$ equal to the right-hand side of Equation \eqref{Eq:FiniteStoppingDistance}. If the policy and population state are evaluated once more at the midpoint after $T$ updates, the same proposition applies with $d=(b-a)2^{-(T+1)}$. Thus, Proposition \ref{Thm:finite_bounds} adds a geometric bound on the interaction to the policy and population bounds that hold for any method.

We finally relate $\delta_H$ and $\delta_K$ in Equation \eqref{Eq:FinitePolicyPopulationErrors} to the errors after $H$ Q-learning updates and $K$ population observations. Suppose
\begin{equation}
\label{Eq:FiniteQError}
\norm{\widehat Q_{H,m}-Q_{\lambda,m}^*}_\infty
\leq\eta_H,
\end{equation}
and obtain $\widehat\sigma$ by applying Equation \eqref{Eq:LogitPolicy} to $\widehat Q_{H,m}$. Then
\begin{equation}
\label{Eq:FiniteQPolicyError}
d_\sigma\left(
\widehat\sigma,
\sigma_\lambda(\cdot\mid\cdot,m)
\right)
\leq
\frac{\eta_H}{\lambda}.
\end{equation}
To verify Equation \eqref{Eq:FiniteQPolicyError}, connect the two Q-functions by a line segment. At a fixed state, let $p$ be the policy along this segment and let $v$ be the difference between the two action-value vectors. Differentiating Equation \eqref{Eq:LogitPolicy} gives
\begin{equation*}
\sum_a|\dot p(a)|
=
\frac{1}{\lambda}
\sum_a p(a)|v(a)-\mathbb E_p[v]|.
\end{equation*}
The last sum is at most $\sqrt{\operatorname{Var}_p(v)}$, which is no larger than $\norm{v}_\infty$. Hence
\begin{equation*}
\sum_a|\dot p(a)|\leq\frac{\norm{v}_\infty}{\lambda}.
\end{equation*}
Integrating this inequality along the segment proves Equation \eqref{Eq:FiniteQPolicyError}.

Suppose also that the $K$ population observations satisfy
\begin{equation}
\label{Eq:FiniteMonteCarloError}
\norm{\widehat s-s^{m,\widehat\sigma}}_1
\leq\eta_K.
\end{equation}
Then Propositions \ref{Thm:finite_bounds1}, \ref{Prop:FiniteSimplicial}, and \ref{Thm:finite_bounds} apply with
\begin{equation*}
\delta_H=\frac{\eta_H}{\lambda},
\qquad
\delta_K=\eta_K,
\end{equation*}
provided $\eta_H/\lambda\leq\bar r$. Bounds of the form in Equation \eqref{Eq:FiniteQError} are available for tabular Q-learning; see \cite{even2003learning,qu2020finite,li2020sample,li2024q}. When the observations are generated under a fixed policy, bounds of the form in Equation \eqref{Eq:FiniteMonteCarloError} follow from finite-state Markov-chain concentration results; see \cite{paulin2015concentration}. The dependence of an invariant distribution on perturbations of the transition matrix is studied in \cite{seneta2006non,shardlow2000perturbation}. If the policy continues to change while one stored trajectory is extended, as in Algorithm \ref{alg:online_simplicial_mfe}, Equation \eqref{Eq:FiniteMonteCarloError} must instead be established for that changing sequence of transition matrices. Lemma \ref{Lemma:PopulationUnderConvergingPolicies} gives the corresponding asymptotic statement.

The values of $\eta_H$ and $\eta_K$, and the probabilities with which Equations \ref{Eq:FiniteQError} and \ref{Eq:FiniteMonteCarloError} hold, depend on the exploration rule and on the mixing properties of the relevant Markov chain. At an adaptively selected $m$, it is enough that these bounds hold conditional on the observations collected before $m$ is selected. If they hold with conditional probabilities at least $1-\epsilon_H$ and $1-\epsilon_K$, Proposition \ref{Thm:finite_bounds1} holds with probability at least $1-\epsilon_H-\epsilon_K$. For at most $T$ scalar evaluations, all bounds hold simultaneously with probability at least
\begin{equation*}
1-T(\epsilon_H+\epsilon_K).
\end{equation*}
The same union-bound argument applies to the finitely many vertices evaluated by the simplicial algorithm. Independence across evaluations is not required.

Equations \ref{Eq:InvariantDistributionPolicyContinuity}--\ref{Eq:FiniteResponseContinuity} require continuity but do not impose global Lipschitz constants. If an application provides explicit Lipschitz or mixing bounds, the functions in these equations can be replaced by the corresponding formulas. In particular, an inequality of the form
\begin{equation*}
b_\lambda(r)\leq C_\lambda r^\alpha
\end{equation*}
for small $r$ gives an explicit rate in Proposition \ref{Thm:finite_bounds1}. Without such a quantitative inequality, Equations \ref{Eq:FiniteContinuityLimits} and \ref{Eq:FiniteDistanceLimit} still imply that the bounds converge to zero as $\eta_H$, $\eta_K$, and $\delta$ converge to zero. Proposition \ref{Prop:FiniteSimplicial} also requires $h\to0$, while Proposition \ref{Thm:finite_bounds} obtains its geometric term by increasing the number of bisection updates.

In practice, $H$ and $K$ can be small during the first evaluations and increased when $\norm{\widehat f_\lambda(m)}$ becomes small. This reduces the policy and population errors where greater accuracy matters most. A smaller $\lambda$ reduces the approximation error in Proposition \ref{Prop:logit-approximation}, but Equation \eqref{Eq:FiniteQPolicyError} shows that it also makes the policy more sensitive to an error in the Q-function. If, without the entropy term, the optimal Q-value exceeds
every other feasible action's Q-value by a positive amount
uniformly over states and interactions, greedy action
selection recovers the optimal policy whenever the sup-norm
Q-function error is less than half that amount. The same
finite-time argument then applies provided the
invariant-distribution assumptions hold for the
unregularized policies.

\section{Simulations}

This section provides the numerical specifications for the inventory, capacity competition, and ridesharing applications. We set the entropy regularization parameter $\lambda=10^{-4}$ in the inventory, ridesharing, and social-learning simulations.

\subsection{Dynamic Inventory Competition Simulations Details} \label{Section:InventorySimulationDetails}

\textit{Demand model.}
The inventory simulations use stockout-based substitution. A retailer's demand is
\begin{equation*}
D(\zeta,M(s))=\lfloor \zeta+\gamma M(s)\rceil,
\end{equation*}
where $\lfloor\cdot\rceil$ denotes the stochastic rounding described below, $\zeta$ is baseline demand, and $\gamma\in(0,1]$ is the fraction of competitors' unmet demand that is reallocated. If $\bar s$ is the stationary distribution of order-up-to levels, the scalar interaction is
\begin{equation*}
M(s)
=
\sum_a\int_\zeta(\zeta-a)_+q(d\zeta)\bar s(a)
\end{equation*}
where $q$ is the distribution of the baseline demand. 
This quantity is the average unmet demand that can be reallocated to other retailers.\footnote{With $N$ retailers and equal redistribution, the demand reallocated to retailer $i$ is $\sum_{j\neq i}(\zeta_j-a_j)_+/(N-1)$. The displayed scalar is the corresponding mean field limit; see \cite{olsen2014markov}.}

The interaction depends on actions. Under a stationary randomized policy $\sigma$, it can be computed from the stationary state distribution as
\begin{equation}
\label{Eq:InventoryInteractionRandomized}
M(s,\sigma)
=
\sum_xs(x)\sum_{a\in\Gamma(x)}\sigma(a\mid x,m)
\int_\zeta(\zeta-a)_+q(d\zeta).
\end{equation}
Equation \eqref{Eq:InventoryInteractionRandomized} is the value of $M$ under the stationary state-action distribution.

With discrete $\zeta$, we use stochastic rounding to preserve continuity
in the scalar interaction. If $\zeta+\gamma M(s)$ lies between two
adjacent integers, demand is equal to these two integers with
probabilities that make its conditional mean equal to
$\zeta+\gamma M(s)$. Thus, the resulting transition probabilities vary
continuously with the scalar interaction. All demand expectations average over both baseline demand and
this randomization.

The state space is $X=\{0,1,\ldots,9\}$. The action space is $A=\{0,1,\ldots,9\}$ and $\Gamma(x)=\{x,x+1,\ldots,9\}$. The discount factor is $\beta=0.95$. Production cost is $(a-x)^2$, shortage cost is $b=2$ per unit, and the unit price is $r=30$. Baseline demand takes values $\zeta\in\{0,0.5,\ldots,9\}$. If $\zeta=z/2$, its probability is proportional to $1/(z+5)$ for $z=0,\ldots,18$. We set $\gamma=1$.

For Algorithm \ref{alg:bisection_q_learning}, the learning rate for an updated state-action pair is $0.003$. The exploration probability begins at $0.9$ and decreases exponentially to $0.05$. Each episode contains $100$ transitions and then restarts from a randomly selected state. The experience-replay buffer contains $500$ transitions and each update uses a batch of $16$. We use $1{,}000$ episodes and $200{,}000$ transitions to estimate the stationary distribution, so $H=100{,}000$ and $K=200{,}000$. The stopping tolerance is $10^{-3}$.

\begin{figure}[H]
    \centering
    \begin{subfigure}[b]{0.4\textwidth}
        \includegraphics[width=\textwidth, height=4cm]{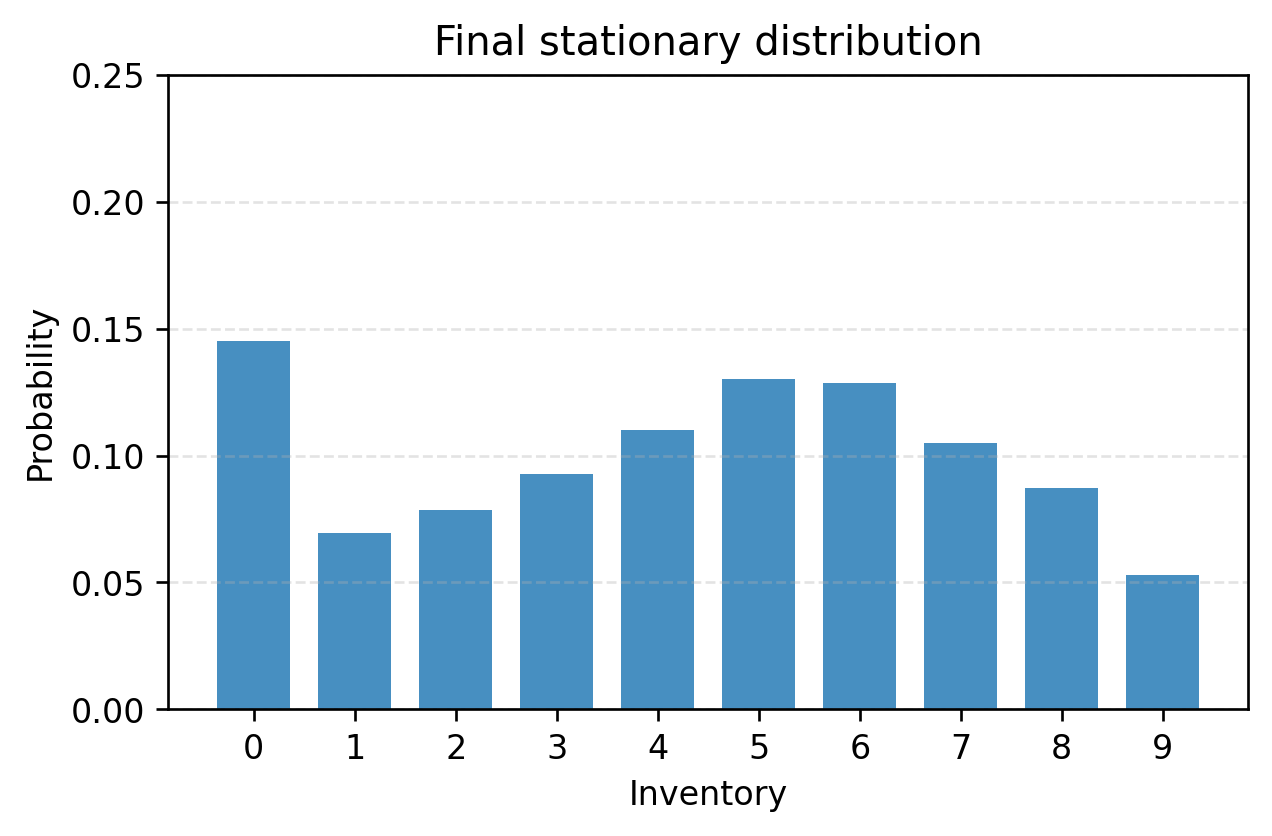}
        \caption{Holding cost = 2}
        \label{fig:Q_holding_cost_2}
    \end{subfigure}
    \hfill
    \begin{subfigure}[b]{0.4\textwidth}
        \includegraphics[width=\textwidth, height=4cm]{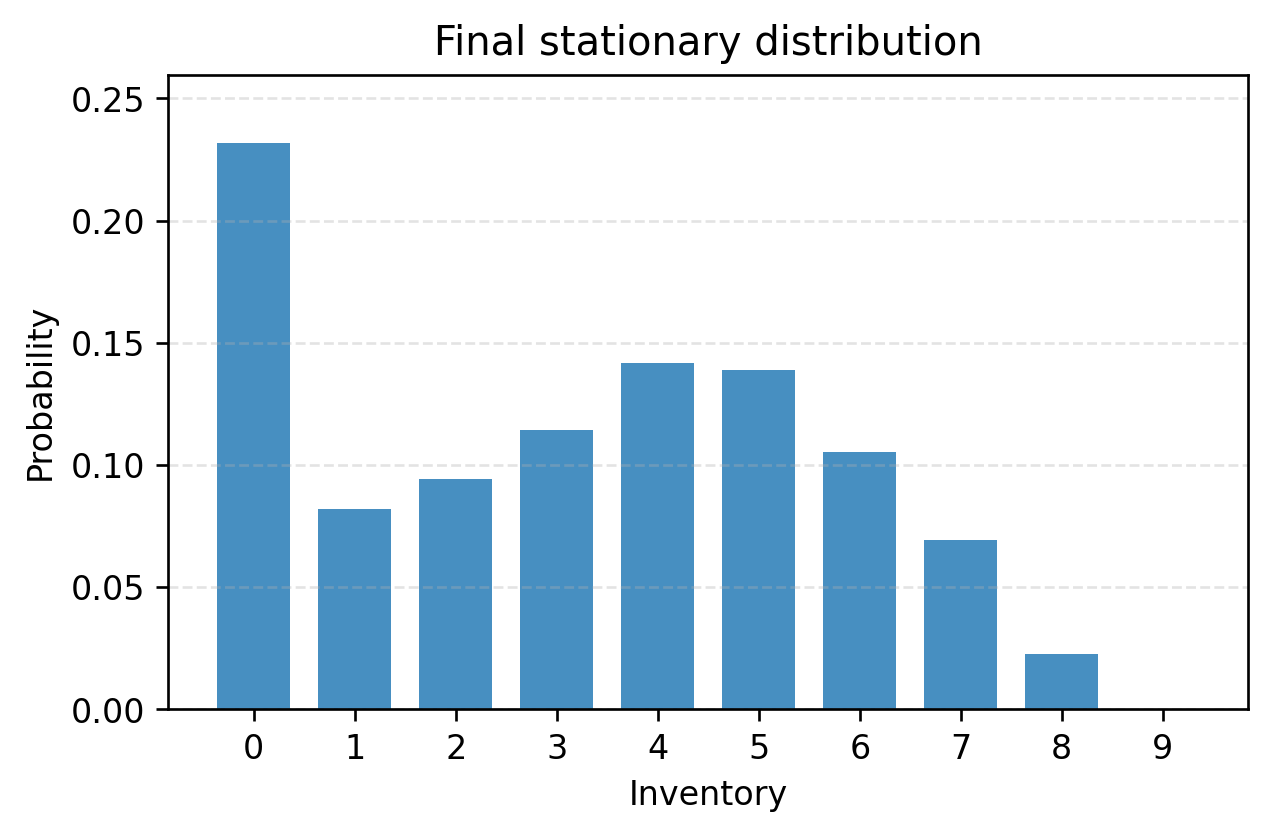}
        \caption{Holding cost = 5}
        \label{fig:Q_holding_cost_5}
    \end{subfigure}
    \vspace{0.5cm}
    \begin{subfigure}[b]{0.4\textwidth}
        \includegraphics[width=\textwidth, height=4cm]{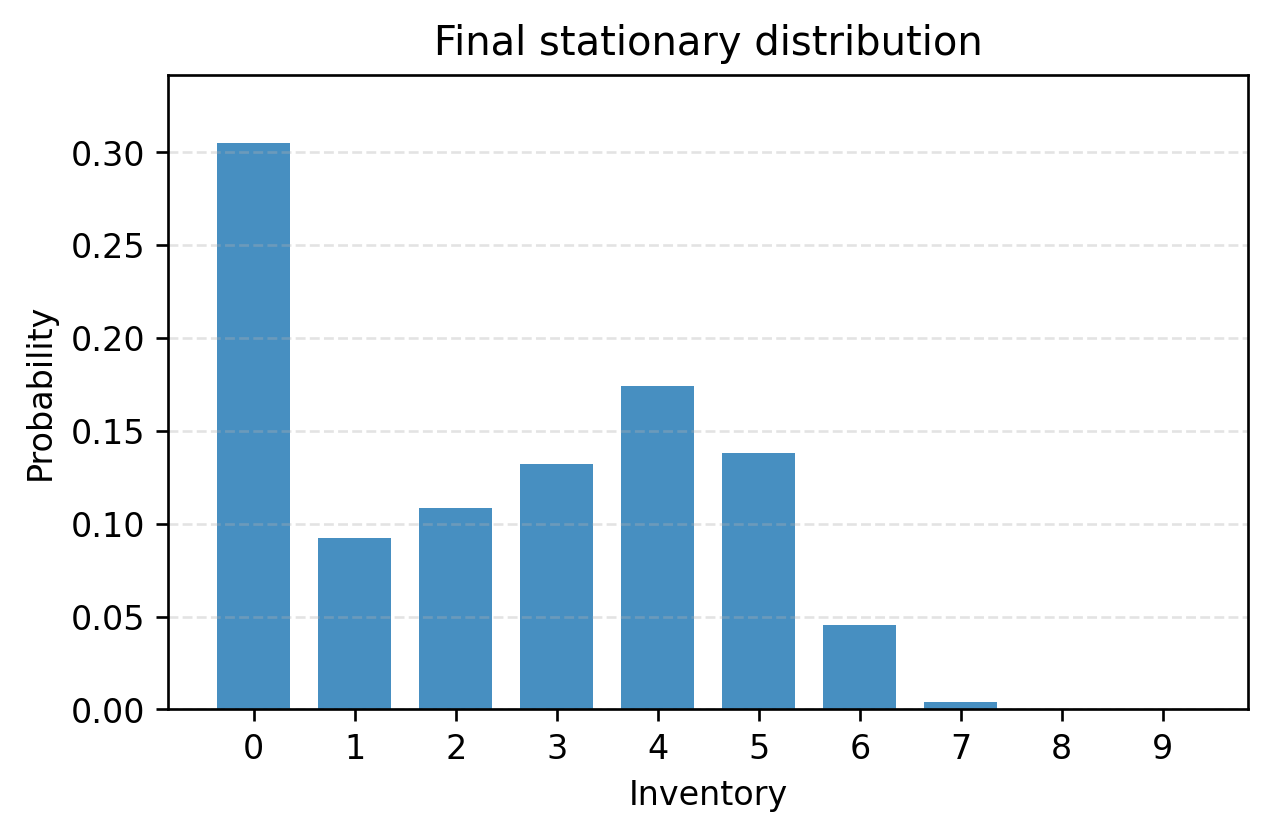}
        \caption{Holding cost = 8}
        \label{fig:Q_holding_cost_8}
    \end{subfigure}
    \hfill
    \begin{subfigure}[b]{0.4\textwidth}
        \includegraphics[width=\textwidth, height=4cm]{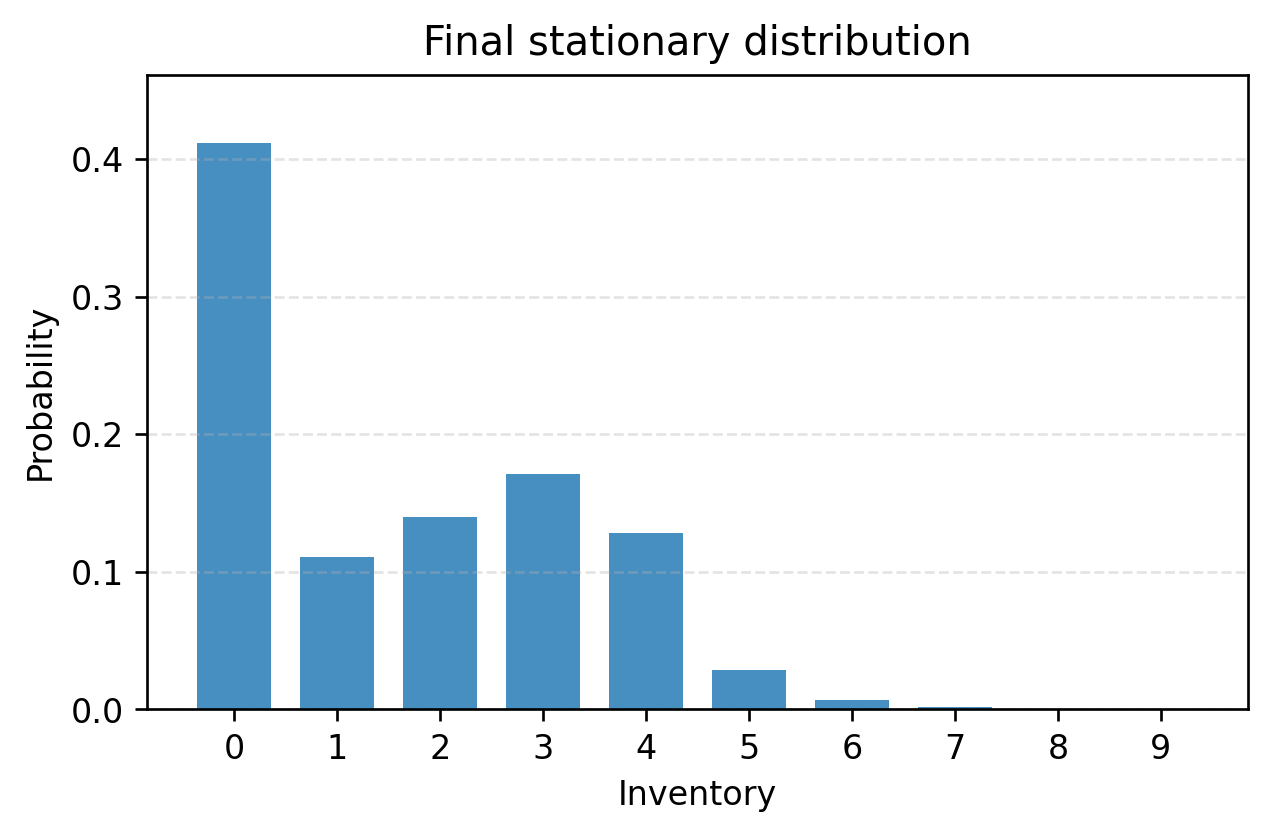}
        \caption{Holding cost = 12}
        \label{fig:Q_holding_cost_12}
    \end{subfigure}
    \caption{Approximate equilibrium inventory distributions under different holding costs using Algorithm \ref{alg:bisection_q_learning}. As holding costs increase, retailers hold less inventory.}
    \label{fig:Q_holding_cost_comparative_statics}
\end{figure}

The constant learning rate and experience replay used in these simulations do not satisfy the decreasing step-size conditions in Assumption \ref{Assumption:ScalarExploration}. They are practical numerical choices for stability. In this sense Figure \ref{fig:Q_holding_cost_comparative_statics} gives results close to those obtained from Algorithm \ref{alg:value_iteration_mfe}.

For the model-based results in Section \ref{Section:Value}, we use the same model parameters. The regularized value function iteration stops when the sup-norm change in the value function is below $10^{-4}$, with at most $1{,}000$ iterations at each scalar value.

For Figure \ref{fig:PlatformRevenueHeatmap}, the retailer receives the fraction $\tau\in[0,1]$ of sales revenue, so the platform's transaction fee is $1-\tau$. Let $\sigma_{h,\tau}(a\mid x)$ be the policy computed for a fixed pair $(h,\tau)$ and let $s_{h,\tau}$ be its invariant state distribution. Platform revenue is
\begin{align*}
\Pi_P(h,\tau)
={}&
\sum_{x\in X}s_{h,\tau}(x)
\sum_{a\in\Gamma(x)}\sigma_{h,\tau}(a\mid x)
\Bigg[
(1-\tau)r\,\mathbb E_\zeta
\min\{a,D(\zeta,M(s_{h,\tau},\sigma_{h,\tau}))\}\\
&\hspace{43mm}+
h\,\mathbb E_\zeta
\bigl(a-D(\zeta,M(s_{h,\tau},\sigma_{h,\tau}))\bigr)_+
\Bigg].
\end{align*}
The first term is commission revenue and the second is holding-fee revenue.

We evaluate
\begin{equation*}
\tau\in\{0.3,0.4,0.5,0.6,0.7\},
\qquad
h\in\{0,1,\ldots,12\}.
\end{equation*}
For each of the $65$ pairs, Algorithm \ref{alg:value_iteration_mfe} computes the policy and invariant distribution and the displayed formula gives platform revenue. Figure \ref{fig:PlatformRevenueHeatmap} reports these values.

\begin{figure}[H]
    \centering
    \begin{subfigure}[b]{0.48\textwidth}
        \includegraphics[width=\textwidth, height=6.5cm]{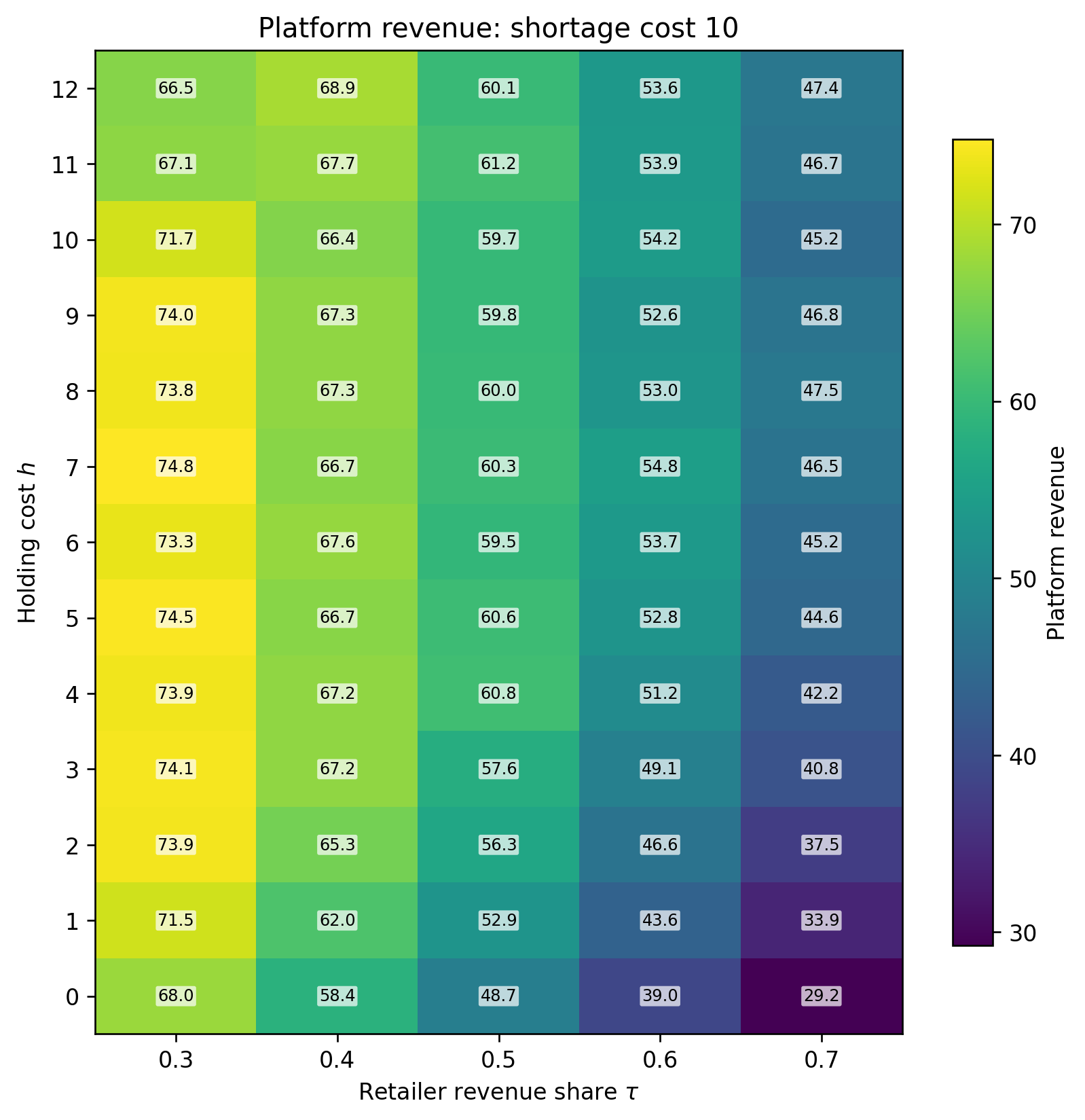}
        \caption{Shortage cost $10$.}
        \label{fig:Shortage}
    \end{subfigure}
    \hfill
    \begin{subfigure}[b]{0.48\textwidth}
        \includegraphics[width=\textwidth, height=6.5cm]{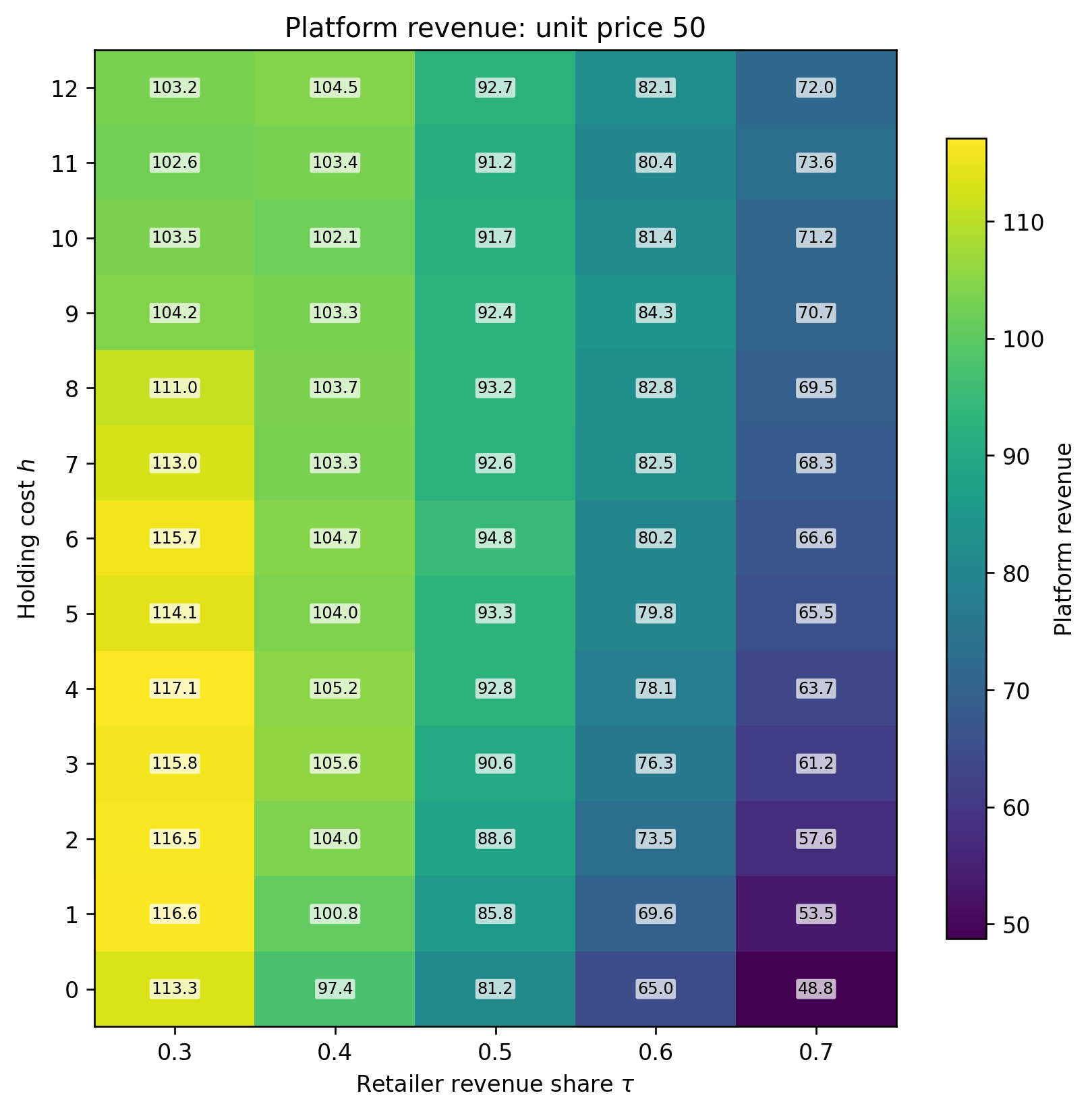}
        \caption{Per-unit price $50$.}
    \end{subfigure}
    \caption{Platform revenue under alternative inventory parameters.}
    \label{fig:heatappendix}
\end{figure}

Across the parameter values considered, higher platform revenue is
generally concentrated toward higher transaction fees and relatively
low holding costs. Figure \ref{fig:heatappendix} shows that this broad
qualitative pattern remains when the shortage cost is raised from $2$
to $10$ or the unit price is raised from $30$ to $50$. 

\subsection{Additional Simulations}

The full code for the simulations will be made available in a public repository.

\subsubsection{Dynamic Oligopoly Models}\label{Sec:OligSimulations}

We use the linear inverse demand function $P(Q)=\alpha-Q$. In the limiting model with many firms and negligible individual market power, each firm treats total production as fixed and produces at capacity. Its payoff is
\begin{equation*}
\pi(x,a,M(s))
=
P\left(\sum_{y\in X}\bar q(y)s(y)\right)\bar q(x)-c(a),
\end{equation*}
and the scalar interaction is
\begin{equation*}
M(s)=\sum_{y\in X}\bar q(y)s(y).
\end{equation*}

\begin{figure}[H]
    \centering
    \begin{subfigure}[b]{0.4\textwidth}
        \includegraphics[width=\textwidth, height=4cm]{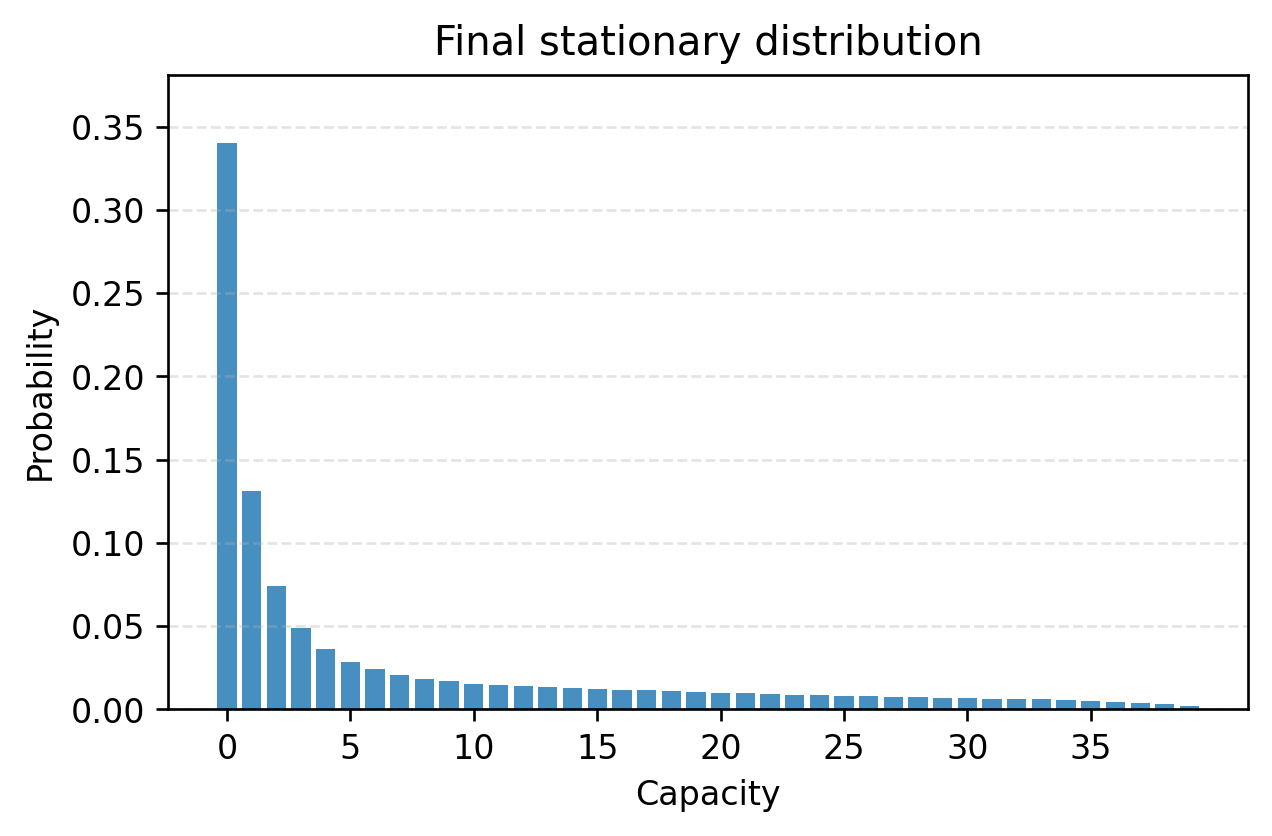}
        \caption{$\alpha=45$. Average production is $6.798052$.}
        \label{fig:D9}
    \end{subfigure}
    \hfill
    \begin{subfigure}[b]{0.4\textwidth}
        \includegraphics[width=\textwidth, height=4cm]{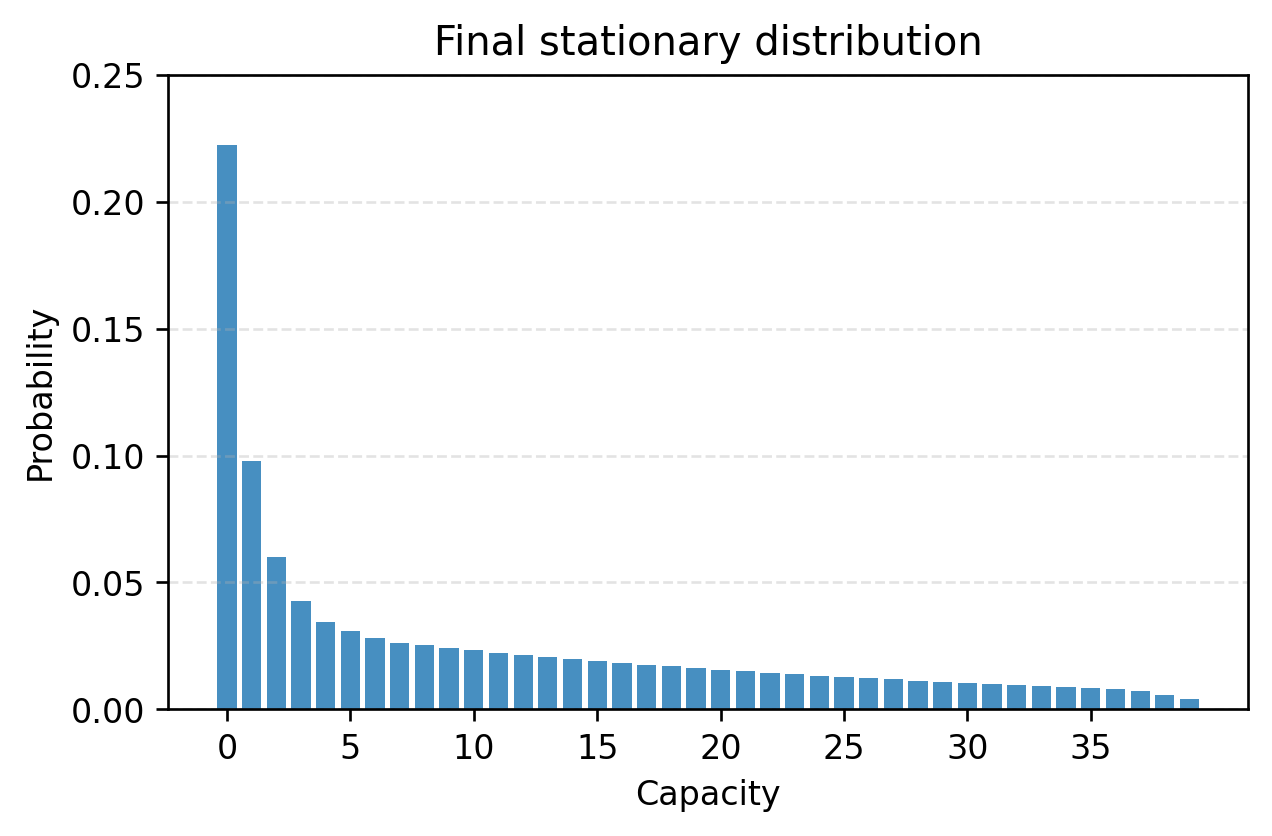}
        \caption{$\alpha=55$. Average production is $10.117766$.}
    \end{subfigure}
    \caption{Approximate equilibrium capacity distributions for two demand intercepts. Higher demand raises equilibrium capacity.}
    \label{fig:combined_figures}
\end{figure}

Capacity levels range from $0$ to $39$, and the investment grid contains the values from $0.05$ to $1$ in increments of $0.05$. We set $c(a)=150a^3$, $\bar q(x)=x$, $\delta=0.51$, and $\beta=0.98$. All finite-action calculations use the entropy parameter $\lambda=10^{-4}$. Figure \ref{fig:combined_figures} compares $\alpha=45$ and $\alpha=55$. The two computations require, respectively, $17$ and $15$ midpoint evaluations of Algorithm \ref{alg:value_iteration_mfe}.

\paragraph{Computational performance.}
For the numerical benchmark, we change only the demand intercept to $\alpha=50$. For a conjectured average production $m$, let $F_\lambda(m)$ denote the average production generated by the invariant distribution under the regularized optimal policy. We compare bisection with Brent's bracketed root-finding method. Both methods solve $m-F_\lambda(m)=0$ and use the same routine for each fixed-$m$ dynamic program. The number of fixed-$m$ evaluations in Table \ref{tab:oligopoly_computation} includes the endpoint calculations used to verify the bracket.

\begin{table}[H]
\centering
\small
\begin{tabular}{lrrrrr}
\hline
Method & $|X|$ & Fixed-$m$ evaluations
& $|m-F_\lambda(m)|$
& $\|sL_{m,\sigma_\lambda}-s\|_\infty$
& Seconds \\
\hline
Bisection & 40  & 30 & $3.9\times10^{-6}$  & $5.6\times10^{-17}$ & 3.12 \\
Brent     & 40  & 23 & $2.1\times10^{-10}$ & $1.1\times10^{-17}$ & 2.42 \\
Bisection & 80  & 32 & $4.1\times10^{-6}$  & $1.4\times10^{-17}$ & 6.23 \\
Bisection & 120 & 22 & $1.4\times10^{-6}$  & $5.6\times10^{-17}$ & 6.65 \\
\hline
\end{tabular}
\caption{Computational performance in the dynamic oligopoly model. Times are from one run of the accompanying code and depend on the computing environment.}
\label{tab:oligopoly_computation}
\end{table}

For $40$ states, bisection gives $m=8.727307934$, while Brent's method gives $m=8.727307917$. Thus, the two bracketed methods find the same equilibrium up to their stopping tolerances. Brent uses fewer fixed-$m$ calculations because it uses previously computed residuals to select the next point, whereas bisection always evaluates the midpoint of the retained interval. The Bellman residual is below $1.2\times10^{-10}$ in every row of Table \ref{tab:oligopoly_computation}.

Increasing the number of capacity states from $40$ to $80$ and $120$ gives equilibrium average capacities $11.402223$ and $12.324483$, respectively. The probability assigned to the upper state decreases from $3.23\times10^{-3}$ with $40$ states to $3.90\times10^{-4}$ with $80$ states and $6.96\times10^{-5}$ with $120$ states. Enlarging the upper capacity bound changes the finite model, so these calculations assess computational performance. The time needed for the calculation increases moderately over this range.

\begin{figure}[H]
    \centering
    \includegraphics[width=0.98\textwidth]{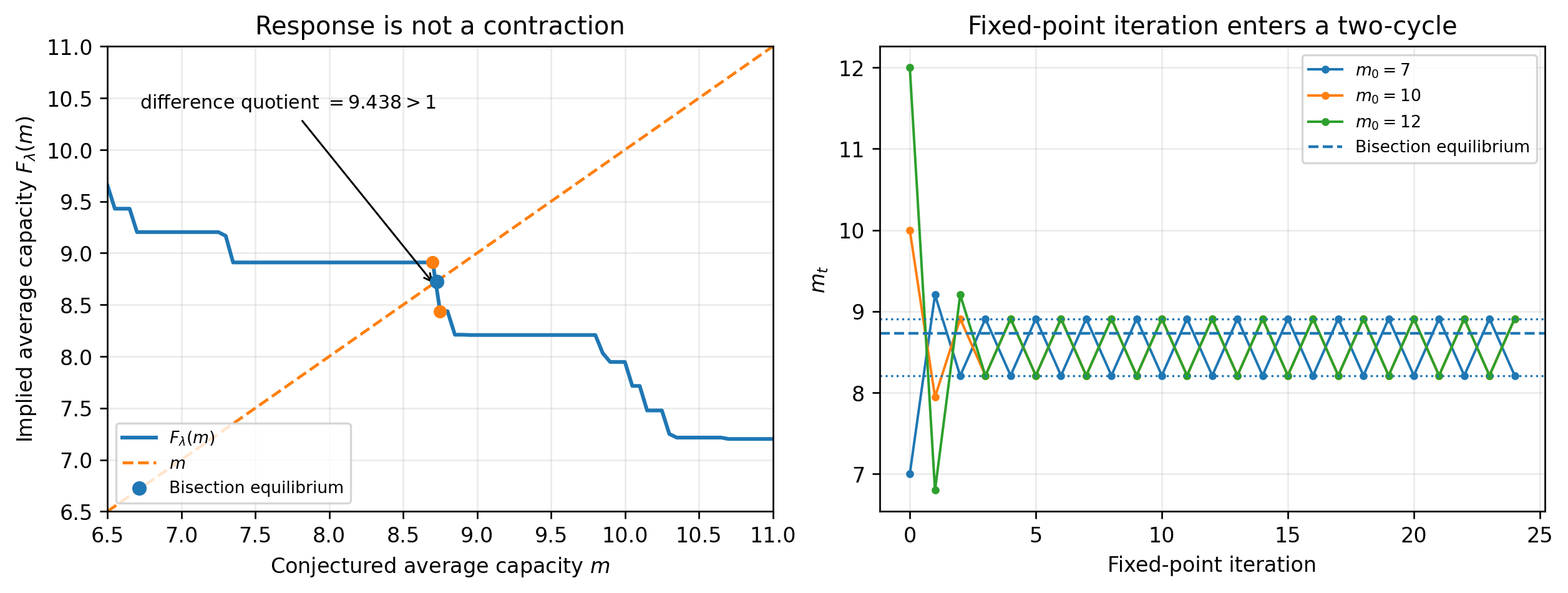}
    \caption{The equilibrium response and direct fixed-point iteration for $\alpha=50$. The left panel compares $F_\lambda(m)$ with the $45$-degree line and displays two nearby points whose difference quotient exceeds one. The right panel applies $m_{t+1}=F_\lambda(m_t)$ from three initial values. The dashed line is the equilibrium computed by bisection, and the dotted lines identify the resulting two-cycle.}
    \label{fig:oligopoly_benchmark}
\end{figure}

The left panel of Figure \ref{fig:oligopoly_benchmark} shows directly that the equilibrium response is not a contraction:
\begin{equation*}
\frac{|F_\lambda(8.75)-F_\lambda(8.70)|}{0.05}
=
9.438170>1.
\end{equation*}
The original $\alpha=45$ calibration has the same difference quotient between $m=3.70$ and $m=3.75$. Nevertheless, fixed-point iteration initialized at $m_0=7$ converges to $6.798052$ when $\alpha=45$. Thus, failure of contraction removes the usual convergence guarantee but does not imply that every fixed-point trajectory must fail.

The $\alpha=50$ calibration provides a direct non-convergence example. Numerically,
\begin{equation*}
F_\lambda(8.210214)=8.909697,
\qquad
F_\lambda(8.909697)=8.210214.
\end{equation*}
Fixed-point iterations initialized at $7$, $10$, and $12$ all enter this two-cycle. At either point, $|m-F_\lambda(m)|$ is approximately $0.6995$, so additional fixed-point iterations do not produce an approximate equilibrium. 

\subsubsection{Dynamic Ridesharing Model} \label{Sec:RideSimulations}

We use $X_1=X_2=\{0,1,2,3\}$. The first component describes the availability or remaining trip duration, and the second describes the request type. The trip durations are $d_j=j$ for $j=1,2,3$. The probability of no request is $M$, where $M$ is the fraction of available drivers, and each of the three request types has probability $(1-M)/3$. The discount factor is $0.95$ and the stopping tolerance is $10^{-4}$.

We compare payoff vectors $[0,1,1.3,5]$ and $[0,1,1.3,10]$. With the larger payoff for the longest trip, drivers reject some type-2 requests while waiting for type 3. This raises the fraction of available drivers and lowers the request probability faced by each available driver. Algorithms \ref{alg:value_iteration_mfe} and \ref{alg:bisection_q_learning} produce closely matching numerical results.

\begin{figure}[H]
    \centering
    \begin{subfigure}[b]{0.37\textwidth}
        \includegraphics[width=\textwidth, height=3.7cm]{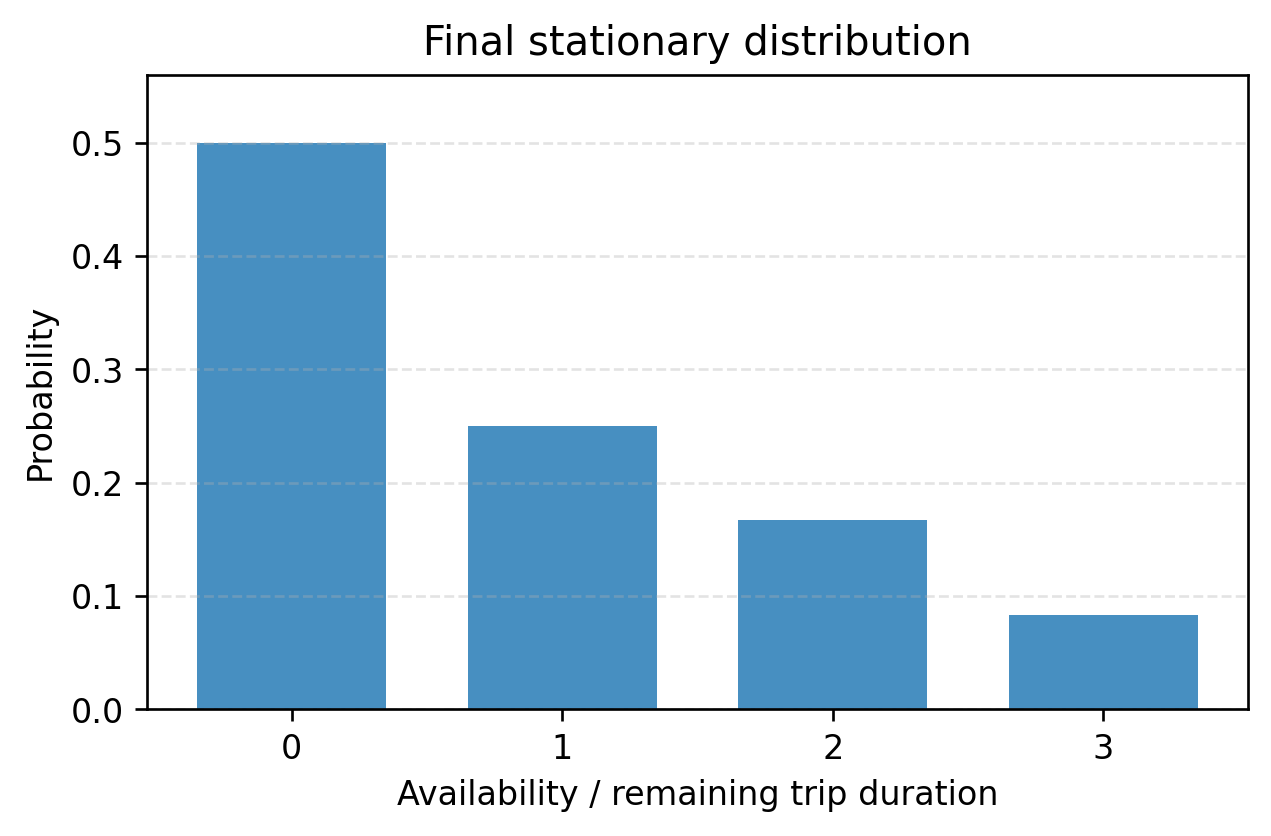}
        \caption{Long-trip payoff = 5}
        \label{fig:long5}
    \end{subfigure}
    \hfill
    \begin{subfigure}[b]{0.37\textwidth}
        \includegraphics[width=\textwidth, height=3.7cm]{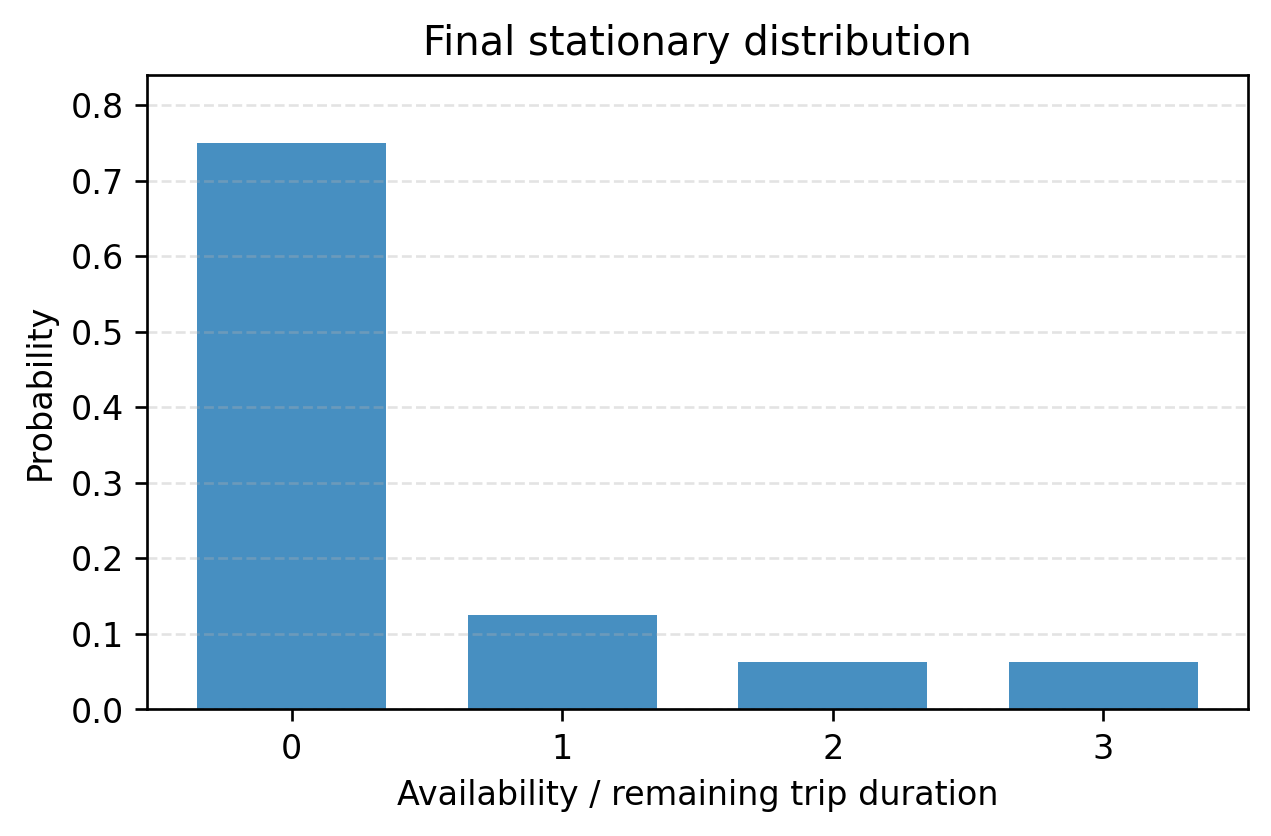}
        \caption{Long-trip payoff = 10}
        \label{fig:long10}
    \end{subfigure}
    \caption{Approximate equilibrium driver behavior under two long-trip payoffs. With payoff $10$, drivers reject some type-2 requests while waiting for the most valuable trip; with payoff $5$, they accept every request.}
    \label{fig:ridesharing}
\end{figure}

\end{document}